\documentstyle[psfig]{mn2e}

\begin{document}

\title[Debris in low-mass planetary systems]
{Herschel imaging of 61 Vir:
implications for the prevalence of debris in low-mass planetary systems}
\author[M. C. Wyatt et al.]
  {M. C. Wyatt$^1$\thanks{Email: wyatt@ast.cam.ac.uk},
   G. Kennedy$^1$,
   B. Sibthorpe$^2$,
   A. Moro-Mart\'{i}n$^3$,
   J.-F. Lestrade$^4$,
\and
   R. J. Ivison$^2$,
   B. Matthews$^5$,
   S. Udry$^6$,
   J. S. Greaves$^7$,
   P. Kalas$^{8,9}$,
   S. Lawler$^{10}$,
\and
   K. Y. L. Su$^{11}$,
   G. H. Rieke$^{11}$,
   M. Booth$^{12,5}$,
   G. Bryden$^{13}$,
   J. Horner$^{14}$,
\and
   J. J. Kavelaars$^5$,
   D. Wilner$^{15}$%,
%% B. Zuckerman$^{16}$
\\
  $^1$ Institute of Astronomy, University of Cambridge, Madingley Road,
  Cambridge CB3 0HA, UK\\
  $^2$ UK Astronomy Technology Centre, Royal Observatory, Blackford Hill,
  Edinburgh EH9 3HJ, UK\\ 
  $^3$ Centro de Astrobiolog\'{i}a (CSIC-INTA), 28850 Torrej\'{o}n de Ardoz,
  Madrid, Spain\\
  $^4$ Observatoire de Paris, 77 Avenue Denfert Rochereau, 75014, Paris, France\\
  $^5$ Herzberg Institute of Astrophysics, 5072 West Saanich Road, Victoria V9E 2E7,
  BC, Canada\\
  $^6$ Observatoire de l'Universit\'{e} de Gen\`{e}ve, 51 Ch. des Maillettes,
  CH-1290 Versoix, Switzerland\\
  $^7$ Scottish Universities Physics Alliance, University of St. Andrews,
  Physics \& Astronomy, North Haugh, St Andrews KY16 9SS, UK\\
  $^8$ Department of Astromomy, University of California, 601 Campbell Hall,
  Berkeley, CA 94720, USA\\
  $^9$ SETI Institute, 515 North Whisman Road, Mountain View, CA 94043, USA\\
  $^{10}$ University of British Columbia, Department of Physics and Astronomy,
  6244 Agricultural Road, Vancouver, BC V6T 1Z1, Canada\\
  $^{11}$ Steward Observatory, University of Arizona, Tucson, AZ 85721, USA\\
  $^{12}$ Department of Physics \& Astronomy, University of Victoria, Elliott Building,
  3800 Finnerty Rd, Victoria, BC, V8P 5C2, Canada\\
  $^{13}$ Jet Propulsion Laboratory, California Institute of Technology, 4800 Oak Grove Drive,
  Pasadena, CA 91109, USA\\
  $^{14}$ Department of Astrophysics and Optics, School of Physics, University of
  New South Wales, Sydney 2052, Australia\\
  $^{15}$ Harvard-Smithsonian Center for Astrophysics, 60 Garden Street, Cambridge,
  MA 02138, USA %\\
%  $^{16}$ Department of Physics and Astronomy, University of California, Los Angeles,
%  CA 90095, USA
}

\maketitle

%%%%%%%%%%%%%%%%%%%%%%%%%%%%%%%%%%%%%%%%%%%%%%
%%%%%%%%%%%%%%%%%%%%%%%%%%%%%%%%%%%%%%%%%%%%%%
%%%%%%%%%%%%%%%%%%%%%%%%%%%%%%%%%%%%%%%%%%%%%%
\begin{abstract}
This paper describes Herschel observations of the nearby (8.5pc) G5V
multi-exoplanet host star 61 Vir at 70, 100, 160, 250, 350 and 500 $\mu$m
carried out as part of the DEBRIS survey.
These observations reveal emission that is significantly extended
out to a distance of $>15$arcsec with a morphology that can be fitted by a
nearly edge-on ($77^\circ$ inclination) radially broad (from 30AU out to at least 100AU)
debris disk of fractional luminosity $2.7 \times 10^{-5}$, with two additional (presumably
unrelated) sources nearby that become more prominent at longer wavelengths.
Chance alignment with a background object seen at 1.4GHz provides potential for
confusion, however the star's 1.4arcsec/year proper motion allows archival
Spitzer 70$\mu$m images to confirm that what we are interpreting as \textit{disk} emission
really is circumstellar.
Although the exact shape of the disk's inner edge is not well constrained, the region
inside 30AU must be significantly depleted in planetesimals.
This is readily explained if there are additional planets outside those already
known (i.e., in the 0.5-30AU region),
but is also consistent with collisional erosion.
We also find tentative evidence that the presence of detectable debris around nearby stars
correlates with the presence of the lowest mass planets that are detectable in
current radial velocity surveys.
Out of an unbiased sample of the nearest 60 G stars, 11 are known to have planets, of
which 6 (including 61 Vir) have planets that are all less massive than Saturn, and 4
of these have evidence for debris.
The debris toward one of these planet-hosts (HD20794) is reported here for the first time.
This fraction (4/6) is higher than that expected for nearby field stars (15\%), and
implies that systems that form low-mass planets are also able to retain bright
debris disks.
We suggest that this correlation could arise because such planetary systems are
dynamically stable and include regions that are populated with planetesimals in the
formation process where the planetesimals can remain unperturbed over Gyr timescales.
\end{abstract}


%%%%%%%%%%%%%%%%%%%%%%%%%%%%%%%%%%%%%%%%%%%%%%
%%%%%%%%%%%%%%%%%%%%%%%%%%%%%%%%%%%%%%%%%%%%%%
%%%%%%%%%%%%%%%%%%%%%%%%%%%%%%%%%%%%%%%%%%%%%%
\section{Introduction}
\label{s:intro}
Main sequence stars, like the Sun, are often found to be orbited
by circumstellar material that can be categorised into two groups,
planets and debris, the latter of which is comprised of asteroids,
comets and the dust derived from them (e.g., Wyatt 2008; Krivov 2010).
Although there are 11 examples of nearby stars that are known to
have both planets and debris (e.g., Moro-Mart\'{i}n et al. 2010), there
is as yet no evidence for any correlation between the two types
of material (Greaves et al. 2004; K\'{o}sp\'{a}l et al. 2009), and
the properties of the debris disks around stars that have planets
are not found to be significantly different to those of stars without
known planets (Bryden et al. 2009).
This is usually explained as a consequence of the spatial separation
between the planets (which are typically found within a few AU) and
debris (which is typically found at 10s of AU).

Despite this spatial separation, it is still
expected that outer debris can be affected by the gravitational
perturbations of close-in planets (Mustill \& Wyatt 2009), and that
such debris can have a significant dynamical effect on the evolution of
interior planets, e.g., through promoting migration (Kirsh et al. 2009) or
by triggering instabilities (Tsiganis et al. 2005; Levison et al. 2011).
The delivery of debris to the inner planets may also be an important
source of volatiles to the planets (e.g., Horner \& Jones 2010).
Furthermore it is reasonable to expect that the initial conditions in
protoplanetary disks that favour the formation of certain types
of inner planetary system might also result in (or exclude) specific
types of debris disk (e.g., Wyatt, Clarke \& Greaves 2007; Raymond et al. 2012).
Thus the search for any correlation between the two phenomena
continues (e.g., Maldonado et al. 2012), so that light may be shed on the
interaction between planets and debris, as well as on the formation mechanism
for the two components.

This paper describes observations carried out as part of the key programme
DEBRIS (Disc Emission via a Bias-free Reconnaissance in the Infrared/Submillimetre)
on the \textit{Herschel Space Observatory}\footnote{{\it Herschel}
is an ESA space observatory with science instruments
provided by European-led Principal Investigator consortia and with important
participation from NASA.} (Pilbratt et al. 2010).
DEBRIS is an unbiased survey searching for evidence of circumstellar
dust at 100 and 160 $\mu$m toward the nearest $\sim 80$ stars of
each spectral type A,F,G,K,M (see Phillips et al. 2010 for a description of
the sample).
The first results have already shown that Herschel observations have the
potential to detect disks down to much fainter levels than previous observations,
and moreover have the resolution to resolve the disks at far-IR wavelengths
(Matthews et al. 2010; Churcher et al. 2011; Kennedy et al. 2012).

Several of the stars in the sample are known planet hosts;
others may be found to host planets in the future.
As such, the survey is ideally suited to determining
whether any correlation exists between planets and debris.
Here we focus on Herschel observations of the star 61 Vir (HD115617), which is a main
sequence G5 star.
At $8.55 \pm 0.02$ pc (van Leeuwen 2007) 61 Vir is the 8th nearest 
main sequence G star to the Sun.
This star exemplifies the merits of an unbiased survey, since the star
was relatively unremarkable at the time the survey was conceived, but
has since been found by radial velocity surveys to host 3 planets at
0.05, 0.218, 0.478AU with minimum masses (i.e., $M_{\rm{pl}}\sin{i}$) of 5.1, 18.2,
22.9 $M_\oplus$ respectively (Vogt et al. 2010), and has subsequently been the
subject of several studies on the secular dynamics of its planetary system
(Batygin \& Laughlin 2011; Greenberg \& van Laerhoven 2012)
and on its formation mechanism (Hansen \& Murray 2012).

Thus 61 Vir is one of the first of a growing number of systems around which only
low-mass planets are known, which we define here as sub-Saturn mass.
Such low-mass planets have only been discovered recently, often in multiple planet
systems, either by high precision radial velocity measurements (Lovis et al. 2006;
Mayor et al. 2011), or by transiting studies with Kepler (Borucki et al. 2011;
Lissauer et al. 2011).

Previous Spitzer observations showed that 61 Vir hosts a debris
disk with a fractional luminosity of around $2 \times 10^{-5}$
(Bryden et al. 2006), but the location of the emission was uncertain, with
estimates ranging from 8.3AU (Trilling et al. 2008) and 4-25AU
(Lawler et al. 2009) to 96-195AU (Tanner et al. 2009).
Our observations confirm the existence of a bright debris disk, and
moreover the resolution of the images permit the disk structure to be ascertained.
This allows us to consider the relationship between the disk and the planetary
system, and the implications for the system's formation and evolution.
Given that there are other systems with debris in low-mass planetary
systems (e.g., Beichman et al. 2011), we also consider whether this
discovery heralds an emerging correlation between the presence of
debris and low-mass planets (e.g., as predicted by Raymond et al. 2011).

The layout of the paper is as follows.
The observations are described in \S \ref{s:obs},
including both the Herschel observations and archival Spitzer
and radio observations that are needed to disentangle circumstellar from
extragalactic emission.
Modelling of the Herschel observations is presented in \S \ref{s:mod}
to derive the radial distribution of dust.
We find that the known planets probably do not play a primary role in stirring the
debris, and discuss the implications of the debris for the evolution of the planetary
system in \S \ref{s:pl}.
Finally a statistical analysis of an unbiased sample of nearby stars is given
in \S \ref{s:stat} to consider the possibility of a disk-planet correlation.
Conclusions are presented in \S \ref{s:conc}.

%%%%%%%%%%%%%%%%%%%%%%%%%%%%%%%%%%%%%%%%%%%%%%
%%%%%%%%%%%%%%%%%%%%%%%%%%%%%%%%%%%%%%%%%%%%%%
%%%%%%%%%%%%%%%%%%%%%%%%%%%%%%%%%%%%%%%%%%%%%%
\section{Observations}
\label{s:obs}

%%%%%%%%%%%%%%%%%%%%%%%%%%%%%%%%%%%%%%%%%%%%%%
%%%%%%%%%%%%%%%%%%%%%%%%%%%%%%%%%%%%%%%%%%%%%%
\subsection{Herschel}
\label{ss:hersch}

%%%%%%%%%%%%%%%%%%%%%%%%%%%%%%%%%%%%%%%%%%%%%%
\subsubsection{PACS}
\label{sss:pacs}
Herschel DEBRIS observations of 61 Vir were taken on 2010 August 10.
These consist of simultaneous 100 and 160~$\mu$m images taken with PACS
(Photodetector and Array Camera \& Spectrometer, Poglitsch et al. 2010),
using the small scan-map mode.
Scan-map observations have eight repeats in two scan directions
that differ by $40^\circ$, at a scan rate of 20 arcsec s$^{-1}$.
Eight 3-arcmin scan legs were performed per map with a 2-arcsec
separation between legs.
The total observing time was 890s.
An identical observation was obtained on 2011 July 3, this time with 
simultaneous 70 and 160~$\mu$m images.
The Herschel observations are summarised in Table~\ref{tab:obs}.

\begin{table}
  \caption{\emph{Herschel} observations of 61 Vir.}\label{tab:obs}
  \begin{tabular}{llll}
    ObsId & Date & Instrument & Duration \\
    \hline
    1342202551 & 2010 August 10 & PACS 100/160 & 445s \\
    1342202552 & 2010 August 10 & PACS 100/160 & 445s \\
    1342212412 & 2011 January 9 & SPIRE 250/350/500 & 721s \\
    1342223608 & 2011 July 3 & PACS 70/160 & 445s \\
    1342223609 & 2011 July 3 & PACS 70/160 & 445s \\
  \end{tabular}  
\end{table}

The data were reduced using a near-standard pipeline with the Herschel
Interactive Processing Environment (HIPE, version 7.0 build 1931, Ott 2010).
The data were pre-filtered to remove low-frequency (1/f) noise using
a boxcar filter with a width of 66~arcsec at 70 and 100~$\mu$m and 102~arcsec
at 160~$\mu$m.
Different filter widths were tested to confirm that this did not significantly
affect conclusions about disk fluxes or morphology.
The point spread function (PSF) of the PACS instrument includes power
on large scales (e.g., 10\% beyond 1~arcmin) that is removed by the filtering
process.
For point sources this can be readily accounted for in the calibration
by comparison of the aperture fluxes of bright stars with known fluxes.
For extended sources like 61 Vir, the calibration is slightly more
complicated, as described below.

The pixel scale in the maps was set to 1~arcsec at 70 and 100~$\mu$m
and 2~arcsec at 160~$\mu$m.
Since this is smaller than the native pixel scale, noise in
neighbouring pixels is correlated, and to derive the uncertainty
on any given measurement (such as flux within an aperture) the noise
per pixel in the resulting map must be interpreted appropriately
(Fruchter \& Hook 2002; Kennedy et al. 2012).
Because the DEBRIS survey comprises many stars, we can confirm that this
correction has been applied correctly, because a histogram of
(observed-predicted)/uncertainty for all observations is consistent with
a Gaussian of unity dispersion as expected.
If the histogram had been wider, for example, it would indicate that
the uncertainties had been underestimated.

%%%%%%%%%%%%%%%%%%%%%%%%%%%%%%%%%%%%%%%%%%%%%%
\begin{figure*}
  \begin{center}
    \vspace{-0.1in}
    \begin{tabular}{ccc}
      \hspace{-0.45in} \psfig{figure=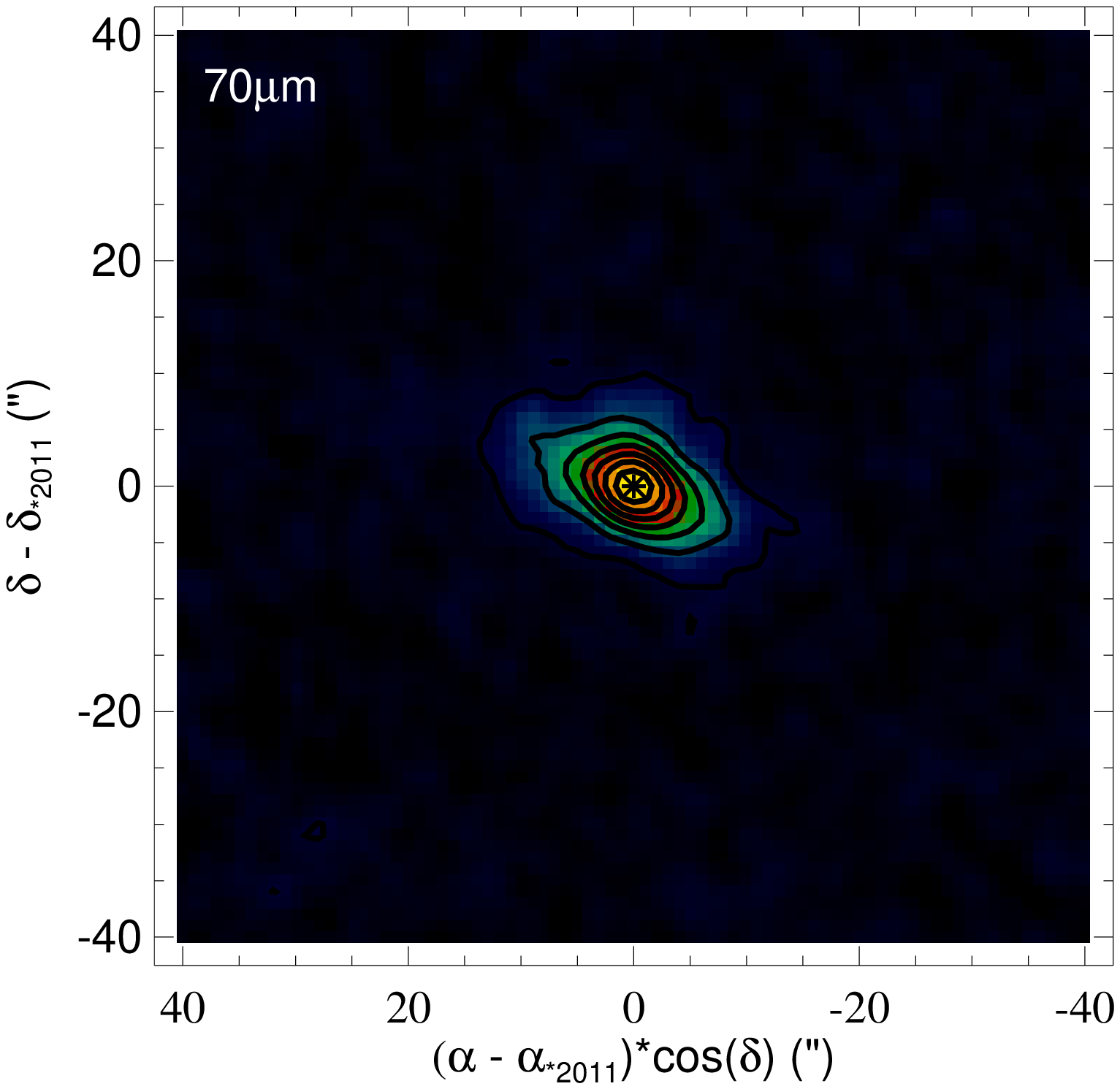,height=2.4in} &
      \hspace{-0.75in} \psfig{figure=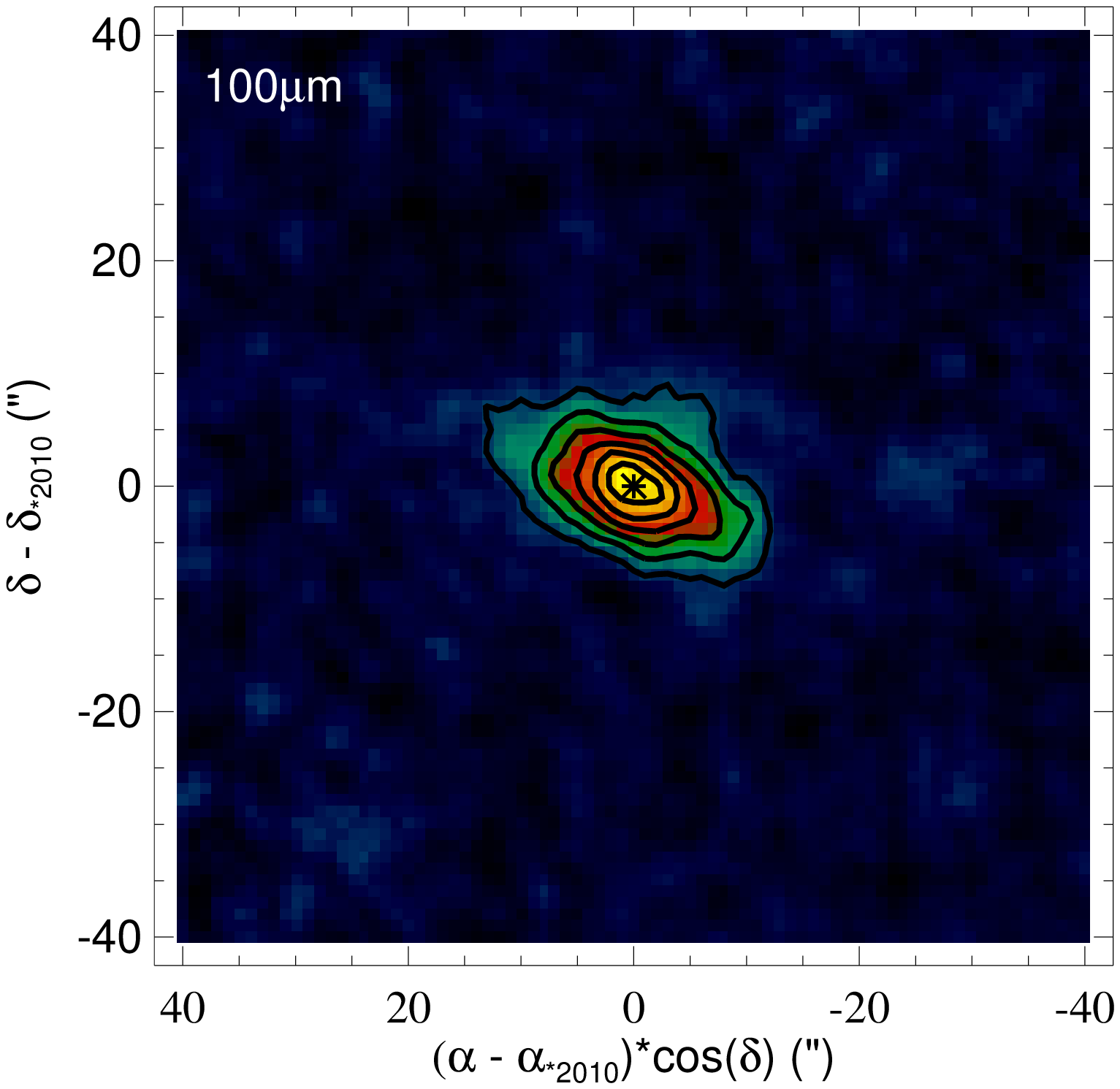,height=2.4in} &
      \hspace{-0.75in} \psfig{figure=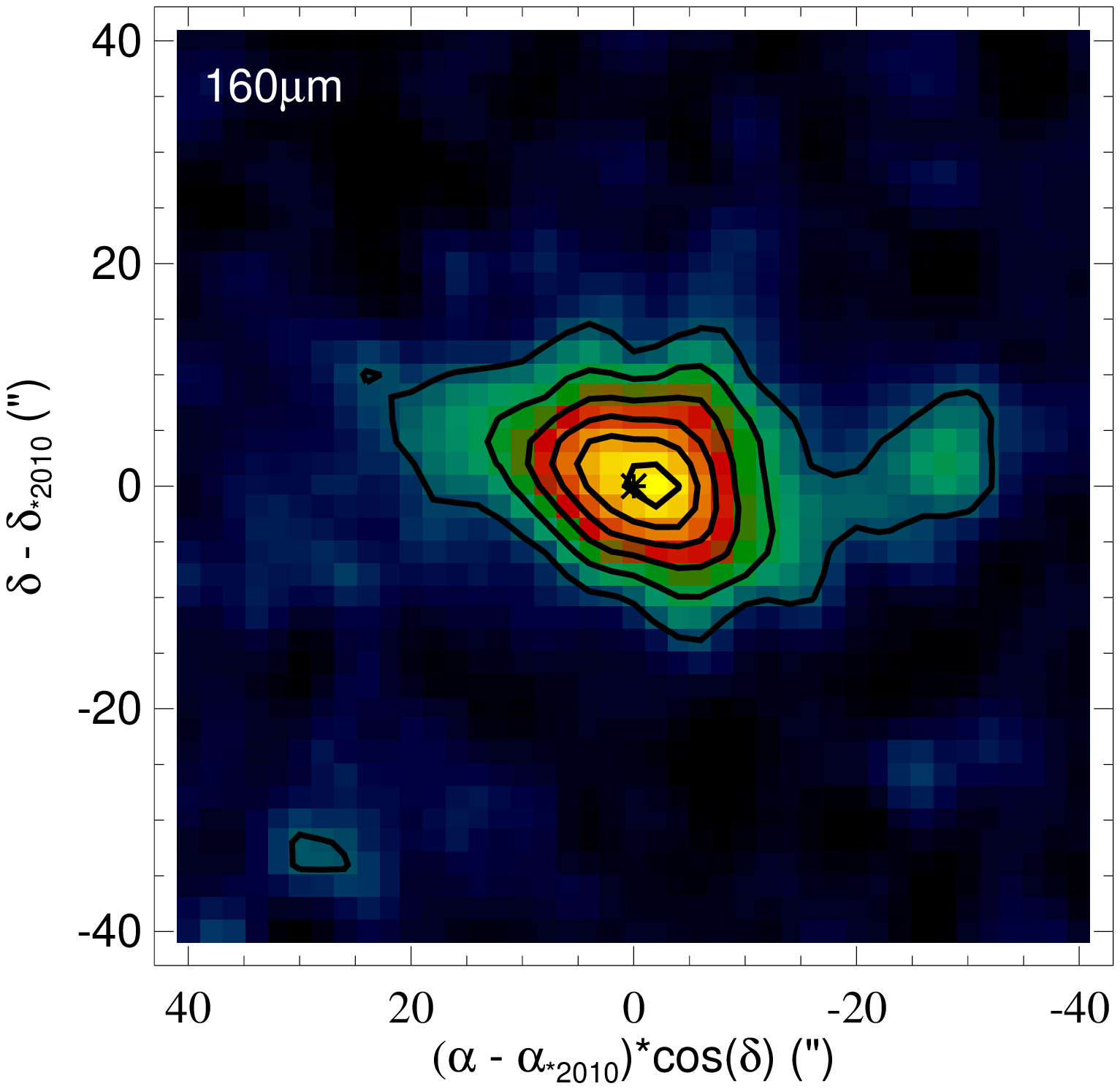,height=2.4in} \\[-0.1in]
      \hspace{-0.45in} \psfig{figure=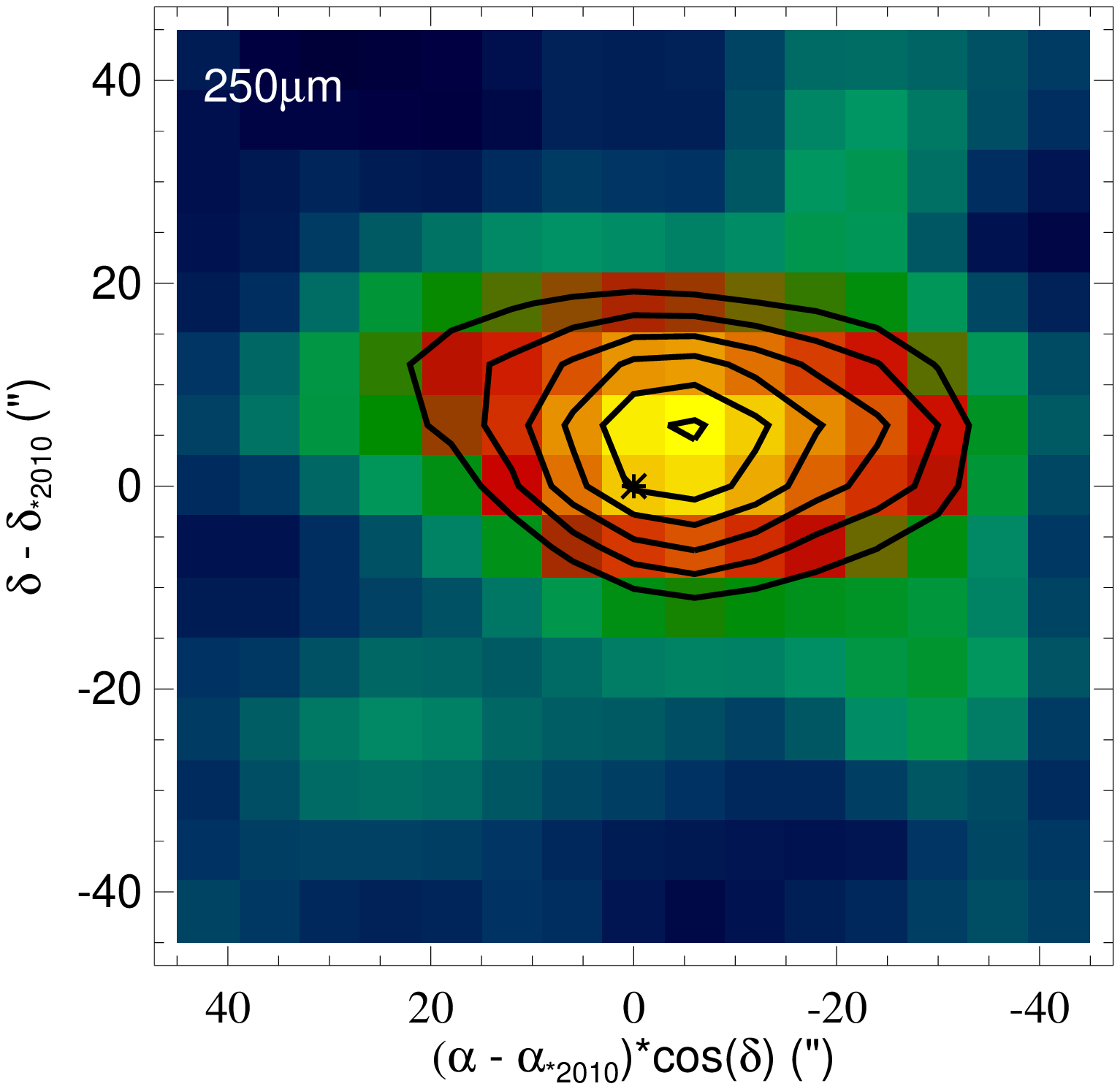,height=2.4in} &
      \hspace{-0.75in} \psfig{figure=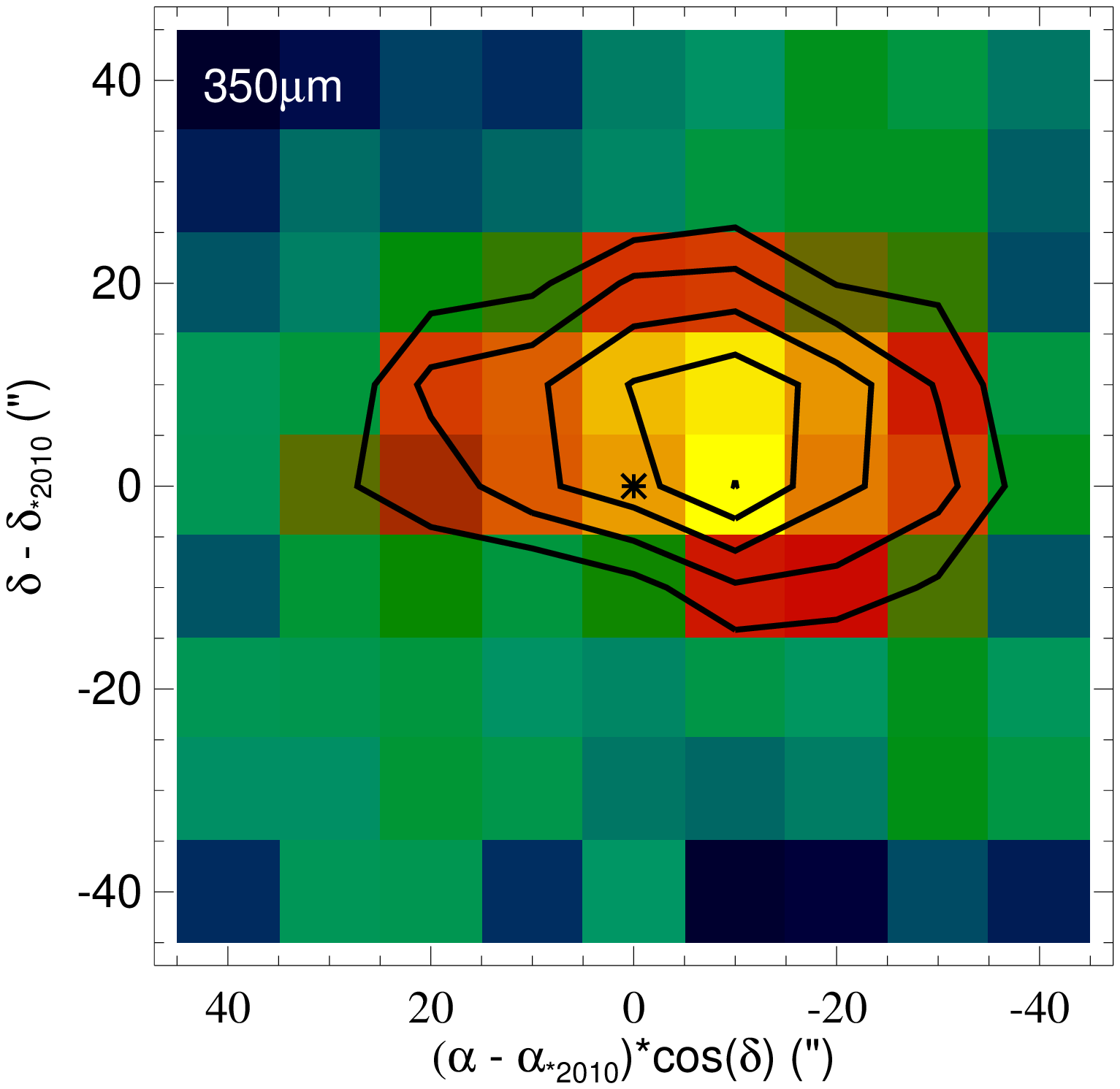,height=2.4in} &
      \hspace{-0.75in} \psfig{figure=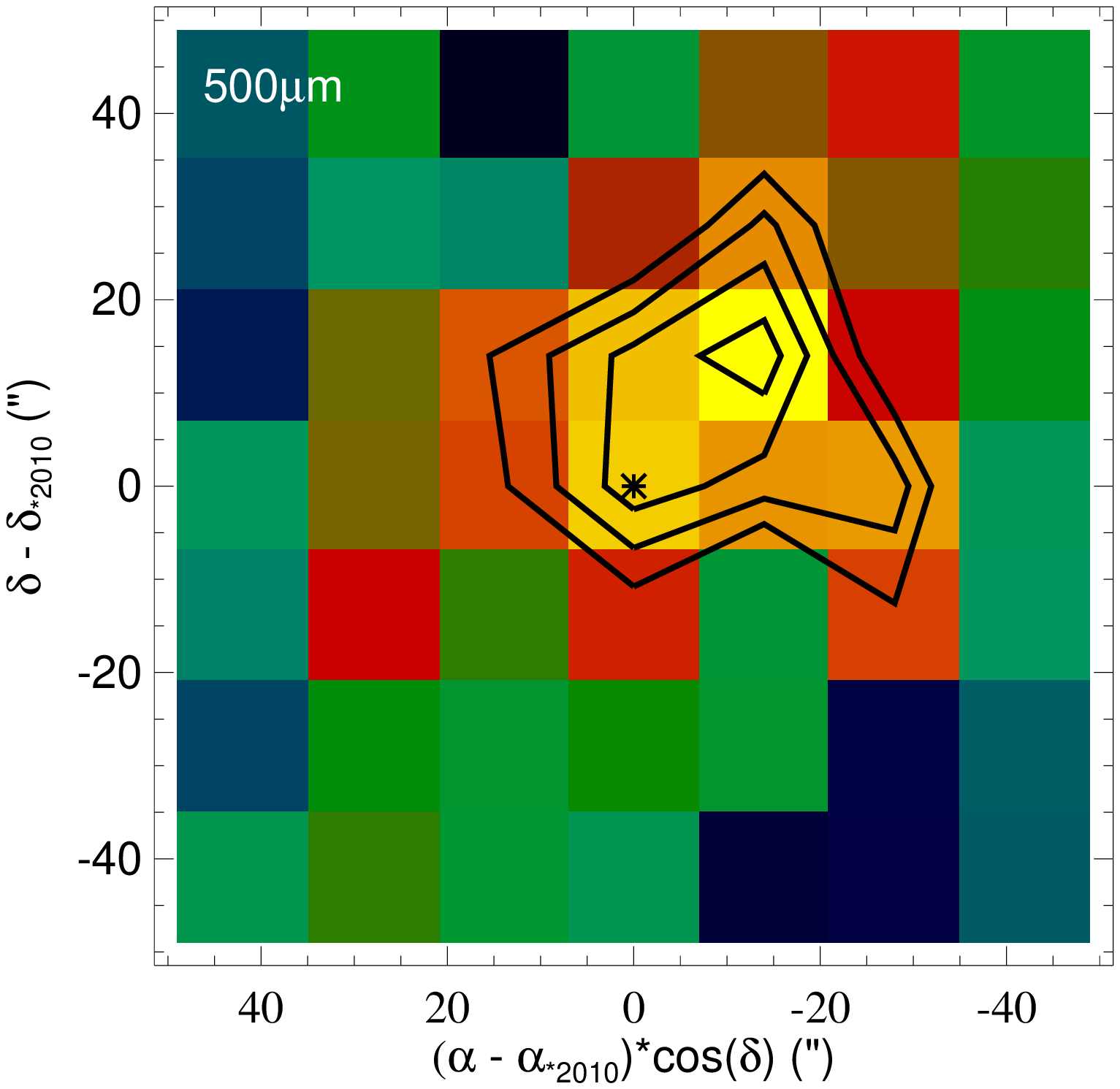,height=2.4in}
    \end{tabular}
    \caption{DEBRIS PACS and SPIRE images of 61 Vir, all centred on the star at the
    epoch of the observation, the location of which is shown with an asterisk:
    \textbf{(top left)} 70~$\mu$m,
    \textbf{(top middle)} 100~$\mu$m,
    \textbf{(top right)} 160~$\mu$m,
    \textbf{(bottom left)} 250~$\mu$m,
    \textbf{(bottom middle)} 350~$\mu$m,
    \textbf{(bottom right)} 500~$\mu$m.
    Contours shown in black are those of the presented images at levels of
    4, 13, 22, 31, 40, 49, 58 times $\sigma_{70}=0.028$ mJy/pixel at 70~$\mu$m,
    4, 9, 14, 19, 24, 29 times $\sigma_{100}=0.029$ mJy/pixel at 100~$\mu$m,
    4, 8, 12, 16, 20, 24 times $\sigma_{160}=0.052$ mJy/pixel at 160~$\mu$m,
    3, 4, 5, 6, 7, 8 times $\sigma_{250}=4.8$ mJy/beam at 250~$\mu$m,
    2, 3, 4, 5, 6 times $\sigma_{350}=6.2$ mJy/beam at 350~$\mu$m, and
    2, 2.5, 3, 3.5 times $\sigma_{500}=6.4$ mJy/beam at 500~$\mu$m,
    where $\sigma_{\lambda}$ is the pixel-to-pixel uncertainty in the images
    (which is measured from the images and so includes confusion noise),
    and the colour scale goes from $-2\sigma_{\lambda}$ to the maximum brightness.}
   \label{fig:herschel}
  \end{center}
\end{figure*}

Fig.~\ref{fig:herschel} shows the 70, 100 and 160~$\mu$m images cropped to the
region $\pm 40$~arcsec from the star.
The peak emission at 70 and 100~$\mu$m was found respectively to be offset (in different
directions) by 1.8 and 1.6~arcsec from the nominal location of the star
(13h 18m 23.52s, -18d 18m 51.6s).
Since this is consistent with the expected pointing uncertainty of Herschel
of 2~arcsec ($1\sigma$), a value that in turn is consistent with the
pointing offsets seen for stars detected in the DEBRIS sample, 
we applied a pointing correction on the assumption that the star was
at this peak location.
As the two 160~$\mu$m images were taken simultaneously with those at
70 and 100~$\mu$m, this re-registering process also determines the centre
of the 160~$\mu$m maps that were then coadded to give the presented image.

\begin{figure}
  \begin{center}
    \vspace{-0.3in}
    \begin{tabular}{c}
      \hspace{-0.5in} \psfig{figure=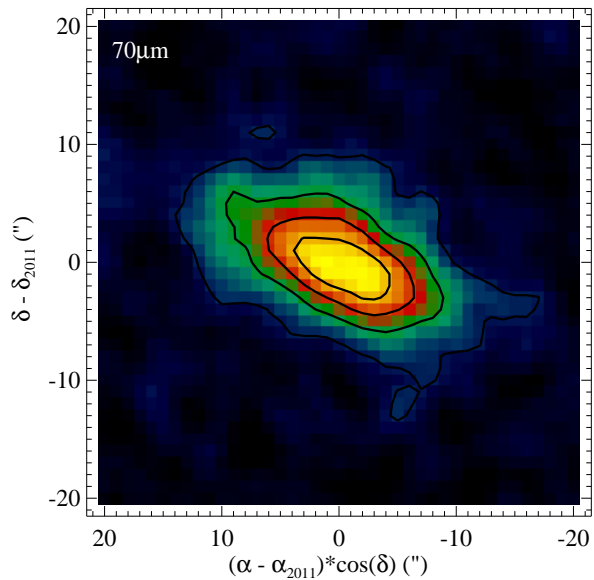,height=3.3in}
    \end{tabular}
    \caption{Star-subtracted 70~$\mu$m PACS image of 61 Vir.
    Contours are the same as in Fig.~\ref{fig:herschel}.}
   \label{fig:70ss}
  \end{center}
\end{figure}

The images at all three PACS wavelengths show a very similar
morphology, of emission that is peaked on the star, but that is
extended symmetrically about the star along a position angle of $\sim 65^\circ$.
This suggests that the star is surrounded by a disk that is inclined to our
line of sight.
After subtracting a PSF scaled to the stellar photospheric emission
(see Fig.~\ref{fig:70ss} for the resulting image at 70~$\mu$m),
elliptical Gaussian fits to the 3 images find for the FWHM of the
major and minor axes and for the position angle:
$15.9 \pm 0.36$", $7.7 \pm 0.2$", $64 \pm 1.3^\circ$ at 70~$\mu$m,
$17.4 \pm 0.7$", $8.7 \pm 0.3$", $65 \pm 1.6^\circ$ at 100~$\mu$m, and
$24.8 \pm 2$", $14.8 \pm 1$", $75 \pm 4.5^\circ$ at 160~$\mu$m.
Given that the FWHM of the PSF is 5.6, 6.7 and 11.4~arcsec at these wavelengths,
the emission is extended along both the major and the minor axes.
Quadratic subtraction of the FWHM of the PSF suggests a disk inclination of 
around $20 \pm 2^\circ$ from edge-on at 70 and 100~$\mu$m and $26 \pm 7^\circ$
at 160~$\mu$m (see \S \ref{ss:plane} for a discussion of the
implications for the planet masses);
the fact that the long wavelength emission is more extended (after this
quadratic subtraction) than that at 70 and 100~$\mu$m suggests that the 
disk may be radially broad (since longer wavelengths tend to probe
cooler material).
The uncertainty in the level of photospheric emission
to be subtracted from the images has a negligible effect on the inferred disk structure;
e.g., the resulting uncertainties in the parameters of the elliptical Gaussian fit are
smaller than those quoted above.

%%%%%%%%%%%%%%%%%%%%%%%%%%%%%%%%%%%%%%%%%%%%%%

The total fluxes were calculated from the images using aperture photometry.
A circular 20~arcsec (170AU) radius aperture was employed since, at least for 70 and 100~$\mu$m,
it was found that there was no flux present on larger scales.
The fluxes derived in this way need to be revised because a significant fraction
of the flux has been filtered out in the reconstruction of the map. 
This correction factor has been derived for point sources in the DEBRIS survey to be
16, 19, 21\% at 70, 100 and 160~$\mu$m respectively, based on how the inferred
fluxes compare with those predicted from the photosphere models for those stars for
which purely photospheric emission is confidently detected,
and we consider the uncertainty in our correction factor to be $\sim 5$\%
(i.e., the correction factor at 70~$\mu$m is $16 \pm 5$\%).
These values are consistent with the aperture correction for a 20~arcsec aperture 
quoted by the PACS team\footnote{http://herschel.esac.esa.int/twiki/bin/view/Public}
of 17, 20 and 26\%.
Although this correction factor was derived for point sources, it should also
apply to resolved sources, as long as the aperture size is larger than the
extended source, a condition that is tested for the 61 Vir maps in \S \ref{s:mod}.
The derived fluxes are
$198 \pm 3$ mJy at 70~$\mu$m, 
$152 \pm 6$ mJy at 100~$\mu$m, and 
$133 \pm 6$ mJy at 160~$\mu$m
(noting that these uncertainties need to be increased due to the systematic
effects outlined in the next paragraph).

The uncertainties in the previous paragraph have been derived from
the distribution of fluxes obtained by placing apertures of the same size
at different locations in the map.
Given the uneven coverage of the map resulting from the observing strategy,
the centres of these apertures were constrained
to lie in regions with uncertainties within 10\% of that of the region around
61 Vir.
Additional uncertainties of 1.4, 1.6 and 3.5\% for repeatability in the
different wavebands\footnote{http://herschel.esac.esa.int/twiki/bin/view/Public},
and the 5\% calibration uncertainty mentioned above, also need to be applied.
A quadratic combination of uncertainties results in $\pm 11$, $\pm 10$,
$\pm 17$ mJy at 70, 100 and 160~$\mu$m respectively.

These fluxes can be compared with those determined by MIPS.
The 70~$\mu$m MIPS bandpass is very similar to that of PACS,
and the MIPS flux has been quoted as $185.6 \pm 16.6$~mJy
(Bryden et al. 2006) and $195 \pm 8$~mJy (Trilling et al. 2008),
but since the source is resolved the flux is expected to be higher
than early estimates, with $224 \pm 8$~mJy quoted most recently
(Lawler et al. 2009).
In \S \ref{sss:mips} we reanalyse the MIPS data to give $237 \pm 10$~mJy
in a 36 arcsec radius aperture.
Although the PACS flux is lower than that measured by MIPS,
the values are consistent within plausible statistical and systematic errors
(e.g., cross-calibration between the instruments).
The 160~$\mu$m MIPS flux is given as between 89 and 141 $\pm 20$~mJy
(Tanner et al. 2009), where different values are obtained depending 
on how the observations are corrected for a spectral leak.
These fluxes are very close to that measured by PACS, and 
given the issues with the leak, the MIPS flux is not used in the
subsequent analysis.

There are two additional lower level features in the PACS images.
There is clearly a source offset $\sim 27$~arcsec to the west of the star.
This is not detected at 70~$\mu$m (a $3\sigma$ upper limit of 6mJy), it is present
above the $3\sigma$ level at 100~$\mu$m (flux of $10 \pm 2$ mJy), and is more
confidently detected at 160~$\mu$m (flux of $19 \pm 2$ mJy).
Given the low signal-to-noise, there is no evidence that this emission
is extended, and we attribute it to a background source.
There is also a protrusion to the north of the star which
occurs in the 4$\sigma$ contour at 70 and 100~$\mu$m, and in the
12$\sigma$ contour at 160~$\mu$m.
For the shorter wavelengths this level of perturbation is consistent with
that expected from noise superimposed on a symmetrical disk, but 
the deviation at 160~$\mu$m cannot be explained in this way. 
We considered whether this is caused by the shape of the PACS beam, which at
such low levels is tri-lobed (noting that the telescope position angles
for the two 160~$\mu$m images differ by just $6^\circ$ and so the PSF for the
combined image is still tri-lobed).
However, preliminary modelling showed that an asymmetry of this magnitude
could not be caused by the beam shape, and we concluded that this is a
real feature that originates either in a background source, or in a cold spot
in the disk.
The presence of such an extra component explains why the elliptical Gaussian fitting
resulted in a higher inferred inclination from the 160~$\mu$m image ($26^\circ$) than
at shorter wavelengths ($20^\circ$).

%%%%%%%%%%%%%%%%%%%%%%%%%%%%%%%%%%%%%%%%%%%%%%
\subsubsection{SPIRE}
\label{sss:spire}

%%%%%%%%%%%%%%%%%%%%%%%%%%%%%%%%%%%%%%%%%%%%%%
\begin{figure*}
  \begin{center}
    \vspace{-0.1in}
    \begin{tabular}{ccc}
      \hspace{-0.45in} \psfig{figure=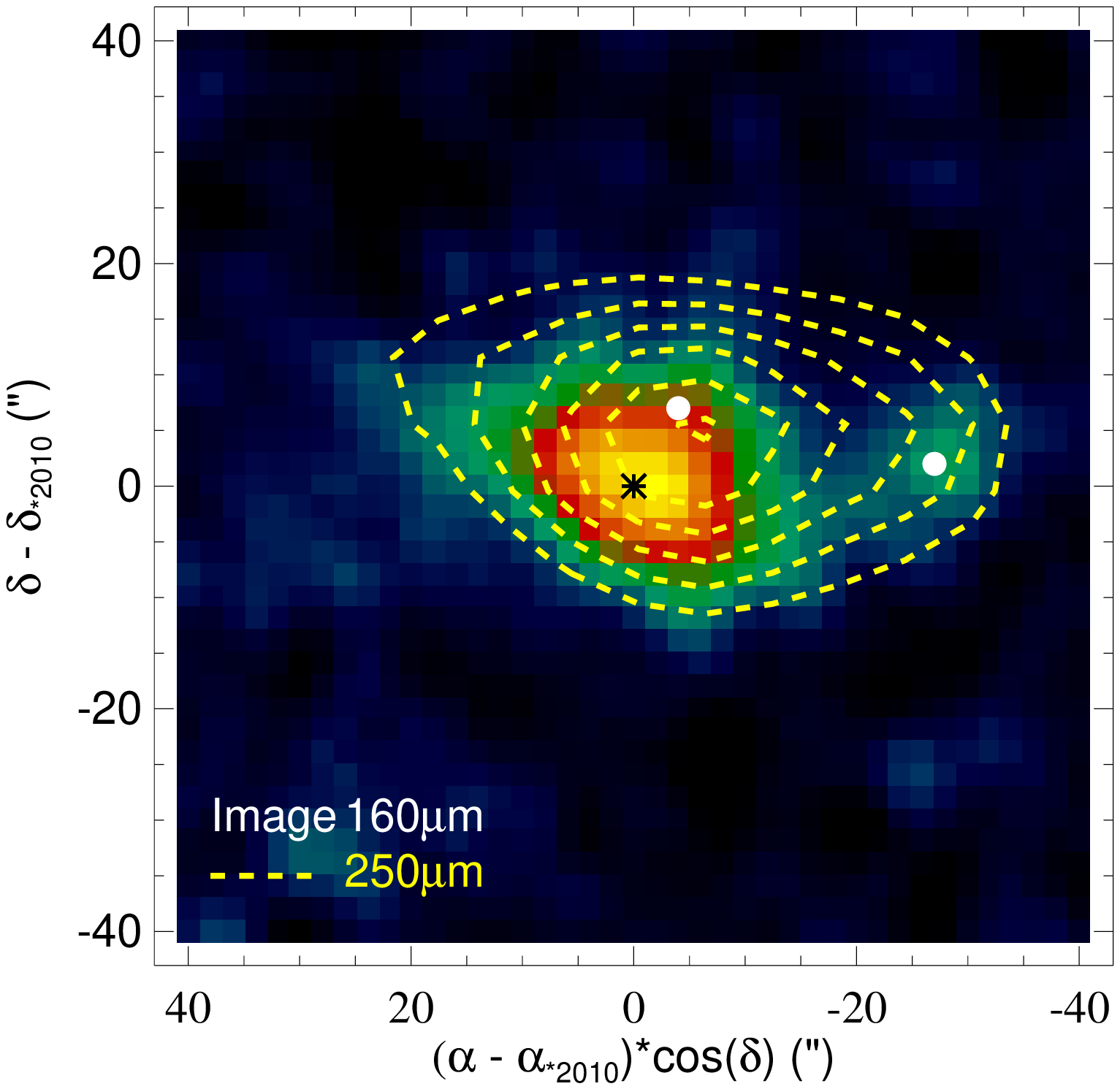,height=2.4in} &
      \hspace{-0.75in} \psfig{figure=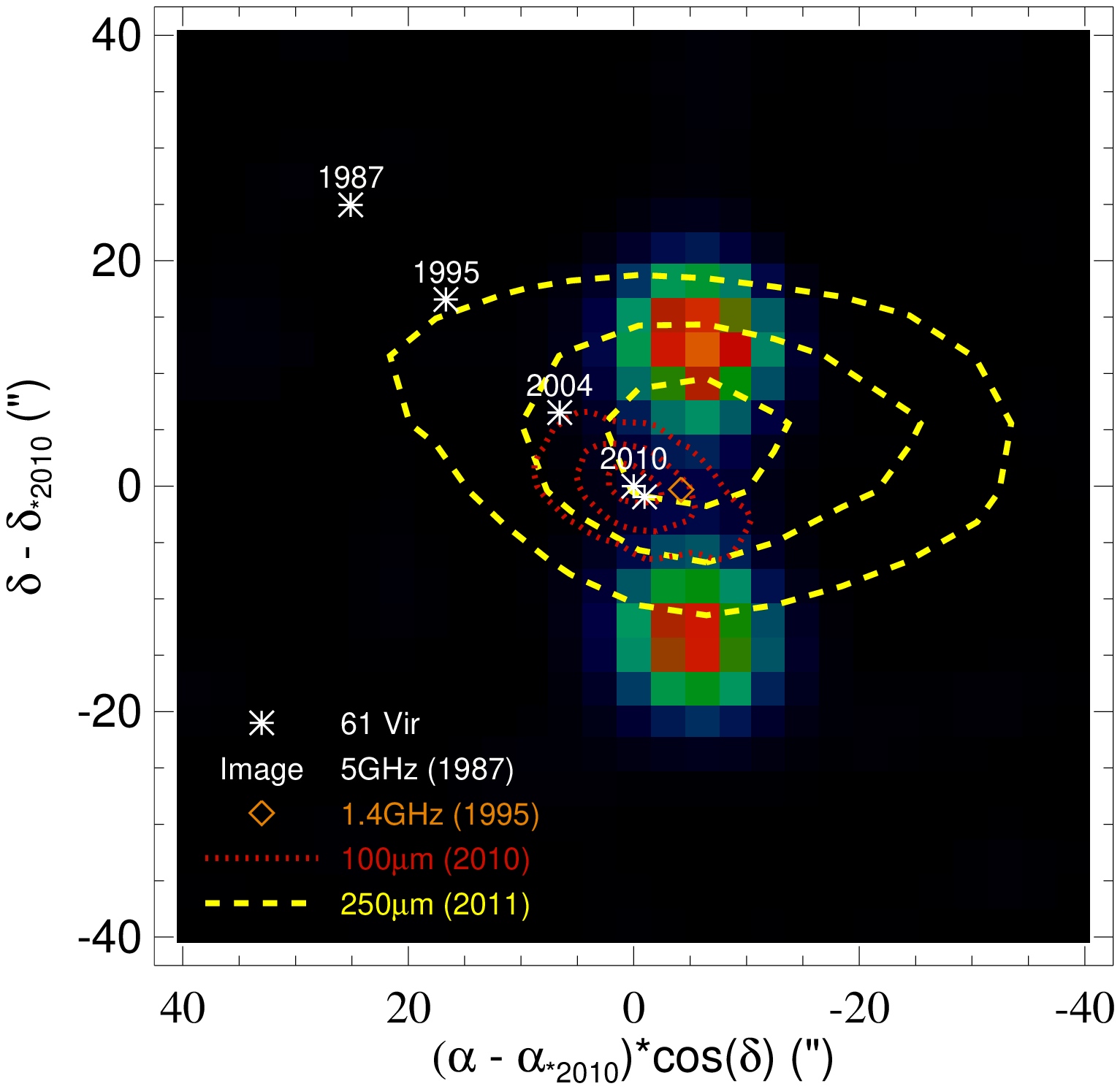,height=2.4in} &
      \hspace{-0.75in} \psfig{figure=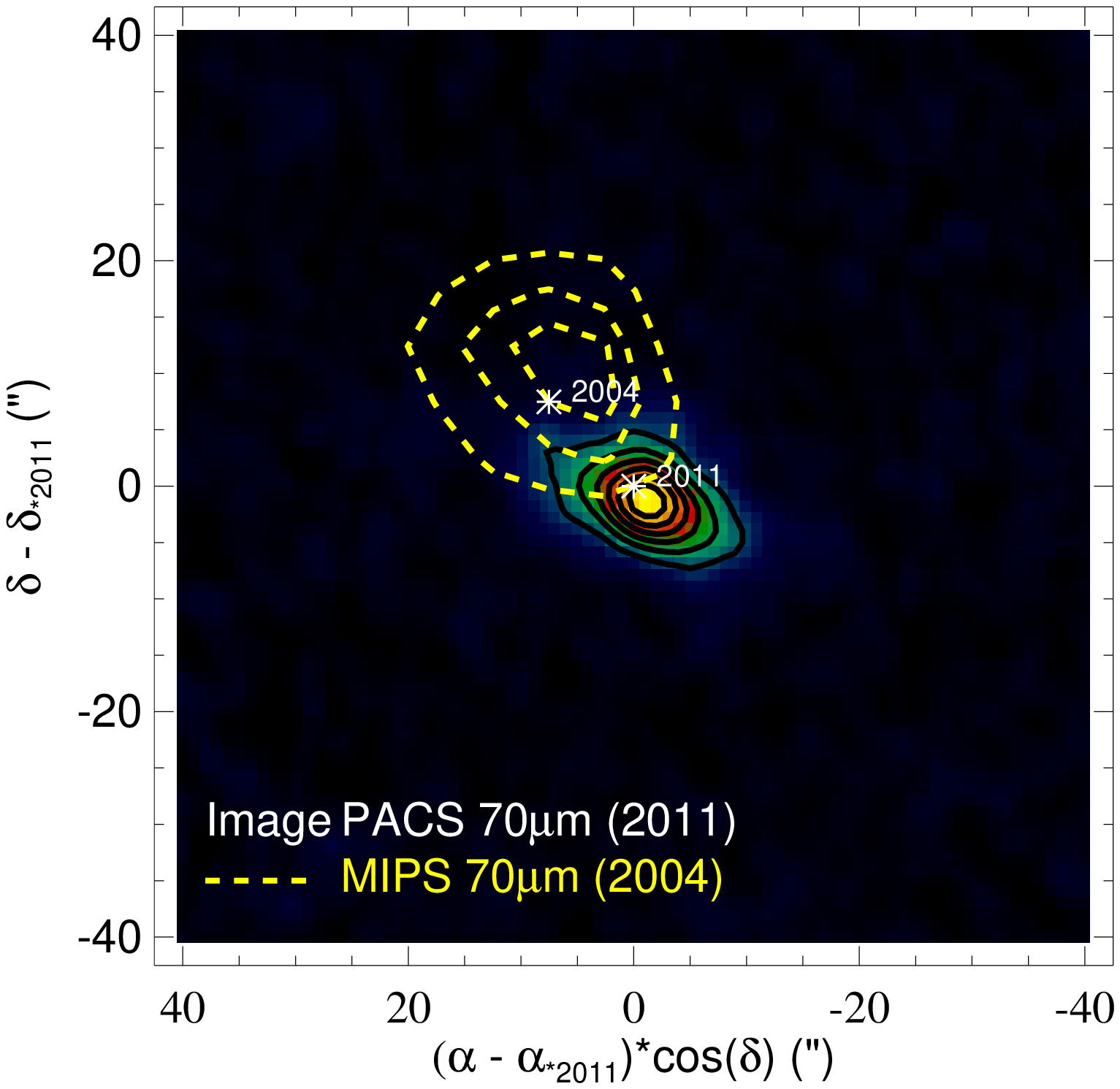,height=2.4in}
    \end{tabular}
    \caption{
    Disentangling the confusion toward 61 Vir.
    \textbf{(Left)} Superposition of the 250~$\mu$m contours onto the 160~$\mu$m
    image suggests that the 250~$\mu$m morphology is comprised of a circumstellar
    component (centred on the asterisk) and two offset sources (centred on the white dots).
    \textbf{(Middle)} The high proper motion of 61 Vir (see asterisks for the location
    of the star at the different observed epochs) confirms that double-lobed radio emission
    that is centred just 5~arcsec from 61 Vir at the epoch of the Herschel observations is
    background and not comoving with the star.
    The presented image shows 5~GHz observations from 1987, and the brown diamond shows
    the centroid of an extended 1.4~GHz source detected in 1995.
    \textbf{(Right)} The high proper motion of 61~Vir confirms that 
    the extension of the 70~$\mu$m emission seen in 2011 with PACS (image reproduced from
    Fig.~\ref{fig:herschel}) was also seen in the archival Spitzer MIPS 70~$\mu$m image
    from 2004, which is shown with yellow contours at 50\%, 75\% and 90\% of the peak value
    at 6.84~mJy/pixel.
    The high proper motion of the star confirms that this extension is comoving with the
    star, and so validates the circumstellar interpretation for the majority of the emission
    seen at 70-160~$\mu$m.}
   \label{fig:confusion}
  \end{center}
\end{figure*}

Follow-up observations were obtained on 2011 January 9 with SPIRE
(Spectral \& Photometric Imaging REceiver, Griffin et al. 2010) using
small-map mode, resulting in simultaneous 250, 350 and 500~$\mu$m images
(see Table~\ref{tab:obs}).
The data were reduced using HIPE (version 7.0 build 1931), adopting the natural
pixel scale of 6, 10, 14 arcsec at 250, 350 and 500~$\mu$m
respectively;
the beam size at these wavelengths is 18.2, 24.9, 36.3~arcsec.
The resulting images are shown in Fig.~\ref{fig:herschel}.

Emission is confidently detected at 250 and 350~$\mu$m in
the vicinity of the star, and is also found to be
extended beyond the beam at these wavelengths.
Surprisingly though, the emission is extended along a different position
angle to the extension seen by PACS, and furthermore the peak is
found to be offset significantly ($\sim 7$~arcsec) from the star
(see Fig.~\ref{fig:herschel}).
For example, a 2D Gaussian fit to the 250 and 350~$\mu$m images
finds FWHM for the major and minor axes and position angles and
centres of:
43", 22", $82^\circ$, at (5.7"E, 4.9"N), and
48", 28", $85^\circ$, at (6.9"E, 5.3"N).

Since the pointing accuracy of Herschel is still 2 arcsec for the SPIRE
observations, such a large offset is highly unlikely to be caused by
a pointing error.
Rather we prefer to attribute the observed morphology, both the offset
centre and the different position angle, to the superposition of emission
from at least two sources.
That is, the circumstellar component seen at all PACS wavelengths
and the background source 27"W of 61 Vir seen most prominently at 160~$\mu$m
with a spectrum rising toward longer wavelengths, and probably also a
source to the N of the star.
This superposition is illustrated in Fig.~(\ref{fig:confusion}, left).

This superposition makes it problematic to derive fluxes for the circumstellar
component at SPIRE wavelengths.
In this section we present total fluxes measured within apertures
large enough to encompass the emission from both sources, noting that
a significant fraction of this is expected to arise from the background object(s).
Within a 40~arcsec radius aperture, the emission at 250, 350 and 500~$\mu$m is
$120 \pm 21$ mJy, $55 \pm 17$ mJy, and $20 \pm 12$ mJy.
Again, these uncertainties have to be enhanced by a 7\% repeatability and calibration
uncertainty (see SPIRE observer's manual).
In \S \ref{s:mod} an alternative approach is used to separate the flux from
the different components by modelling the image as a circumstellar disk with offset
point sources.

%%%%%%%%%%%%%%%%%%%%%%%%%%%%%%%%%%%%%%%%%%%%%%
%%%%%%%%%%%%%%%%%%%%%%%%%%%%%%%%%%%%%%%%%%%%%%
\subsection{Ancillary data}
\label{ss:ancillary}

%%%%%%%%%%%%%%%%%%%%%%%%%%%%%%%%%%%%%%%%%%%%%%
\subsubsection{VLA}
\label{sss:vla}
To assess the possibility that there is extragalactic emission in the vicinity
of 61 Vir we searched the literature for previous radio observations.
VLA observations on 1995 January 15 taken as part of the NRAO VLA Sky Survey
(Condon et al. 1998) detected a $50.9 \pm 1.9$ mJy
source at 1.4~GHz at a location (13h 18m 23.22s, -18d 18m 51.9s), with positional
uncertainty of $\sim 1$~arcsec.
This source was noted to be extended beyond the 45arcsec beam,
with a fitted (deconvolved) FWHM of $33.2 \pm 2.0$~arcsec along a major axis at a
position angle of $3.7 \pm 2.4^\circ$ (compared with a FWHM of $<18.4$ arcsec along
the minor axis).

A search of the VLA archive found a 5~GHz scan of the region dating back to 1987 February
24 with $\sim 10$~arcsec resolution (in DnC configuration).
These data, imaged using uniform weighting (robust=-5), are shown in
Fig.~(\ref{fig:confusion}, middle) indicating a double-lobed structure. 
Fitting this as two Gaussians gives 8.0~mJy for the southerly lobe
(13h 18m 23.14s, -18d 19m 04.7s) and 8.8~mJy for the
northerly lobe (13h 18m 23.13s, -18d 18m 39.3s), both of which have deconvolved
sizes of around 7x4~arcsec;
similar results are obtained from images reduced with natural weighting.
The peak of the 1.4~GHz emission lies, to within the pointing uncertainty, directly
between the lobes.

At the epoch of the first (5~GHz) radio observations the star was offset
$\sim 39$~arcsec NE of the centre of the radio emission.
The high proper motion of 61 Vir, $-1.07$~arcsec/yr in RA and $-1.06$~arcsec/yr in Dec,
means that the star had moved 11.9~arcsec to the SW by the time of the second
(1.4~GHz) radio observations.
Thus the lack of change in the centroid of the radio emission from 1987 to 1995
implies that the radio emission was not co-moving with the star
(see Fig.~\ref{fig:confusion});
it can thus be assumed to be a background object.
Indeed, the morphology and spectral index of the radio emission are typical for a
non-thermal jet, which most likely originates from an AGN
(Condon et al. 1998) that is directly between the lobes.

The stellar proper motion has the unfortunate consequence that
the star's location at the epoch of the Herschel observations
(see \S \ref{sss:pacs})
was just $\sim 4$~arcsec E of the putative AGN.
Thus it must be considered whether the emission seen with
Herschel is associated with the extragalactic radio emission
rather than being circumstellar.
It is not possible to predict the far-IR brightness of the extragalactic emission
from its radio flux, since the far-IR/radio correlation does not hold for
radio-loud AGN.
%However, the circumstellar nature of the PACS emission is confirmed in
%\S \ref{sss:mips}.

%%%%%%%%%%%%%%%%%%%%%%%%%%%%%%%%%%%%%%%%%%%%%%
\subsubsection{MIPS}
\label{sss:mips}
The possibility that the emission interpreted as circumstellar is in fact
a background object is ruled out by archival MIPS observations taken 
at both 24 and 70~$\mu$m on 2004 June 21, that we have re-reduced 
here using an updated pipeline.
Since the star contributes the majority of the 24~$\mu$m emission, this
image does not provide information on the morphology of any circumstellar
emission.
However, the 70~$\mu$m image can be compared directly with that obtained
at the same wavelength with Herschel 7 years later.

The MIPS 70~$\mu$m observation has lower resolution (FWHM of 19~arcsec) than that
of PACS (FWHM of 5.6~arcsec), but the MIPS contours (e.g., those shown in
Fig.~\ref{fig:confusion} right) are significantly extended beyond the beam,
a point already noted in Lawler et al. (2009).
Both 70~$\mu$m images include the flux from the point-like star, though at 48.5~mJy
this contributes only around 20\% of the total emission, and the morphology seen
by MIPS does not readily identify this as a distinct component as does the higher
resolution PACS image.
However, the MIPS 70~$\mu$m emission is seen to be extended along a position angle
that is very similar to that seen in the PACS images, i.e., $56 \pm 5^\circ$, with a 2D
Gaussian resulting in FWHM of $24.5 \pm 0.6$" and $20.8 \pm 0.5$" along the major and
minor axes, respectively.

As previously noted the Herschel pointing accuracy is 2~arcsec;
to allow a proper comparison, the astrometry of the PACS image in
Fig.~(\ref{fig:confusion}, right) is that given by the Herschel pointing
(i.e., the small correction used to recentre the image in Fig.~\ref{fig:herschel}
has not been applied).
The pointing accuracy of Spitzer
\footnote{http://irsa.ipac.caltech.edu/data/SPITZER/docs/
spitzermission/missionoverview/spitzertelescopehandbook/12/}
is 1~arcsec,
and again this has not been corrected for in Fig.~(\ref{fig:confusion}, right),
explaining the 2.7~arcsec offset of the peak from the stellar location at this epoch.
We conclude that the stellar proper motion allows us to confirm that the excess
emission is centred on the star, and that both epochs see the same emission morphology
that is comoving with the star;
i.e., it is not possible to interpret this emission as a stationary background object.

Given the lower resolution of the Spitzer beam, the MIPS 70~$\mu$m image is not used
in subsequent modelling of the dust distribution, although
the total flux derived from the image using a 36~arcsec radius aperture
was reported in \S \ref{sss:pacs} as $237 \pm 10$~mJy, where
it was compared with that derived from PACS.
Virtually all the flux at this wavelength is within 20~arcsec radius, since smaller
aperture radii measure 223mJy (16~arcsec radius) and 233mJy in (20~arcsec radius).
%%% 7.674 unc in small aperture which was (16/35)^2 smaller, though aperture correction was
%%% 2.047 cf 1.2278, so unc should be 10.1? 

Photometry on the 24~$\mu$m image was used in the SED modelling.
The total flux derived from fitting a PSF to the 24~$\mu$m image is
$457.8 \pm 4.6$~mJy, compared with a stellar flux of $441.4 \pm 8$~mJy, with no evidence
for extension beyond the 5.5~arcsec beam.
Thus this re-reduction has resulted in a modest upwards revision of the flux from
the $\sim 450$~mJy reported previously (Bryden et al. 2006; Trilling et al. 2008),
and consequently there is now marginal evidence, taking uncertainties in the
photospheric contribution into account, for a small $\sim 1.8\sigma$ excess at
this wavelength.
This flux is consistent with that seen at 24~$\mu$m by IRS (see \S \ref{ss:excess} and
Fig.~\ref{fig:sed}).
There is no evidence for any flux at the location of the putative AGN, which was
13~arcsec from the star at the time of the observations;
isolated point sources above 0.51~mJy could have been detected in the MIPS image with
$>3\sigma$ confidence, but given its proximity to the star the limit on the 24~$\mu$m
AGN flux is $\sim 5$ times higher.
%%% Bryden et al. (2006): 451.1mJy observed, 491.0mJy expected
%%% Trilling et al. (2008): 449mJy observed, 433 expected.

%%%%%%%%%%%%%%%%%%%%%%%%%%%%%%%%%%%%%%%%%%%%%%
\subsubsection{STIS}
\label{sss:stis}
Archival HST/STIS images were searched for evidence either for scattered light from the
debris disk or for extragalactic emission.
The data were obtained as part of a calibration program (CAL/STIS-8419, PI Charles Proffitt),
and were not optimized for a debris disk search.
For example, images were obtained with a $1100\times588$ pixel subarray
($5.58''\times2.98''$) and contemporaneous images for the subtraction of the point spread
function were not obtained.
61 Vir was observed on 2000 March 11 and again on 2000 May14.
In both observations 61 Vir was occulted at two positions behind a tapered focal plane
occulting wedge.
At the first position the wedge is $1.0''$ wide, and 160 frames with 7 seconds
integration each were obtained.
At the second position the wedge is $1.8''$ wide, and 56 frames of 7 seconds each were
acquired.
The only difference between the two epochs of observation, aside from wedge width and
total exposure time, is that in 2000 March 11 the telescope was oriented such that north is
$102.44^\circ$ counterclockwise from the +Y axis of the detector, whereas in 2000 May 14 north is
$51.19^\circ$ clockwise from the +Y axis.
For PSF subtraction we registered the two epochs of data (for a given wedge position) and
subtracted 61 Vir from itself.
The difference in telescope position angle means that a linear feature symmetric about the star,
such as a debris disk midplane, would be offset from itself by $26.4^\circ$.

\begin{figure}
  \begin{center}
    \vspace{-0.3in}
    \begin{tabular}{c}
      \psfig{figure=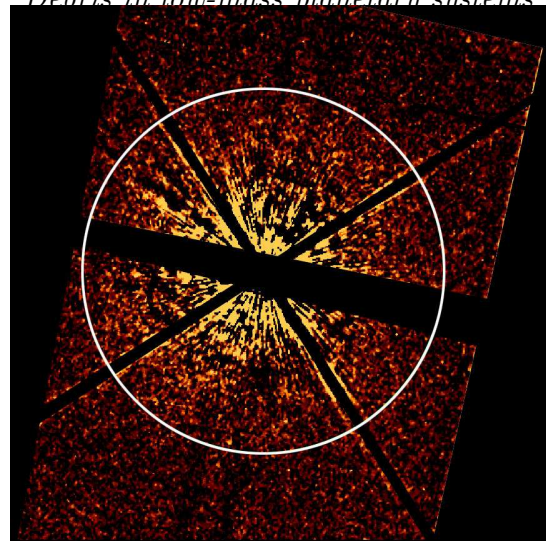,height=2.8in}
    \end{tabular}
    \caption{HST/STIS optical image of 61 Vir.
    North is up, east is left, and the white circle indicates $10''$ radius.
    We show the data with a $3 \times 3$ pixel median filter to improve sensitivity
    to faint extended sources.
    A software mask blocks orthogonal diffraction spikes and the occulting bar
    that differs from a horizontal orientation by $12^\circ$.
    With a position angle of $65^\circ$, the midplane of the 61 Vir debris disk is
    offset from the bar by another 13 degrees clockwise, but is not detected in this
    image.}
   \label{fig:stis}
  \end{center}
\end{figure}

Fig.~\ref{fig:stis} shows our PSF-subtracted image with 61 Vir occulted behind the $1.8''$ wide
occulting wedge position.
Nebulosity from circumstellar dust is not detected.
The observations made with the $1.0''$ wide occulting wedge position are more sensitive closer
to the star, and from these we estimate an approximate 3$\sigma$ sensitivity for an
extended source at 3.7$''$ radius of 20.0 mag arcsec$^{-2}$ in the VEGAMAG V-band.
At radii $\geq10''$ the data for both occulting wedges have an approximate 3$\sigma$ limiting
sensitivity of 21.5 mag arcsec$^{-2}$.
When combined with the far-IR images, these limits set useful constraints on the albedos of the dust
grains in the disk that are discussed in \S \ref{sss:mod1}, \ref{sss:mod2} and \ref{sss:mod3}.
Unfortunately, given the epoch of the STIS observations,
the field of view does not encompass the possible extragalactic source that is
currently west of 61 Vir.

%%%%%%%%%%%%%%%%%%%%%%%%%%%%%%%%%%%%%%%%%%%%%%

%%%%%%%%%%%%%%%%%%%%%%%%%%%%%%%%%%%%%%%%%%%%%%
%%%%%%%%%%%%%%%%%%%%%%%%%%%%%%%%%%%%%%%%%%%%%%
%%%%%%%%%%%%%%%%%%%%%%%%%%%%%%%%%%%%%%%%%%%%%%
\section{Modelling}
\label{s:mod}
Here we consider the structure of the disk required to explain both the
images presented in \S \ref{s:obs}, and the spectrum of the disk emission.
To do so, first the spectrum of the star is determined (\S \ref{ss:star}),
then additional data needed to constrain the shape of the disk's
emission spectrum is described (\S \ref{ss:excess}), before 
modelling of all the observational data is undertaken in \S \ref{ss:mod}.

%%%%%%%%%%%%%%%%%%%%%%%%%%%%%%%%%%%%%%%%%%%%%%
%%%%%%%%%%%%%%%%%%%%%%%%%%%%%%%%%%%%%%%%%%%%%%
\subsection{The star}
\label{ss:star}
The main sequence G5 star 61 Vir has an age estimated as $4.6 \pm 0.9$ Gyr using
gyrochronology (Wright et al. 2011; Vican 2012), a mass of $0.88M_\odot$ and a metallicity
of [Fe/H]$=-0.02 \pm 0.01$ (Sousa et al. 2008).
%%% (Lachaume et al. 1999; Cayrel de Strobel et al. 2001; Wright et al. 2004).
%%% NB 4.58Gyr from gyrochronology (Wright et al. 2011; Vican et al. submitted)
%%% and 7.35Gyr from chromospheric activity (Wright et al. 2004)
Its spectrum was determined by fitting PHOENIX models from the Gaia grid
(Brott \& Hauschildt 2005; Hauschildt \& Baron 2009) to the observational data
available for the star at wavelengths shorter than 12~$\mu$m where we can be
confident that there is minimal contribution from the disk.
This fit includes fluxes from Tycho2 (B$_{\rm{T}}$,V$_{\rm{T}}$), 2MASS (J,H,K$_{\rm{s}}$), AKARI (9~$\mu$m),
IRAS (12~$\mu$m) (see Kennedy et al. 2012 for more details on this procedure).

A best fit was found with $T_{\rm{eff}}=5602$K scaled to a luminosity of $0.84L_\odot$.
Photospheric fluxes based on this model have a $1\sigma$ uncertainty of $\pm 2$\%.
The same process was applied to other stars from the DEBRIS survey (Kennedy et al.
2012), and a comparison of the predicted fluxes to those observed at longer
wavelengths for stars where a disk is not present provides confidence that this
uncertainty is reasonable.
The predicted stellar spectrum is shown in blue on Fig.~\ref{fig:sed}, and fluxes
predicted in the Herschel bands, with no colour correction, are
50.9, 25.1, 10.0, 4.0, 2.0, 1.0 mJy at
70, 100, 160, 250, 350 and 500 $\mu$m, respectively,
while for MIPS 24~$\mu$m the stellar flux is predicted at 441.4~mJy.

%%%%%%%%%%%%%%%%%%%%%%%%%%%%%%%%%%%%%%%%%%%%%%
%%%%%%%%%%%%%%%%%%%%%%%%%%%%%%%%%%%%%%%%%%%%%%
\subsection{Excess spectrum}
\label{ss:excess}
\begin{figure}
  \begin{center}
    \begin{tabular}{c}
      \psfig{figure=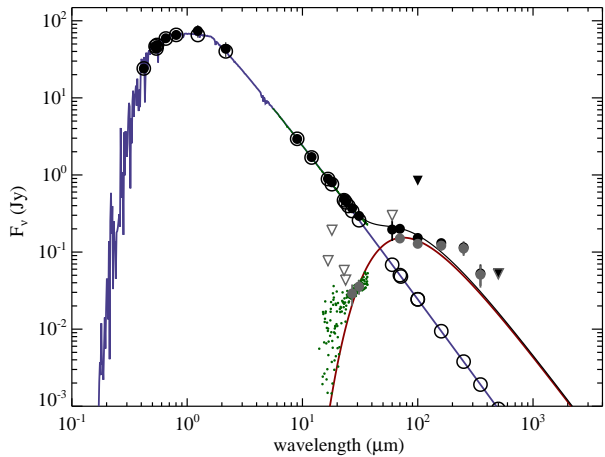,height=2.4in} 
    \end{tabular}
    \caption{
    Spectral energy distribution of 61 Vir.
    Total fluxes are black circles, open circles are predicted photospheric fluxes,
    grey circles are photosphere subtracted (i.e., excess) fluxes;
    upper limits are plotted as downward pointing triangles (grey and open for excess
    fluxes and black and solid for total fluxes).
    Data points from the pre-binned IRS spectrum are shown in green beyond 10~$\mu$m.
    The photospheric model is shown with a blue line, and is a fit to data $<12$~$\mu$m.
    The longer wavelength excess fluxes are fitted as a 68K black body of fractional
    luminosity $2.7 \times 10^{-5}$ that is shown in red.
    The total flux from the models (photosphere plus black body) is shown with a black
    solid line.}
   \label{fig:sed}
  \end{center}
\end{figure}

The spectrum of the excess (i.e., the observed flux less that expected from the photosphere)
is plotted in Fig.~\ref{fig:sed}.
In addition to the fluxes from \S \ref{s:obs}, this plot 
includes the IRAS 60~$\mu$m flux measured using
SCANPI\footnote{http://scanpiops.ipac.caltech.edu:9000} of
$210 \pm 60$~mJy, and importantly also includes the Spitzer IRS
spectrum (AOR 15998976) that was first presented in Lawler et al. (2009).
Here we use the more recent CASSIS reduction (The Cornell AtlaS of Spitzer/IRS
Sources, Lebouteiller et al. 2011), except that the spectra from the individual
modules were aligned and then the whole spectrum was scaled so that the flux in the
8-9~$\mu$m range matched that of our model stellar spectrum (\S \ref{ss:star}).
% 5% is normal I think, but Lawler got 1sigma=8%
This is because the 8\% calibration accuracy of IRS (Lawler et al. 2009)
is lower than that of our photospheric model, and the shape of the IRS spectrum
indicates that there is no contribution from the disk at such wavelengths
(e.g., Lawler et al. 2009 report an excess of $-3.6 \pm 1.4$mJy from the
8.5-12~$\mu$m range).
To aid the modelling, the IRS spectrum was averaged over 7 bins $2-4$~$\mu$m
wide, with the uncertainty in the measurement taken from the distribution of
measurements within that bin summed quadratically with the 2\% calibration
uncertainty of the spectrum (Beichman et al. 2006).
For example, this process finds a flux of $270 \pm 10$~mJy in the range 30-34~$\mu$m,
which is 24.4~mJy higher than the predicted photospheric flux, and can be compared
with a $30.8 \pm 2.0$~mJy excess reported in Lawler et al. (2009). 

%%%%%%%%%%%%%%%%%%%%%%%%%%%%%%%%%%%%%%%%%%%%%%

The overall shape of the excess spectrum resembles that of a black body.
Indeed the whole spectrum can be fitted reasonably well by a single temperature at 68K
with a fractional luminosity $f = 2.7 \times 10^{-5}$ (see Fig.~\ref{fig:sed}).
This fit also highlights where the spectrum departs from a black body.
The most prominent discrepancy occurs at $>160$~$\mu$m, where the observed fluxes are
higher than those of the black body model.
While such discrepancies could indicate that the disk spectrum contains multiple
temperature components, or indeed features associated with specific dust compositions,
it is more likely that these longest wavelength fluxes are contaminated with what
is assumed to be background emission, since they include all flux within 40 arcsec
of the star (see \S \ref{sss:spire});
these contaminating sources are accounted for in the modelling of \S \ref{ss:mod}.

One thing that is clear from the SED, however, is that there is no evidence for
a hot ($\gg 68$K) component suggesting that the inner regions are relatively empty.
This is supported by the star subtracted 70$\mu$m image (Fig.~\ref{fig:70ss}).
Although an inner hole is not evident as a decrease in surface brightness in
the central pixels, the resolution is such that the central pixels are expected to
include flux from the outer regions of the disk.
However, the image would be expected to be much more centrally peaked if there was dust
close to the star, and this point is quantified through modelling in \S \ref{ss:mod}.

%%%%%%%%%%%%%%%%%%%%%%%%%%%%%%%%%%%%%%%%%%%%%%
%%%%%%%%%%%%%%%%%%%%%%%%%%%%%%%%%%%%%%%%%%%%%%
\subsection{Models}
\label{ss:mod}
The aim of the modelling is to arrive at an axisymmetric circumstellar disk
structure that fits all of the images in Fig.~\ref{fig:herschel} as well as the
spectrum in Fig.~\ref{fig:sed}.
Clearly this will not be possible at the longer wavelengths without also including
background sources.
Given the interpretation in previous sections, the simplest morphology for the
background sources is that there are two point sources, one $\sim 27$~arcsec to the
west, and another $\sim 10$~arcsec to the north.
Thus, in the models all images include such sources at offsets from the star that are free
parameters (but that are the same for all wavelengths), with fluxes at each wavelength
that are also free.
All of these parameters are determined by a best fit to the images, however we concluded
that the quality of the 500~$\mu$m observations was not sufficient to derive fluxes for the
individual background sources at this wavelength.
The models also include emission from the point-like star at the level derived
for the photosphere (\S \ref{ss:star}).

Different models are tried for the disk profile, discussed in more detail below.
However, there are some common elements between all models, and how the models are
assessed, that are described here.
First, no attempt is made to constrain the grain properties or size distribution, since
in the absence of spectral features this will be degenerate with other assumptions.
Rather it is assumed that the emission spectrum of the disk at any given radius resembles
a modified black body that is defined by its temperature, $T$, and the wavelength beyond
which the spectrum is modified, $\lambda_0$, and the modification to the spectral slope
$\beta$ (i.e., the spectrum has an additional factor $\propto (\lambda_0/\lambda)^\beta$
for wavelengths $\lambda>\lambda_0$).
The parameters $\lambda_0$ and $\beta$ are left free, but given the level of confusion at
sub-mm wavelengths they are poorly constrained, as discussed further below. 
The temperature is assumed to be $T = f_{\rm{T}} T_{\rm{bb}}$, where $T_{\rm{bb}} = 266 r^{-1/2}$
is the temperature (in K) that black bodies would attain at a distance $r$ (in AU) for this
stellar luminosity, and $f_{\rm{T}}$ is a free parameter. 
It was not found to be necessary to change the $-1/2$ power law radial dependence of
temperature in the models.
In models that include multiple components, the different components were
sometimes allowed to have different properties ($f_{\rm{T}}$, $\lambda_0$, $\beta$).

The emission from different annuli in the disk model was then determined by the
vertical geometrical optical depth distribution $\tau(r)$, that had different assumptions
in the different models.
An image of the model was made by integrating the flux along the line-of-sight of
different pixels given 2 free parameters that define the orientation of the orbital plane
(the inclination to our line-of-sight, $I$, and the position angle, $PA$);
if the pixels undersampled the disk structure a denser grid of pixels was used,
and then rebinned, to ensure that all of the disk flux was captured.
The background sources and the star were added to individual pixels, and the
whole image convolved with an appropriately oriented telescope PSF (e.g., Kennedy et al. 2012).

To quantify the success of the model, the model images were subtracted from the
observed images, then divided by the $1\sigma$ pixel-to-pixel uncertainty given in
Fig.~\ref{fig:herschel}, following which the resulting image was squared.
This was then used to derive a $\chi^2$ by summing over pixels near the
disk\footnote{Near here means those pixels likely to contain non-zero flux.
These were chosen by taking an observed image and setting all pixels over $3\sigma$
to 1, and setting those below $3\sigma$ to 0, then convolving with the PSF and retaining
pixels $>0.1$.
This procedure yields a mask that is somewhat larger than the disk, ensuring that
models are also penalised for being too extended.}.
A $\chi^2$ was also derived from the SED, but only including the wavelengths where
the photometry was not derived from one of the Herschel images, i.e., including
the AKARI 18~$\mu$m, MIPS 24~$\mu$m, and the IRS spectrum (but not the MIPS 70~$\mu$m
as previously noted).

The $\chi^2$ for the different images and that from the SED were then summed
using the following weighting.
Each photometric data point from the literature received a relative weight of 1.
Images were weighted by the area above the $3\sigma$ flux level divided by the area
above the same flux level of the PSF scaled to the image peak.
With this prescription, an unresolved image has the same weight as a photometric data point,
and resolved images are weighted by the number of resolution elements that the source covers.

Since the location of the star in the images is not well constrained,
this location is also left as a free parameter for the 3 observations
(100/160, 70/160, and 250/350/500).
The free parameters of the model are thus:
6 for the location of the star,
4 for the offsets of the two background sources,
12 for the fluxes of the background sources at the 6 wavelengths,
2 for the orientation of the disk plane,
3 for the dust properties ($f_{\rm{T}}$, $\lambda_0$, $\beta$),
plus those describing the spatial distribution.

With so many parameters, finding the best fit by grid methods is prohibitive.
Instead, we initially find reasonably fitting models by varying parameters by hand, which
are further refined with least squares minimisation.
The minimisation step means that our models at least lie at a local minimum in parameter
space, but does not exclude the possibility of a better fit in a different region of
parameter space.
Nevertheless, since the models provide a reasonable fit to the observations, we believe
that they are at least representative of the variety of models that can fit the data.
However, we prefer not to trust the uncertainties in the model parameters returned by the
minimisation process, and quote these based on other considerations as outlined below.

%Lawler's dust models give 4-25AU from IRS and MIPS70 (120-47K), whereas
%those from Tanner et al. gave 96-195AU (55-45K)

%%%%%%%%%%%%%%%%%%%%%%%%%%%%%%%%%%%%%%%%%%%%%%
\subsubsection{Single extended component}
\label{sss:mod1}

%%%%%%%%%%%%%%%%%%%%%%%%%%%%%%%%%%%%%%%%%%%%%%
\begin{figure}
  \begin{center}
    \vspace{-0.1in}
    \begin{tabular}{c}
      \hspace{-0.1in} \psfig{figure=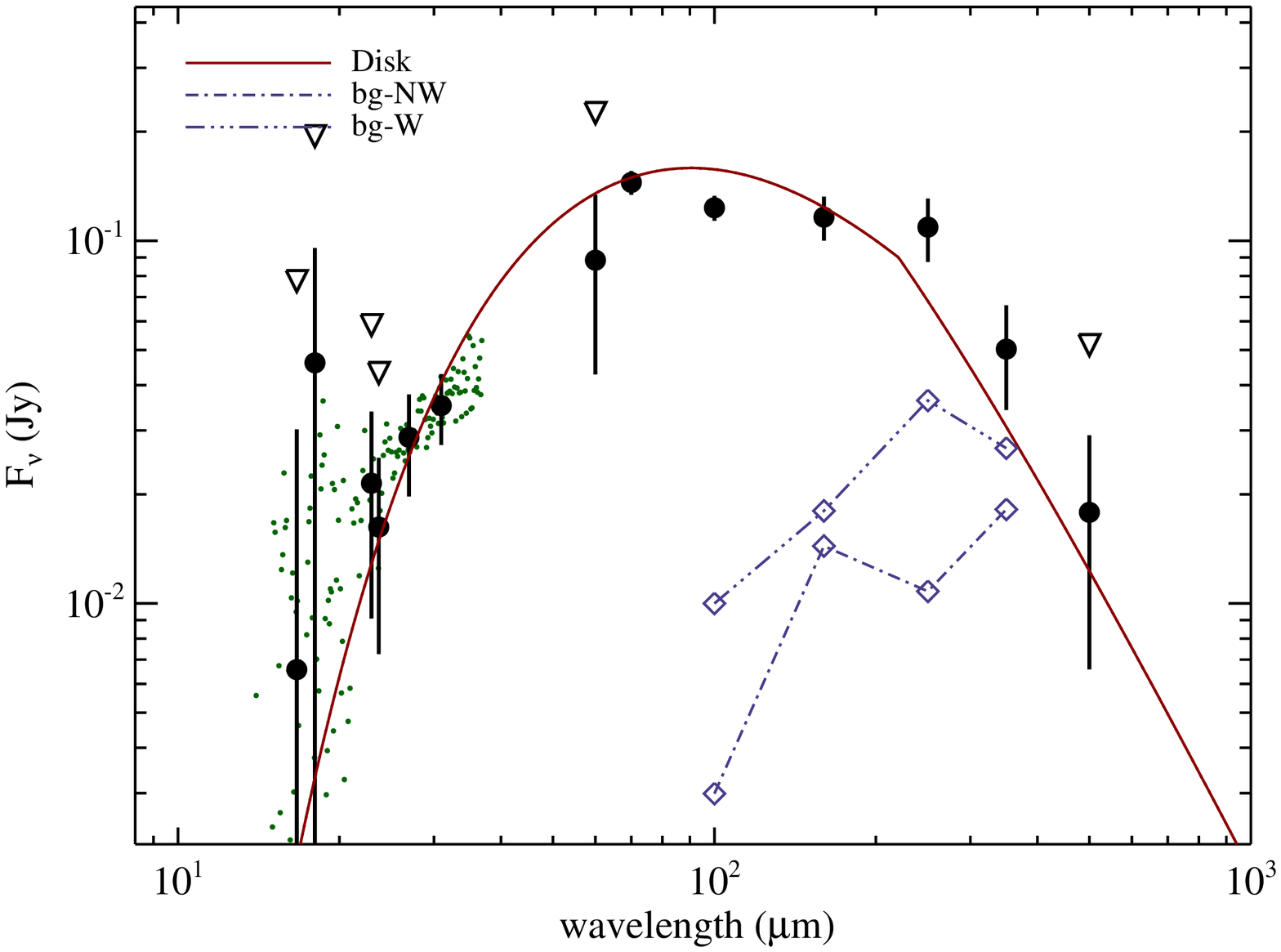,height=2.0in} \\
      \hspace{-0.1in} \psfig{figure=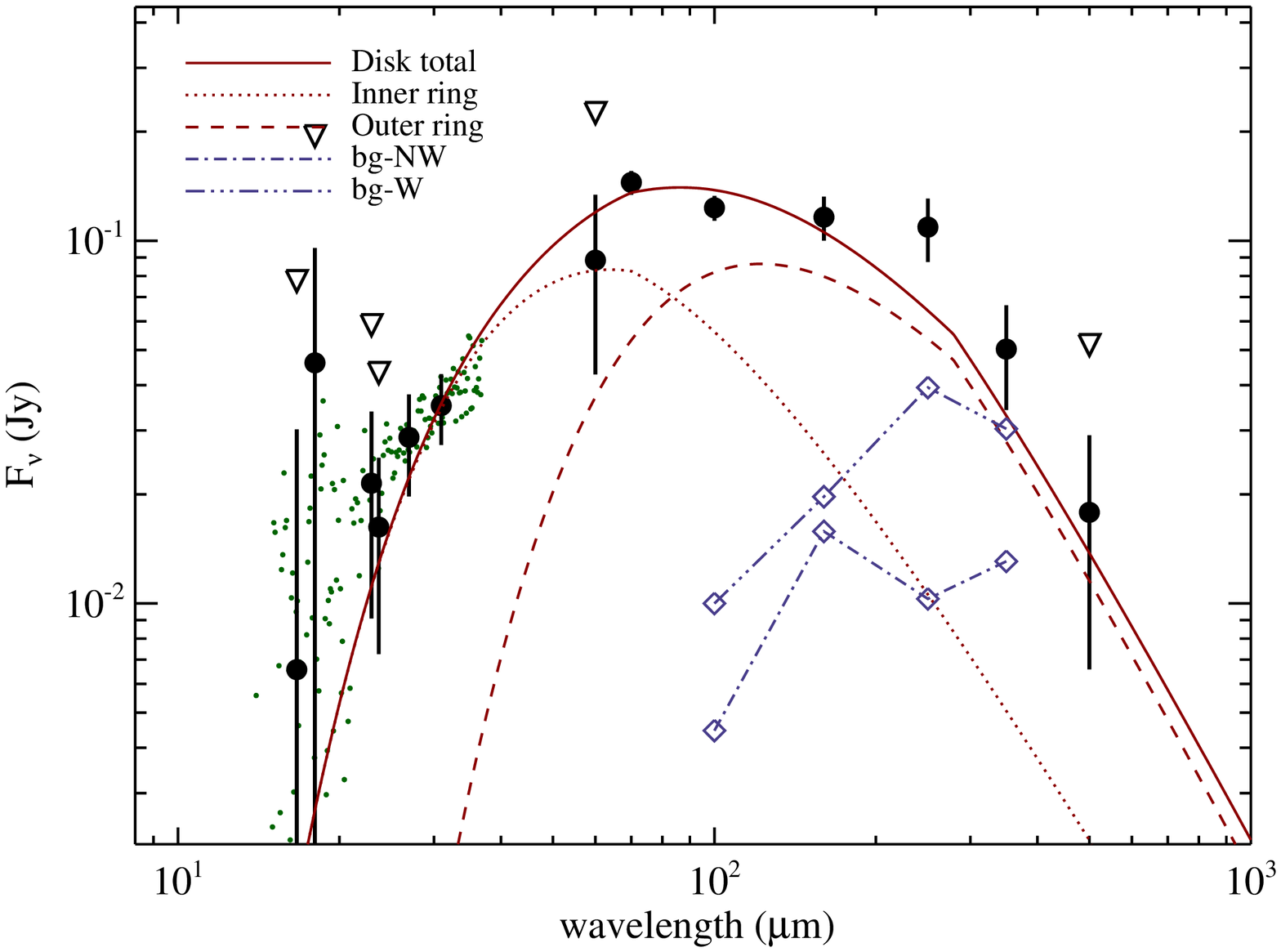,height=2.0in} \\
      \hspace{-0.1in} \psfig{figure=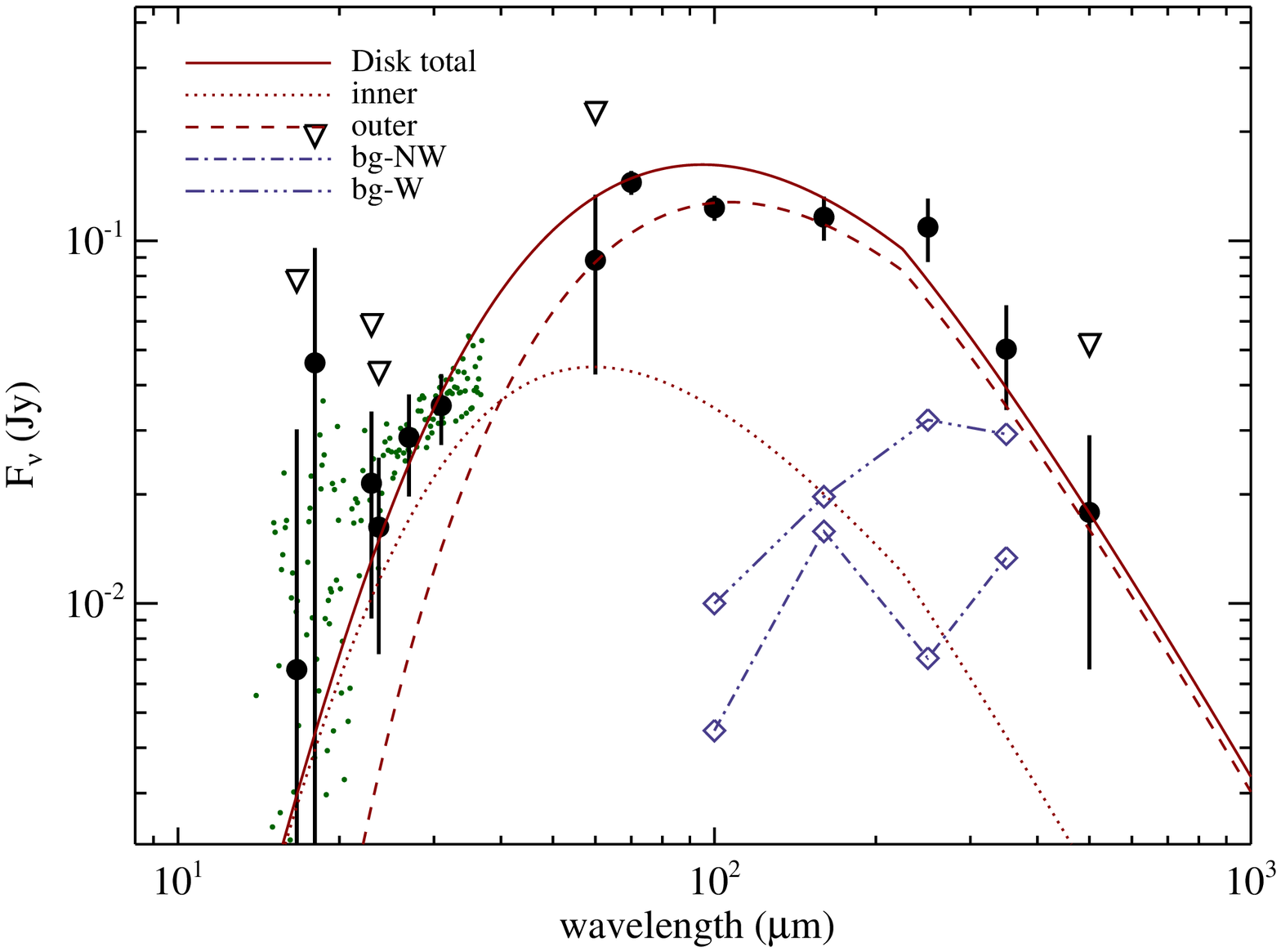,height=2.0in} 
    \end{tabular}
    \caption{Model SEDs of 61 Vir.
    The top panel shows the continuous 29-350AU dust distribution model
    (\S \ref{sss:mod1}), the middle panel shows the 40 and 90AU ring
    model (\S \ref{sss:mod2}), and the bottom panel shows the collisionally
    evolved pre-stirred 1-350AU model (\S \ref{sss:mod3}).
    Fluxes shown in black are those observed less that of the photosphere, i.e.,
    these are the same fluxes plotted in grey on Fig.~\ref{fig:sed}, except that
    here detections that are below $3\sigma$ are plotted both at their measured value
    (with their $1\sigma$ uncertainty) and as triangles to indicate the $3\sigma$
    upper limit.
    The model disk fluxes are shown with a thick red line, and contributions of the
    two components of the bottom two models have also been separated out with dotted
    and dashed lines, for the innermost and outermost component respectively.
    The inferred spectra of the background sources are shown in blue.}
    \label{fig:sedmodels}
  \end{center}
\end{figure}

%%%%%%%%%%%%%%%%%%%%%%%%%%%%%%%%%%%%%%%%%%%%%%
\begin{figure*}
  \begin{center}
    \vspace{-0.1in}
    \begin{tabular}{ccc}
      \hspace{-0.45in} \psfig{figure=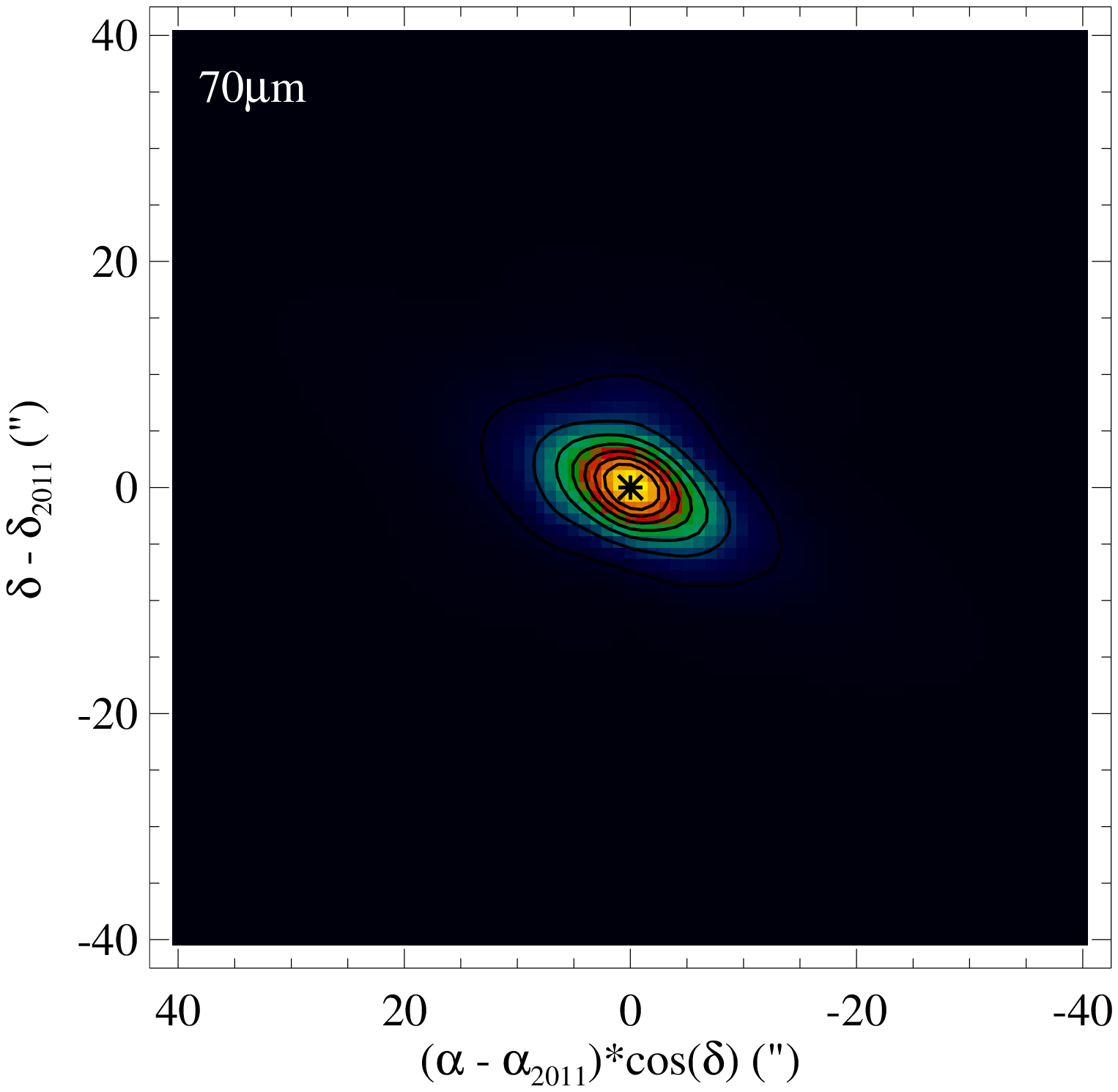,height=2.4in} &
      \hspace{-0.75in} \psfig{figure=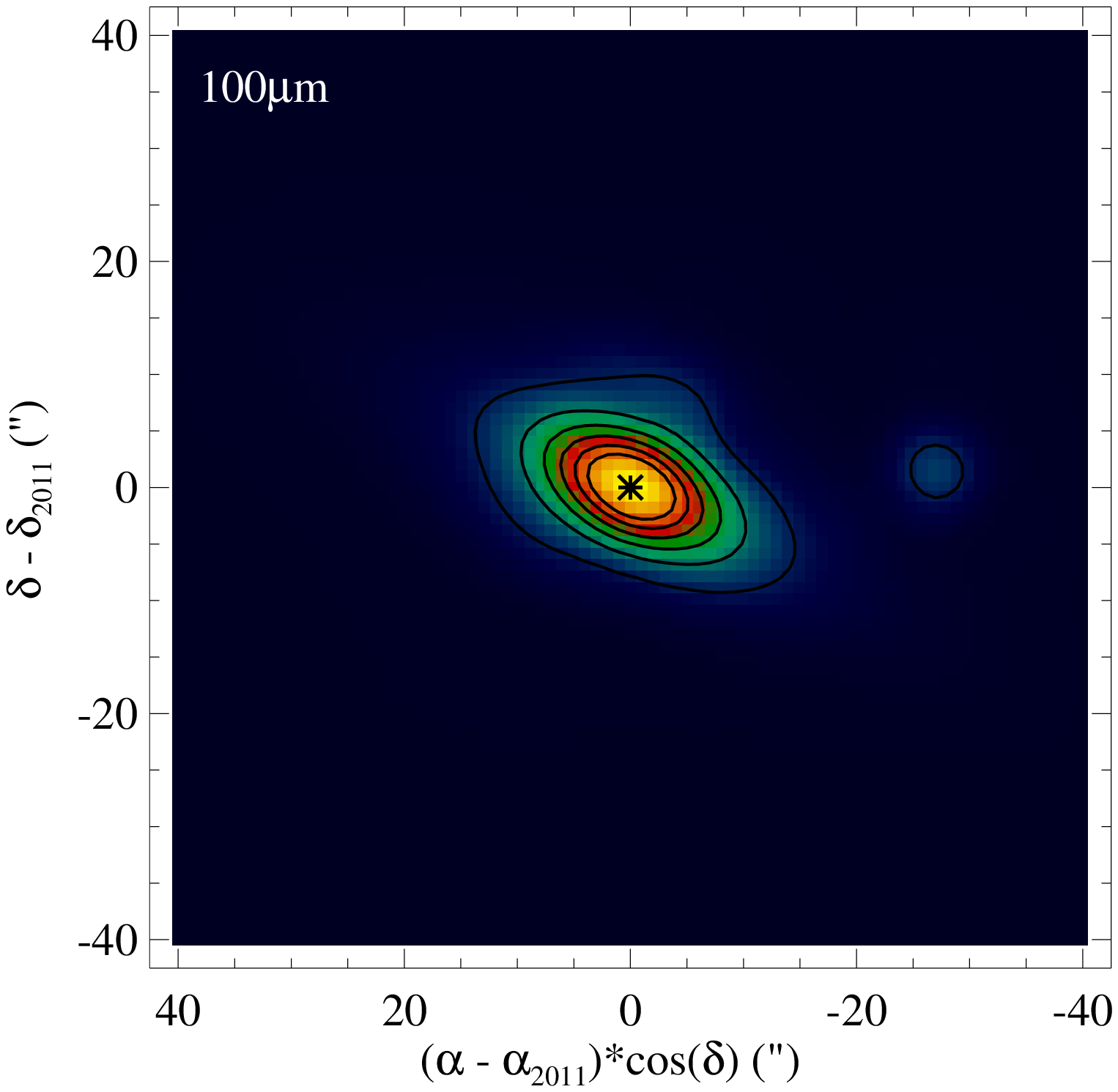,height=2.4in} &
      \hspace{-0.75in} \psfig{figure=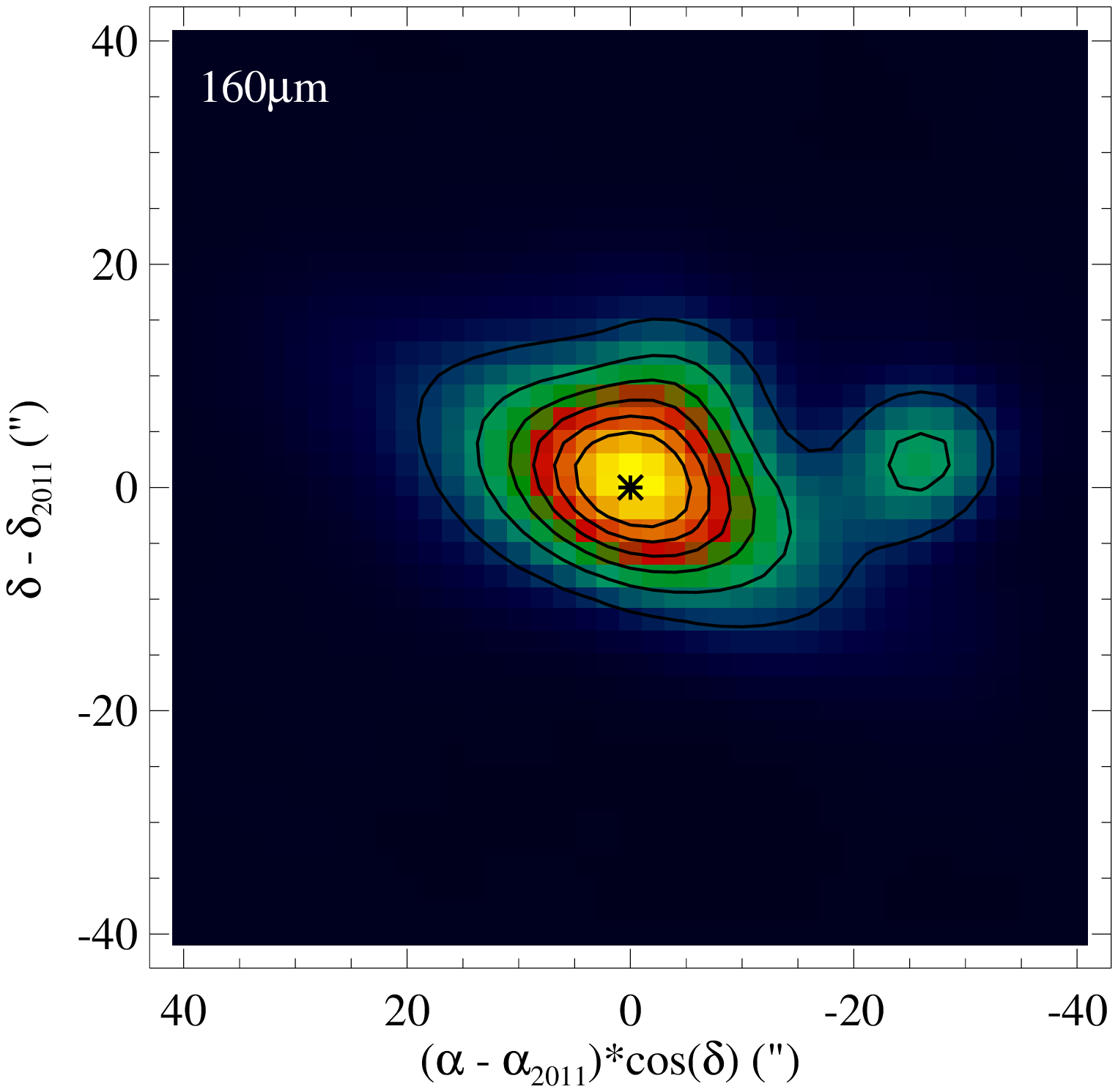,height=2.4in} \\[-0.1in]
      \hspace{-0.45in} \psfig{figure=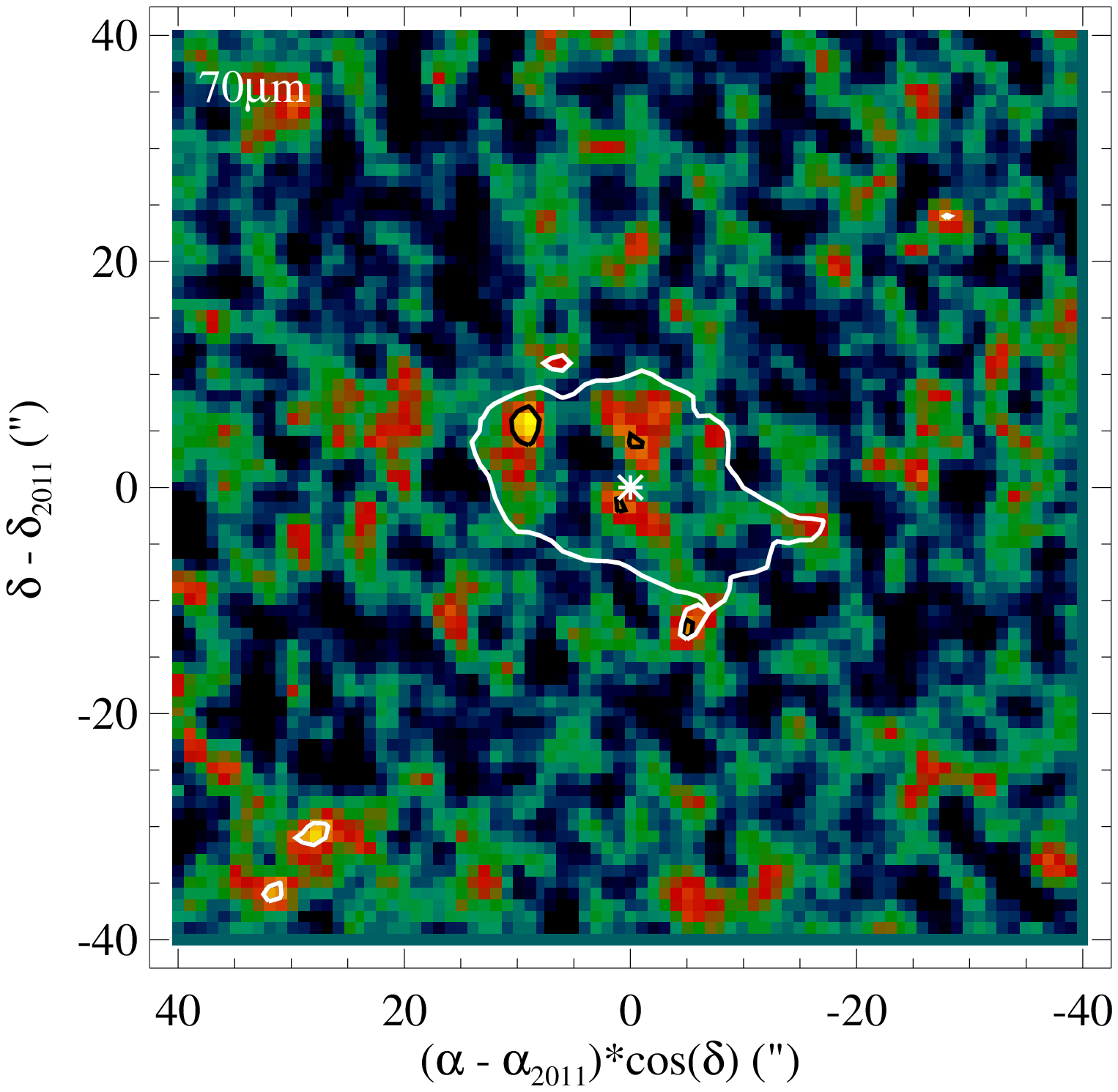,height=2.4in} &
      \hspace{-0.75in} \psfig{figure=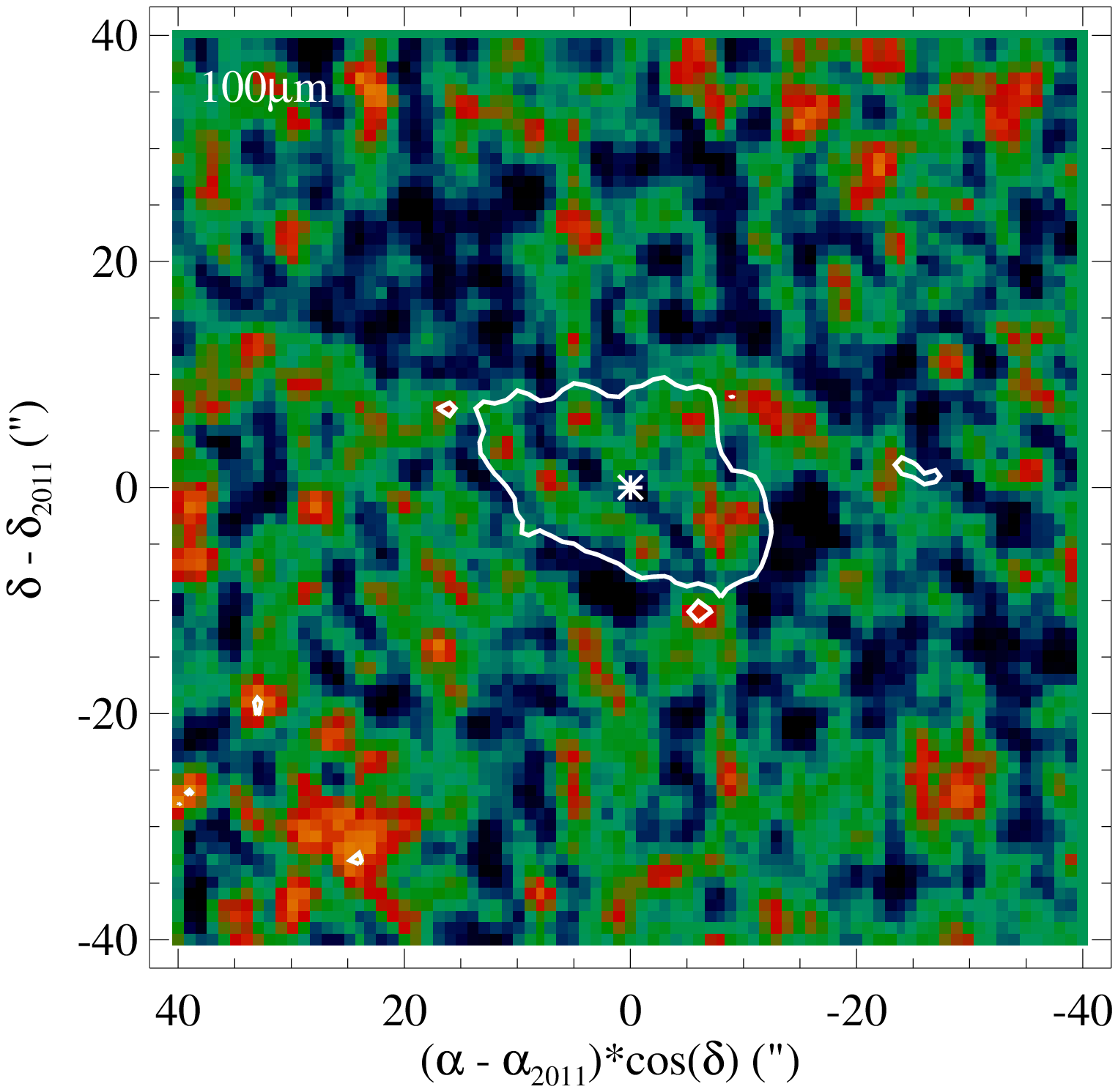,height=2.4in} &
      \hspace{-0.75in} \psfig{figure=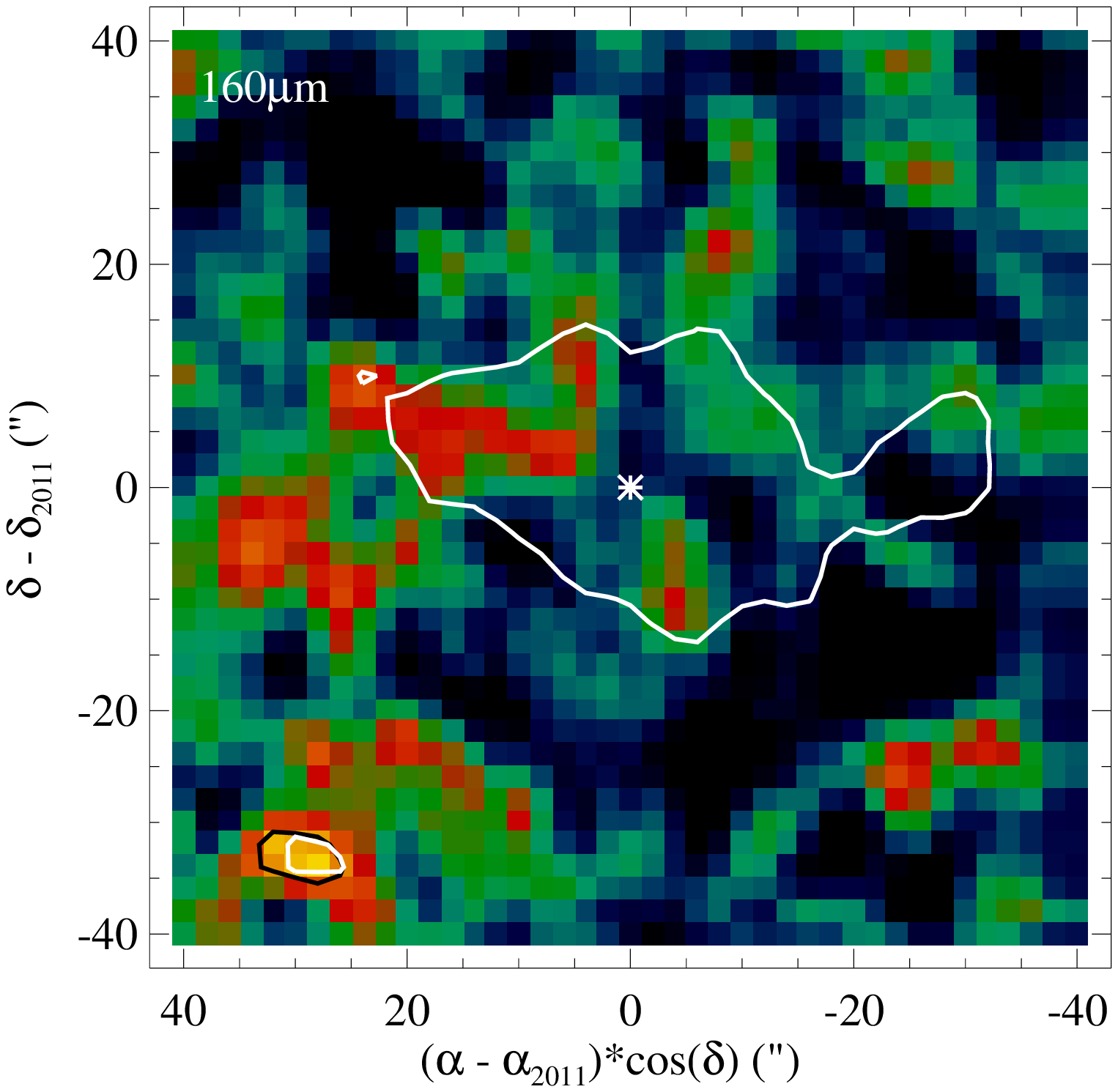,height=2.4in}
    \end{tabular}
    \caption{Model images of the PACS 70~$\mu$m (left), 100~$\mu$m (middle) and
    160~$\mu$m (right) emission distribution toward 61 Vir using the single extended
    component model of \S \ref{sss:mod1} which has a continuous dust distribution
    from 29-350AU.
    The top row of images can be compared directly with Fig.~\ref{fig:herschel}, since
    these are model observations, and employ the same contours and colour scale.
    The bottom row are the residual images (observed minus model), with a colour scale
    that goes from $-2$ to $+6$ times the $1\sigma$ uncertainty in each observation, and
    contours are shown in black at $+4\sigma$.
    The white contour shows the region where the observed flux is above
    $4\sigma$ (from Fig.~\ref{fig:herschel}).}
    \label{fig:mod1}
  \end{center}
\end{figure*}

The simplest model tried had a radial profile given by $\tau(r) = \tau_0 (r/r_{\rm{min}})^{p}$
between inner and outer radii $r_{\rm{min}}$ and $r_{\rm{max}}$.
The best fit found was for $r_{\rm{min}}=29$AU, $r_{\rm{max}}=350$AU, $p=-1.1$, $\tau_0=6.9\times10^{-5}$,
$I=77^\circ$, $f_{\rm{T}}=1.9$, $\lambda_0=220$~$\mu$m, $\beta=1$.
The background sources were found to be offset by (-27",+2") and
(-4",+7") with fluxes resembling cool emission with temperatures of $<40$K
(see Fig.~\ref{fig:sedmodels}), and so are consistent with
extragalactic emission.
The resulting model images are shown in Fig.~\ref{fig:mod1}, that can be compared directly with
Fig.~\ref{fig:herschel}, since the figures employ the same colour scale and contours.
Fig.~\ref{fig:mod1} also shows the residual image (i.e., observed minus model).
Clearly the model provides an excellent fit to the observations, and the residuals in the
vicinity of the star are consistent with the noise elsewhere on the image;
there is no evidence for any asymmetry in the disk structure.
The resulting reduced $\chi^2$ was 1.51.

Although the algorithm provides uncertainties on the model parameters, the most
instructive way to qualify how well the radial distribution is constrained is to compare
this result with an alternative model that provides a comparable fit to the image
(\S \ref{sss:mod2}).
However, both models come to similar conclusions on the parameters defining the dust
properties.
These show that the factor by which the dust temperature exceeds that of
black bodies in thermal equilibrium, $f_{\rm{T}}$, is well constrained, since it is
essentially set by the emission spectrum.
This factor is in line with expectations that it should be greater than unity,
because the $\sim 1$~$\mu$m dust grains that dominate the cross-sectional area of a
debris disk absorb starlight efficiently but reemit inefficiently at longer wavelengths
(e.g., Fig. 3 of Bonsor \& Wyatt 2010).
It is also in line with observations of other disks that have been imaged
(e.g., Fig. 9 of Rodriguez \& Zuckerman 2012), since one
would expect disks to be around $f_{\rm{T}}^2$ times larger than would have been anticipated
from their temperature assuming the grains to act like black bodies.
On the other hand the factors $\lambda_0$ and $\beta$ are poorly constrained, since they
depend on the fraction of the emission seen in the 350 and 500~$\mu$m images that comes
from the disk, and the resolution is insufficient to separate the disk emission from
the background sources (particularly that to the north), which moreover may have a
more complicated structure than the 2 point source model that has been assumed;
the circumstellar emission will be less confused in $>5-10$ years once the star
has moved away from the background sources (see Fig.~\ref{fig:confusion} middle).
The uncertainty on the disk inclination is estimated as $\pm 4^\circ$.

This model was also used to predict the corresponding scattered light images, under
the assumption that the grains scatter light isotropically and have a constant albedo
$\omega$ at all optical wavelengths at which they interact with starlight.
These model images were then compared with the constraints on the optical surface
brightness set by the STIS images in \S \ref{sss:stis} to set constraints on $\omega$.
The scattered light in this model is brightest at its inner edge at around 3.7arcsec,
constraining the albedo to $\omega<0.31$.
This is not a definitive statement about the grain albedo, since other disk structures
also fit the thermal emission and result in slightly different constraints on the albedos
(see \S \ref{sss:mod2} and \ref{sss:mod3}).
However, we note that $\omega<0.31$ is consistent with the grain albedos inferred for
other debris disks imaged in scattered light around solar type stars
(e.g., Backman et al. 2009, Krist et al. 2010, Golimowski et al. 2011),
as well as for Solar System interplanetary dust particles (Kelsall et al. 1998).

%%%%%%%%%%%%%%%%%%%%%%%%%%%%%%%%%%%%%%%%%%%%%%
\begin{figure*}
  \begin{center}
    \vspace{-0.1in}
    \begin{tabular}{ccc}
      \hspace{-0.45in} \psfig{figure=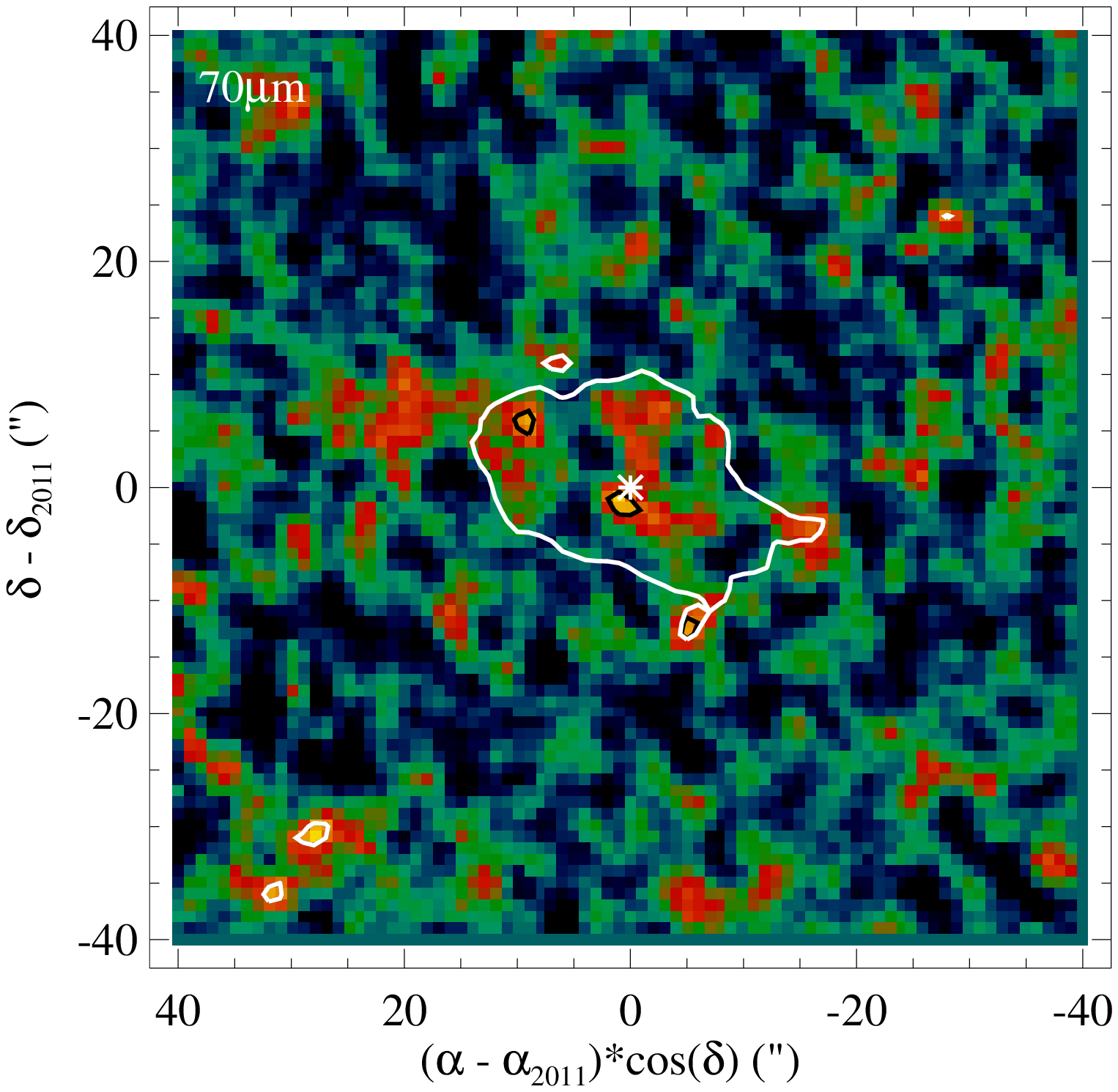,height=2.4in} &
      \hspace{-0.75in} \psfig{figure=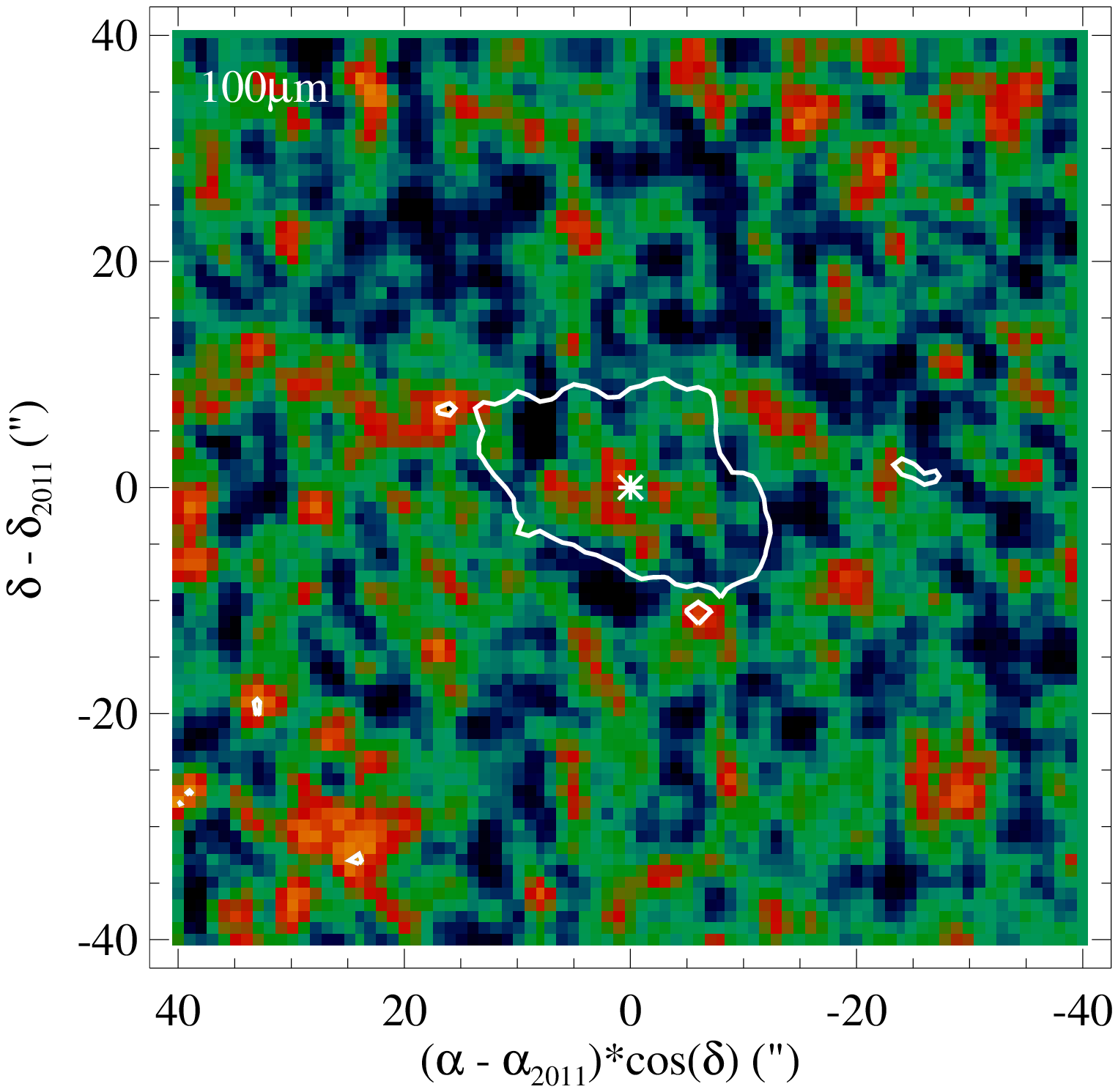,height=2.4in} &
      \hspace{-0.75in} \psfig{figure=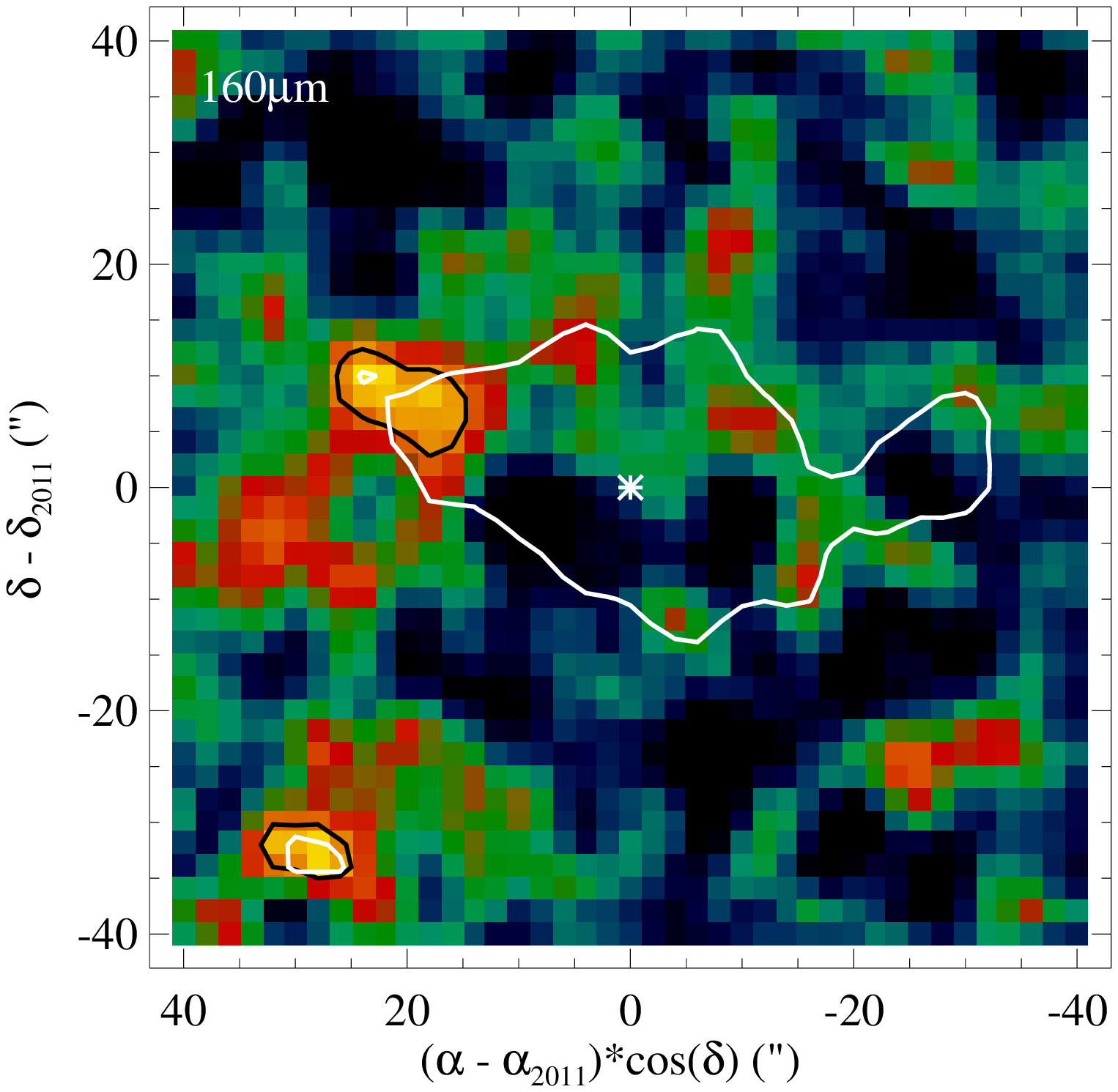,height=2.4in} \\[-0.1in]
      \hspace{-0.45in} \psfig{figure=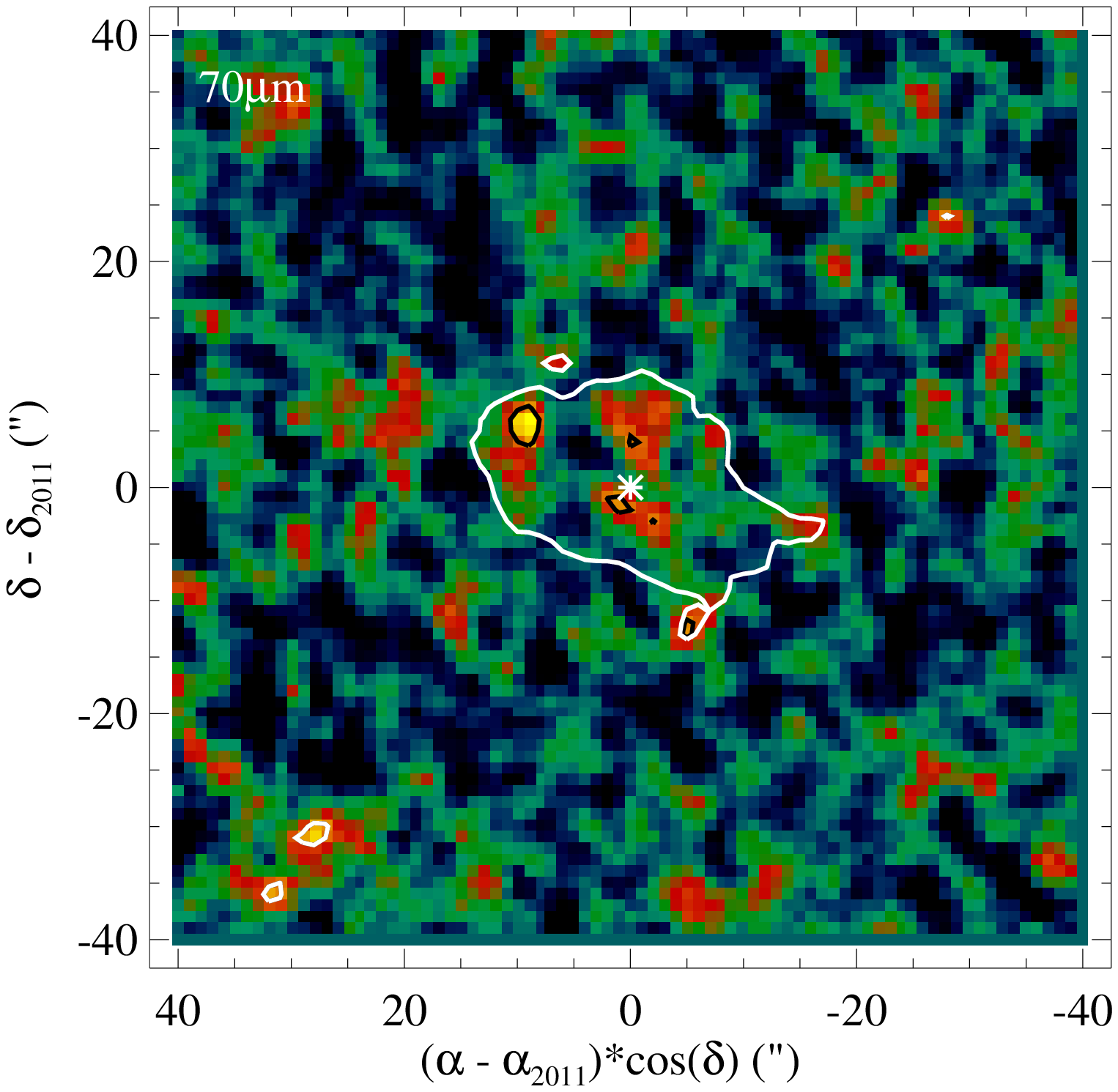,height=2.4in} &
      \hspace{-0.75in} \psfig{figure=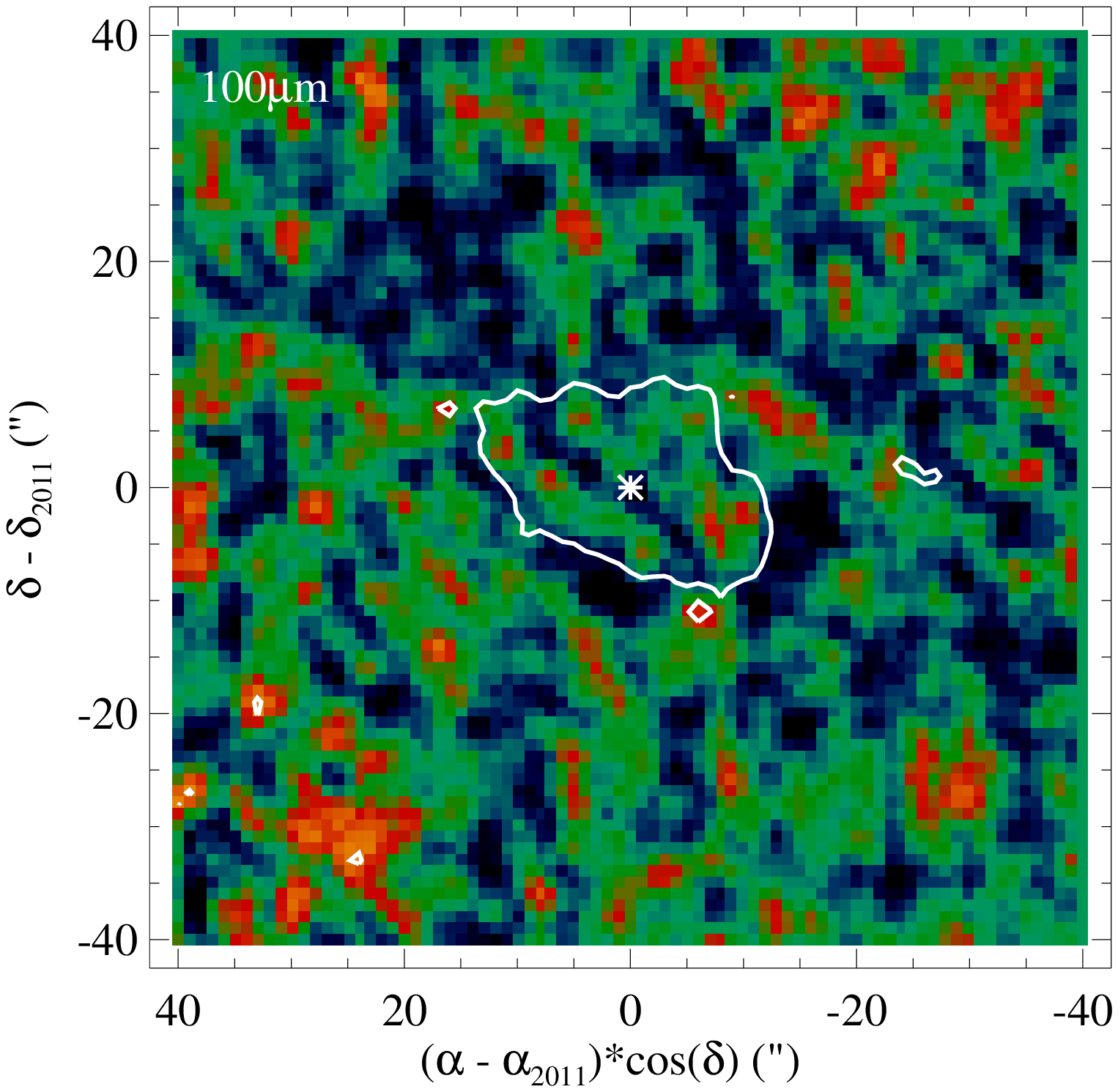,height=2.4in} &
      \hspace{-0.75in} \psfig{figure=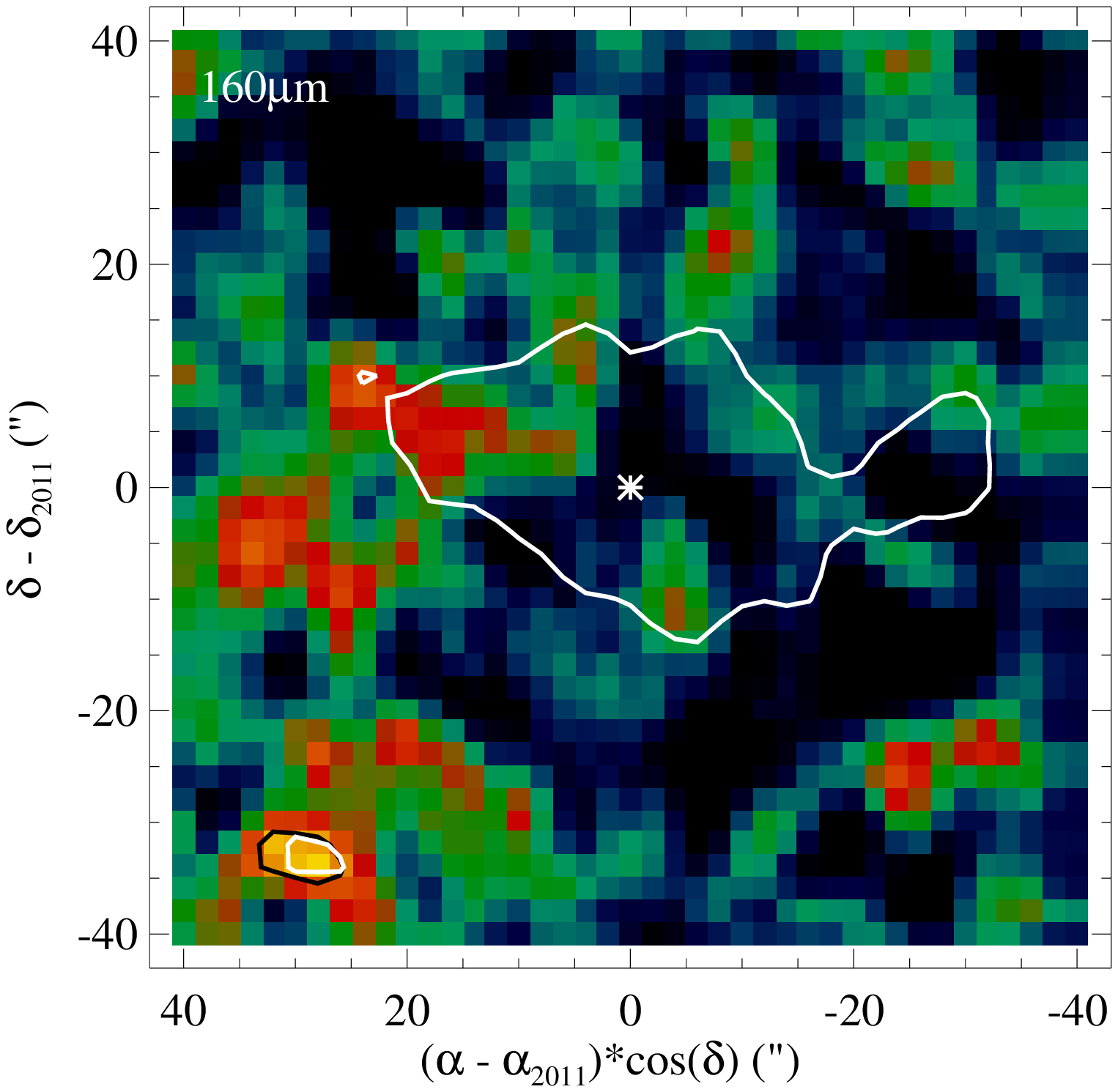,height=2.4in}
    \end{tabular}
    \caption{Residual images (observed minus model) of the PACS 70~$\mu$m (left),
    100~$\mu$m (middle) and 160~$\mu$m (right) emission distribution toward 61 Vir.
    The top row shows the residuals of the best fit model of \S \ref{sss:mod2}
    which has two narrow rings at 40 and 90AU.
    The bottom row shows the residuals of best fit model of \S \ref{sss:mod3} which
    corresponds to a collisionally evolved pre-stirred disk that initially extended
    from 1-350AU.
    The colour scales and contours are the same as Fig.~\ref{fig:mod1}.}
    \label{fig:mod2}
  \end{center}
\end{figure*}

%%%%%%%%%%%%%%%%%%%%%%%%%%%%%%%%%%%%%%%%%%%%%%
\subsubsection{Two radii}
\label{sss:mod2}
The next model had a radial profile given by two narrow rings at radii of
$r_{1}$, $r_{2}$, each of which had a constant optical depth
$\tau_{1}$ and $\tau_{2}$ across a width of 10AU, and with different particle
properties $f_{\rm{T}i}$, $\lambda_{0i}$, $\beta_i$.
The best fit found had $r_{1}=40$AU, $r_{2}=90$AU, $\tau_{1}=1.4 \times 10^{-4}$,
$\tau_{2}=1.7 \times 10^{-4}$, with particle properties
$f_{\rm{T1}}=1.9$, $\lambda_{01}=70$~$\mu$m, $\beta_1=0.53$, and
$f_{\rm{T2}}=1.5$, $\lambda_{02}=280$~$\mu$m, $\beta_2=0.95$,
with background sources with fluxes and locations very similar to those of \S \ref{sss:mod1}
(see Fig.~\ref{fig:sedmodels}).
Again the model provides an excellent fit to the observations (see Fig.~\ref{fig:mod2}),
with residuals that look broadly similar to those of Fig.~\ref{fig:mod1},
and has a total reduced $\chi^2$ of 1.94.

As this model provides a comparable fit to the same set of observations, it illustrates the
uncertainty in the radial profile. 
For example, we should not take too seriously the conclusion of \S \ref{sss:mod1}
that the radial profile extends out as far as 350AU, since the contribution from the
outermost regions beyond 100AU is negligible, even if the fit of a single component model
is improved by including emission in these outer regions.
The larger outer extent is favoured because there is emission out to 20" to the NE at
70-160~$\mu$m.
Fig.~\ref{fig:herschel} shows this is not equally matched to the SW.
Since the two ring model does not include emission at such large distances, this results
in significant ($>4\sigma$) residual emission being present $\sim 20$ arcsec ENE of the
star in Fig.~\ref{fig:mod2}.

The resolution of the images is also not such that we can tell if the distribution between
30 and 100AU is continuous or concentrated in narrower features. 
Similarly the inner edge is not well known, but to explore this we consider a third model
in \S \ref{sss:mod3}.
We do not attribute any physical significance to the fact that the two rings are inferred to
have different $\lambda_0$ and $\beta$, which arises to ensure an appropriate balance of
flux from the two components at all wavelengths.
Rather, again, we consider these parameters to be poorly constrained.
With this model, the non-detection in the STIS images result in a constraint on the
dust albedo of $\omega<0.25$.
This is slghtly more stringent than the model of \S \ref{sss:mod1} as the emission
in this model is more concentrated.

%%%%%%%%%%%%%%%%%%%%%%%%%%%%%%%%%%%%%%%%%%%%%%
\subsubsection{Collisionally evolved pre-stirred disk}
\label{sss:mod3}
This model is the same as that of \S \ref{sss:mod1}, except that we also allow the optical
depth distribution to extend continuously inward of $r_{\rm{min}}$ to an inner edge at
$r_0$ with power-law index $p_0$ (which is positive so that optical depth increases with
radius in this region).
Thus, the disk surface density increases from an inner edge (at $r_0$) to a
maximum (at $r_{\rm{min}}$), then turns over and decreases towards the outer edge
(at $r_{\rm{max}}$).

Given that we have added parameters to a model that already had sufficient parameters to
explain the observations, we do not intend to present a full exploration of this parameter
space.
Rather, the intention here is to test a hypothesis that the planetesimal
disk initially extended all the way from an inner edge just outside the known planets
at $r_0$ (at around 1AU) out to some large distance $r_{\rm{max}}$, and that the inner
regions of this disk (inside $r_{\rm{min}}$) have since been depleted by collisional erosion,
whereas the outer regions (beyond $r_{\rm{min}}$) have yet to reach collisional equilibrium.
If the disk is pre-stirred (see Wyatt 2008), in the sense that collisions between the
planetesimals were destructive from shortly after the star reached the main sequence,
then the inner regions would be expected to attain an optical depth profile with
$p_0=7/3$ (Wyatt 2008; Kennedy \& Wyatt 2010; see \S \ref{sss:plausibility}).
It is this model that we consider here.
The physical interpretation of derived parameters such as $r_{\rm{min}}$ is discussed
in \S \ref{sss:plausibility}, and a discussion of the \textit{self-stirred}
(Kenyon \& Bromley 2010; Kennedy \& Wyatt 2010) and \textit{planet-stirred}
(Mustill \& Wyatt 2009) variants of this model is given in \S \ref{sss:stirring}.

This model has a reduced $\chi^2$ of 1.60, and as with the previous models we find an
excellent fit to the data;
e.g., Fig.~\ref{fig:mod2} shows residuals that are very similar to those of the single
extended component model of Fig.~\ref{fig:mod1}.
The inner edge $r_0$ is not well constrained;
it can extend all the way to the star because the surface density of the
(collisionally depeleted) inner region increases very steeply with radius so little
emission originates from the innermost part of the disk.
The model we present here has $r_0=1$~AU at which radius the optical depth is
$\tau_0=7.4 \times 10^{-9}$.
The turnover occurs at $r_{\rm{min}}=43$~AU, so has moved outwards compared to the inner
edge of the single component model of \S \ref{sss:mod1} to account for the extra inner
emission.
The outer edge lies again at $r_{\rm{max}}=350$~AU (but as noted above could be much closer).
Other parameters are similar to the single component model, e.g., $\lambda_0=225$~$\mu$m,
$\beta=0.7$, $f_{\rm{T}}=1.8$, largely because the model
assumed an optical depth profile in the inner regions that was steep enough to mean
that this component is not a major contributor to the disk emission
(Fig.~\ref{fig:sedmodels}).

Thus, the observations are consistent with a small level of emission being present within
30AU.
To quantify this further, we repeated this modelling, but decreased $p_0$, keeping $r_0$ at 1AU,
until the best fit with the assumed parameters was unacceptable.
This procedure found that distributions with $p_0 \leq 1$ are ruled out, both because
they result in a level of mid-IR emission that exceeds that seen in the IRS observations,
and because the 70~$\mu$m image becomes too centrally peaked.
In other words, the slope of the inner edge is not well constrained, except that we
can say that it rises sharply enough for there to be negligible flux
in the inner regions.
From the IRS spectrum this means that the inner component must have a 10~$\mu$m flux
below around $50$~mJy (i.e., 2\% of the stellar photosphere) and a
32~$\mu$m flux below 54~mJy assuming the outer component does not contribute any
flux at this wavelength (which is unlikely so the true flux limit will be lower).
From the 70~$\mu$m image, the inner component must contribute below around 6~mJy
(the $3\sigma$ point source uncertainty).
The albedo constraint from the STIS non-detection is slightly higher than that of
the previous models at $\omega<0.57$, since the emission in this model is more diffuse.

%%%%%%%%%%%%%%%%%%%%%%%%%%%%%%%%%%%%%%%%%%%%%%
%%%%%%%%%%%%%%%%%%%%%%%%%%%%%%%%%%%%%%%%%%%%%%
%%%%%%%%%%%%%%%%%%%%%%%%%%%%%%%%%%%%%%%%%%%%%%
\section{Implications for the 61 Vir planetary system}
\label{s:pl}

%%%%%%%%%%%%%%%%%%%%%%%%%%%%%%%%%%%%%%%%%%%%%%
%%%%%%%%%%%%%%%%%%%%%%%%%%%%%%%%%%%%%%%%%%%%%%
\subsection{The 61 Vir planetary system}
\label{ss:pl}
The dataset of radial velocity measurements for 61 Vir has expanded since the
observations reported in Vogt et al. (2010), allowing further consideration of the
properties of its planetary system.
Segransan et al. (in prep.) report 142 precise HARPS radial velocities
(averaged values per night) of 61 Vir between $JD=53037$ and 55948
that confirm the existence of the inner two planets, with parameters consistent
with the values of Vogt et al. (2010) (i.e., a $5.1M_\oplus$ planet at 0.05AU
and a $18.2M_\oplus$ planet at 0.218AU).
However, the outermost planet (i.e., the $22.9M_\oplus$ planet at 0.478AU) has
not been confirmed in the HARPS data at this point.
The temporal variation of the $\log{(R'_{HK})}$ activity index (which is also
available from the HARPS spectra) shows that, although the star has been quiet most
of the time, there was a burst of activity between $JD=$54800 and 55220.
Such active periods can influence the interpretation of longer periodicities
in the data, and so affect our ability to extract longer period planets.
As the existence of this planet is not crucial to the conclusions of this
paper, this issue is not discussed further here.

\begin{figure}
  \begin{center}
    \vspace{-0.1in}
    \begin{tabular}{c}
      \psfig{figure=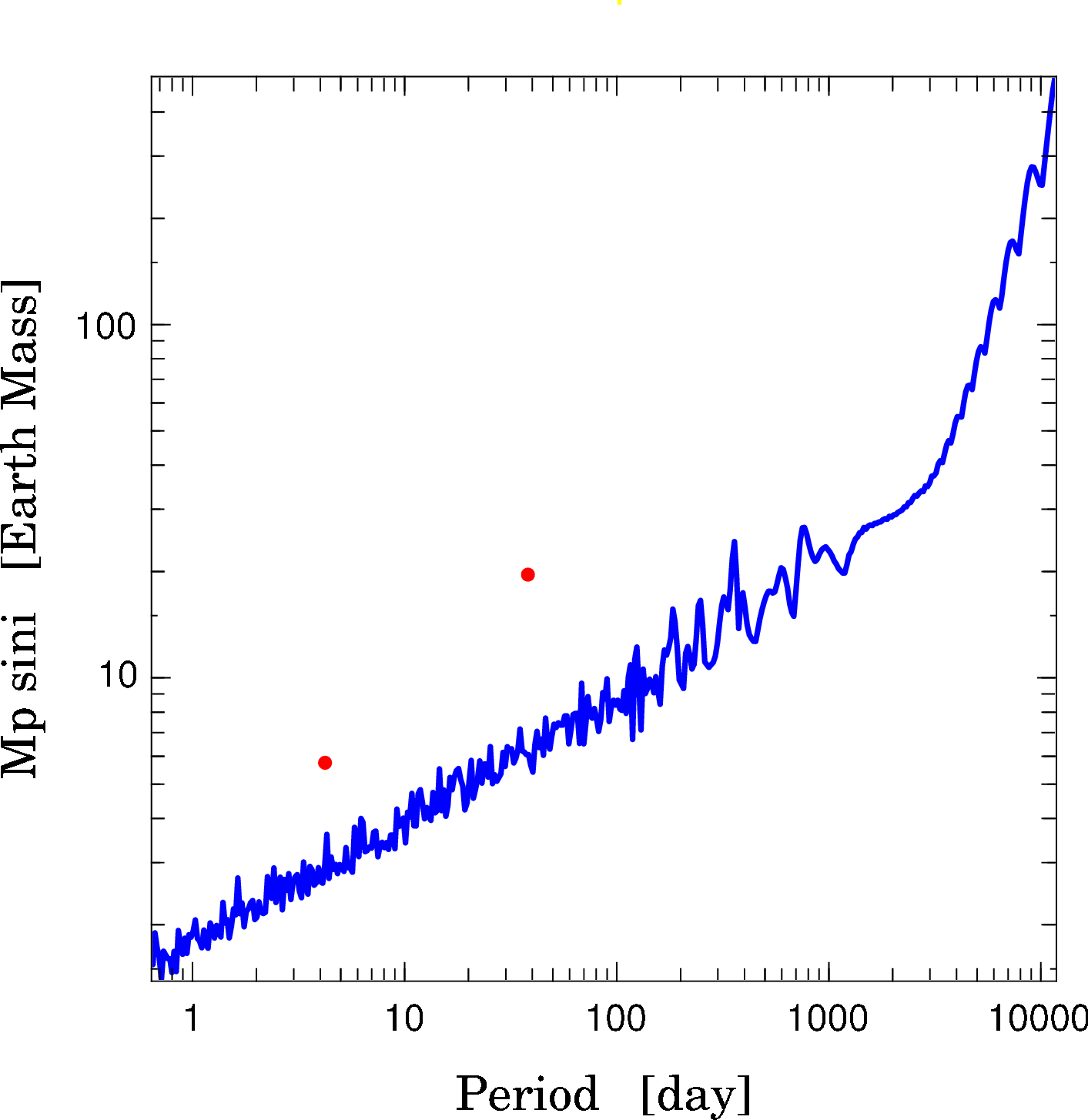,height=2.8in}
    \end{tabular}
    \caption{Detection limit derived from the HARPS radial velocities.
    The blue line indicates the limit above which planets on circular orbits
    would have been detectable in the data (see text for details).
    The red dots indicate the position of the 2 inner planets.}
    \label{fig:harps}
  \end{center}
\end{figure}

Valuable constraints can also be set on the properties of planets
that can be ruled out by the HARPS data (e.g., Mayor et al. 2011).
Since the 2-planet system fits the radial velocities to within $K \approx 1$ (in ms$^{-1}$),
this crudely allows us to rule out planets on circular orbits at semimajor axes
$a_{\rm{pl}}$ (in AU) that are more massive than $M_{\rm{pl}}\sin{i} > 10K\sqrt{a_{\rm{pl}}}$
in $M_\oplus$.
However the limit would be higher than this for planets on periods much longer than the length
over which the observations have been taken ($\sim 8$ years).
Fig.~\ref{fig:harps} shows the limit derived from an analysis of the HARPS data (excluding
that obtained during the activity burst) from which the solution for the 2 inner planets
has been subtracted.
For each point in the mass-period diagram, a corresponding circular orbit is added to
the residuals of the 2-planet model, with 20 phase values (spaced equidistantly between 0 and
$2\pi$).
For each phase, this process is repeated 1000 times with the residuals randomly permuted, and
we consider that the planet is detected only if we see a signal in the resulting
periodogram that is higher than the 1\% false alarm probability limit at all phases.
The detection threshold obtained agrees well with that expected from the crude analysis above,
and shows that the data exclude planets that are more massive than Saturn with semimajor
axes inside 6AU.

%%%%%%%%%%%%%%%%%%%%%%%%%%%%%%%%%%%%%%%%%%%%%%

\subsection{Interaction between known planets and debris}
\label{ss:plane}
Since the planet masses given in \S \ref{ss:pl} were derived from radial velocities
the masses reported are actually $M_{\rm{pl}}\sin{i}$ rather than $M_{\rm{pl}}$, and the
inclination of the planets' orbits to the sky plane ($i$) is unknown.
Our observations determine the inclination of the disk's plane of symmetry to
the sky plane to be $77^\circ$.
If we assume that the planets' orbits are coplanar with the disk then
their masses should be revised modestly upward by 3\% from the values given above.
Such a small change would not affect previous conclusions in the literature about the
dynamics of this system (Vogt et al. 2010; Greenberg \& van Laerhoven 2012).

However, the planets do not have to be coplanar with the disk, since
the two are sufficiently spatially separated that there is little dynamical
interaction between them.
For example, although the planets' secular perturbations will make the orbits of disk
particles precess so that their distribution would end up symmetrical about the
invariable plane of the planetary system, the timescale for this to happen is
at least the secular precession timescale.
At 30AU this is $60$~Gyr considering perturbations from the three Vogt et al. (2010)
planets, and even longer if the outermost planet is excluded.
This also precludes the disk having been stirred by these planets (Mustill \& Wyatt 2009);
i.e., the collisional cascade in the disk must have been ignited by some other mechanism
(see \S \ref{sss:stirring}).
Rather, if there is to be an interaction between the known planets and the disk there
would need to be planets in the intervening 1-30AU region, either to allow disk material
to be scattered to these inner regions (e.g., Bonsor \& Wyatt 2012), with implications
for the habitability of the inner planets (e.g., Horner \& Jones 2010), or for the planets'
gravitational perturbations to influence the intervening planets which in turn influence
the outer regions (e.g., Zakamska \& Tremaine 2004).

The assumption that the planets and the debris disk are coplanar is justified by
the premise that the debris disk and planets were born from the same
protoplanetary disk, which had a single plane of symmetry.
Nevertheless this does not guarantee that the two should remain coplanar,
since the debris disk may have been perturbed by passing stars, and the planets
could have undergone a dynamical interaction that misaligned their orbital
planes (Naoz et al. 2011).
The spin axis of the star can also become misaligned with the protoplanetary
disk under certain conditions (Lai et al. 2011).
%as has been suggested as the explanation for the planets that transit their stars
%indicating a retrograde orbital motion with respect to the stellar spin axis
%(Triaud et al. 2010).
In this respect it is worth noting that the stellar rotation period of
29~days and $v\sin{i}$ of $1.6 \pm 0.5$~km/s
suggest the stellar pole is
inclined $72^\circ$ to our line-of-sight, i.e., consistent with being nearly
perpendicular to the debris disk's symmetry plane.
Thus there is no evidence at the moment to suggest that the angular momentum vectors
of the star, planets and debris disk are misaligned.

%%%%%%%%%%%%%%%%%%%%%%%%%%%%%%%%%%%%%%%%%%%%%%
%%%%%%%%%%%%%%%%%%%%%%%%%%%%%%%%%%%%%%%%%%%%%%
\subsection{Intervening planets?}
\label{ss:selfstirring}
Given that an axisymmetric disk model provides a good fit to the
observations (\S \ref{s:mod}), disk asymmetries cannot be invoked
as evidence for more distant planets as has been the case for other imaged
disks (e.g., Wyatt 2008; Krivov 2010).
However, the emptiness of the inner region does require an absence of
planetesimals, and this emptiness can be readily explained if there are
additional planets in the 1-30AU region, since such planets could dynamically deplete the
region (e.g., Lecar et al. 2001).
In \S \ref{ss:pl} we already used the radial velocity data to set constraints
on the planets that are allowed to exist within 6AU (and have remained undetected,
see Fig.~\ref{fig:harps}).
It is not possible to set hard constraints on more distant planets, except for
generalities such as that the planets cannot reside too close to the inner
edge of the disk (Quillen 2006; Mustill \& Wyatt 2012).
However, in \S \ref{sss:plausibility} we consider the possibility that no
planets exist in the 1-30AU region.

%%%%%%%%%%%%%%%%%%%%%%%%%%%%%%%%%%%%%%%%%%%%%%
%%%%%%%%%%%%%%%%%%%%%%%%%%%%%%%%%%%%%%%%%%%%%%
\subsection{Plausibility of pre-stirred extended disk model}
\label{sss:plausibility}
In \S \ref{sss:mod3} it was suggested that the observed depletion could arise
from collisional erosion, i.e., without requiring any planets to exist in the 1-30AU region.
To confirm that this is a plausible explanation, we show here that the observationally
inferred parameters for this model ($r_{\rm{min}}$, $\tau_0$ and $p$) correspond
to physically plausible parameters within the context of this scenario.

First we consider the regions beyond $r_{\rm{min}}$ which, in this interpretation,
have yet to deplete, so can be used to determine the initial mass distribution
of the debris disk;
we will assume this initially extended all the way in to $r_0$ according to
$\Sigma = \Sigma_{0}(r/r_0)^p$, where $\Sigma$ is the mass surface density
(e.g., in $M_\oplus$ AU$^{-2}$).
The mass surface density simply scales with the optical depth according to the
size distribution, which we assume to be a single power law with the $-3.5$
exponent expected (under certain ideal assumptions) for collisional equilibrium,
from a maximum planetesimal size $D_{\rm{max}}$ (in km), down to the size at which
dust is removed by radiation pressure.
Using equation 15 of Wyatt (2008), for a $0.88M_\odot$, $0.84L_\odot$ star, assuming
a solid density of 2700 kg m$^{-3}$,
\begin{equation}
  \Sigma / \tau = 0.19 D_{\rm{max}}^{1/2}
\end{equation}
in $M_\oplus$~AU$^{-2}$.

For reference in later paragraphs, once we have estimated $D_{\rm{max}}$,
we will compare the mass surface density derived with that of a fiducial
\textit{minimum mass solar nebula} (MMSN), which contains
$1.1M_\oplus$~AU$^{-2}$ of solid material at 1AU,
with surface densities falling off to larger distance as $(r/r_0)^{-3/2}$
(e.g., Weidenschilling 1977).
The best fit model of \S \ref{sss:mod3} has an optical depth at 43AU of
$4.8 \times 10^{-5}$ which is around an order
of magnitude higher than that in the Kuiper belt (see Fig. 13 of Vitense
et al. 2010), and corresponds to a mass surface density at that location of
$2.3 \times 10^{-3} D_{\rm{max}}^{1/2}$ times that in the MMSN.

The plausibility of this interpretation comes down to whether it is
reasonable to expect the transition to collisional equilibrium to occur at 43AU
given the stellar age of $4.6 \pm 0.9$ Gyr.
As well as the aforementioned parameters, the collisional lifetime of the largest
planetesimals in the disk depends on their strength $Q_{\rm{D}}^\star$ and
eccentricity $e$ (which sets collision velocities).
For the assumptions above, this collisional lifetime can be estimated
(see eq. 13 of Wyatt et al. 2007a)
% or eq. 8 of Kennedy \& Wyatt 2010
to be
\begin{equation}
  t_{\rm{c}}(D_{\rm{max}}) = 1.4 \times 10^{-3} r^{7/3}
    [{Q_{\rm{D}}^\star}^{5/6}D_{\rm{max}}^{1/2}e^{-5/3}]\tau^{-1}
  \label{eq:tc}
\end{equation}
in years.
In the inner regions that are assumed to be in collisional equilibrium
this lifetime should be the age of the star, explaining why we set
$p_0=7/3$ in \S \ref{sss:mod3}.

Thus, for the collisional lifetime at 43AU to be equal to the 4.6Gyr age of the star,
the quantity in square brackets in eq.~\ref{eq:tc} must be of order $2.4 \times 10^4$
($\pm 50$\% due to the uncertainties in stellar age and disk optical depth and turnover
radius).
This quantity has been estimated from the detection statistics as a function of age
for the ensemble of debris disks around A stars to be $7.4 \times 10^4$, with a dispersion
of 1dex about this average disk value (Wyatt et al. 2007b).
For sun-like stars the same process gives this quantity as
$2.9 \times 10^{6}$ with a similar dispersion (Kains, Wyatt \& Greaves 2011).
So it appears that, for this interpretation to be valid, the properties of planetesimals
around 61 Vir would have to be broadly similar to those of average A star debris disks, but
would be in the $2\sigma$ minority when compared with debris disks known around sun-like stars.
The sense of this difference is that the planetesimals would have to be 
smaller, or on more elliptical orbits.
We find that the required quantity is obtained for $D_{\rm{max}}=2-5$~km for
$e=0.05-0.1$ for two different strength laws, i.e., that of weak aggregates
(Stewart \& Leinhardt 2009) and that for low temperature
ice (e.g., Vitense et al. 2010).
%For higher eccentricities, such as those present in a scattered disk, an analysis such
%as that of Wyatt et al. (2010) would be required to determine the appropriate collision
%lifetime.

To conclude on the pre-stirred model, we find that the observations can be explained
if the disk is made up of planetesimals up to a few km in size, with a mass surface
density around 0.4\% that of the MMSN.
Despite its low mass, which is comparable with that of the Kuiper belt, the disk in this
model is detectable because the mass is locked up in relatively small planetesimals.
Such a low mass is somewhat at odds with the presence of planets, since if these have even modest
rocky cores, then the solid mass surface density in the inner regions is at least an
order of magnitude above the MMSN.
Even if all of the mass inside 43AU in the hypothetical initial debris disk was
placed in the planet region, that would only correspond to $0.2M_\oplus$. 
One would therefore have to conclude that the disk was initially much more massive,
and that the majority of the mass has been depleted somehow, such as by inwards
migration due to gas drag (or simply never having grown into planetesimals).

Remember though, that this conclusion is based on the premise that the 43AU peak arises
from collisional evolution.
If the inner edge is truncated by some other mechanism, then
the debris disk could include much larger planetesimals and so be much more massive.
For example, if the planetesimals were up to 2000km in diameter the debris disk's
mass surface density would be inferred to be around 10\% of a MMSN at 43AU.
Given the uncertainties, this initial mass is more compatible with the known planets, but
would still require the original protoplanetary disk to be significantly more massive
than the remnant debris disk, and likely also that the planets we know today include a
reasonable fraction of the mass that was originally inside 43AU.

%%%%%%%%%%%%%%%%%%%%%%%%%%%%%%%%%%%%%%%%%%%%%%
%%%%%%%%%%%%%%%%%%%%%%%%%%%%%%%%%%%%%%%%%%%%%%
\subsection{Disk stirring mechanism}
\label{sss:stirring}
In the model discussed in \S \ref{sss:plausibility}, the disk was assumed to be stirred from early on,
since that allowed us to equate the collisional lifetime (eq.~\ref{eq:tc}) with
the stellar age.
Such a model is called \textit{pre-stirred} in Wyatt (2008), and makes no assumptions
about the stirring mechanism.

A variant on the pre-stirred model, that includes a self-consistent consideration
of the stirring mechanism, is the \textit{self-stirred} model.
In this model collisions in an extended planetesimal belt initially occur at
low enough velocity for planetesimal growth to occur
(Kenyon \& Bromley 2010; Kennedy \& Wyatt 2010).
Such models allow planet formation processes to continue throughout the several
Gyr lifetime of the star.
Within such a model a peak in surface brightness naturally corresponds to the location
at which planetesimals have recently grown to around Pluto-size, since at this time
gravitational scattering stirs neighbouring planetesimals resulting in enhanced
dust production.
For the growth of Pluto-size objects, i.e. those large enough to stir the
disk, to have reached 43AU after 4.6Gyr of evolution, this would require
the original protoplanetary disk of 61 Vir to be less than 0.01\% of a MMSN
(see equation 9 of Kennedy \& Wyatt 2010).
Since this would require the disk to be made up of entirely of sub-km-sized
planetesimals, which is contrary to the premise of the model that planetesimals
are growing up to Pluto-size, we do not consider this to be a viable explanation.

Fully formed planets that are well separated from the disk can also stir the disk
through their secular perturbations, i.e., the \textit{planet-stirred} model
(Mustill \& Wyatt 2009).
We already showed in \S \ref{ss:pl} that the known planets cannot have stirred the
disk given the age of the star.
However, planetary perturbations are a plausible stirring mechanism if
there are planets in the 1-30AU region, since the proximity of such
planets to the disk means that their perturbations could affect the disk
within the stellar age.
For example, for a planet with an eccentricity of 0.1 to have stirred the disk at 30AU within
4.6Gyr requires its mass to be $>150a_{\rm{pl}}^{-3}$ (in $M_\oplus$), where
$a_{\rm{pl}}$ is its semimajor axis in AU (see eq. 15 of Mustill \& Wyatt 2009).
Such planets are not ruled out by the current radial velocity observations
(see Fig.~\ref{fig:harps}).

%%%%%%%%%%%%%%%%%%%%%%%%%%%%%%%%%%%%%%%%%%%%%%
%%%%%%%%%%%%%%%%%%%%%%%%%%%%%%%%%%%%%%%%%%%%%%
\subsection{Summary}
\label{sss:summary}
To conclude, we know that there are planets within 1AU, and that there
are planetesimals producing dust beyond 30AU, but the only thing we can
say for certain about the 1-30AU region is that there is a relative
dearth of planetesimals within 30AU compared with the more distant regions.
The presence of planets in the 1-30AU region would naturally explain the
absence of planetesimals, but the emptiness of this region could also
be explained by collisional erosion;
i.e., it could have been full of km-sized planetesimals that were
subsequently depleted through mutual collisions that turned them into dust
that was then removed by radiation forces.
The latter explanation must also invoke a mechanism that stirs the
disk (so that collisions are erosive).
This would be readily explained if there were planets in the
1-30AU region, but could be inherent in the debris disk formation process.

%%%%%%%%%%%%%%%%%%%%%%%%%%%%%%%%%%%%%%%%%%%%%%
%%%%%%%%%%%%%%%%%%%%%%%%%%%%%%%%%%%%%%%%%%%%%%
%%%%%%%%%%%%%%%%%%%%%%%%%%%%%%%%%%%%%%%%%%%%%%
\section{Low-mass planets and debris: is there a correlation?}
\label{s:stat}
The coexistence of planets and debris around the 8th nearest G star prompted us to
reconsider the lack of correlation between the two populations.
In particular, we considered if earlier studies found no correlation
(e.g., Greaves et al. 2004; Moro-Mart\'{i}n et al. 2007; Bryden et al. 2009)
because at that time radial velocity studies were only sensitive to
relatively massive planets.
Here we suggest that a correlation does indeed exist for low-mass planets.

%%%%%%%%%%%%%%%%%%%%%%%%%%%%%%%%%%%%%%%%%%%%%%
%%%%%%%%%%%%%%%%%%%%%%%%%%%%%%%%%%%%%%%%%%%%%%
\subsection{Nearest 60 G star sample}
\label{ss:g60}
To assess the existence of a correlation with low-mass planets
we considered a restricted but unbiased sample comprised of the
nearest 60 G stars (see Phillips et al. 2010).
The sample size was restricted to allow a thorough analysis of individual
stars;
an analysis of a larger sample is deferred to Moro-Mart\'{i}n et al.
(in prep.).

Out of this sample, 11 are reported in the literature to have
planets, all of which were found in radial velocity surveys.
We further subdivided this sample into those that have
at least one massive planet, which we defined as a system with
a planet more massive than $>95M_\oplus$ (i.e., Saturn-mass)
(hereafter called the high-mass planet sample),
and those with only low-mass planets
(hereafter called the low-mass planet sample).
The high-mass planet sample comprises 5 systems
(HD146513, 47 UMa, $\mu$ Ara, 51 Peg, HD190360) and the
low-mass planet sample 6 systems
(HD20794, 61 Vir, HD102365, HD69830, HD136352, HD38858)
that are described further in Table~\ref{tab:g60}.

It may be questioned whether this division is appropriate, given
that longer observations may reveal systems currently classified as having
only low-mass planets to have more massive planets at larger distances.
However, the radial velocities for 61 Vir show that there can be
no planets more massive than Saturn orbiting within 6AU in this system
(\S \ref{ss:pl}).
So if more massive planets do exist they would have to be at larger separation,
and the system would not have been classified as a high-mass planet
system in this study.
In other words, our classification of a low-mass planet system requires
that there are planets detected above the blue line on
Fig.~\ref{fig:harps}, but that none of these is more massive than Saturn
(or this would have been classified as a high-mass planet system),
noting that the exact threshold will vary slightly between stars.

%%%%%%%%%%%%%%%%%%%%%%%%%%%%%%%%%%%%%%%%%%%%%%
\begin{table*}
%\begin{minipage}{160mm}
  \begin{center}
    \caption{Nearest 60 G stars with planets}
    \begin{tabular}{llll}
        \hline
        Star name     & Planets ($a$, $M_{pl}$)      & Debris ($r$, $f$)  & References \\
        \hline
        \multicolumn{2}{l}{\textbf{High-mass planet systems}} & &  \\
        HD147513      & (1.32AU, $400M_\oplus$)
% 12.9pc, G3/G5V, 0.65Gyr, Mayor et al. 2004, A\&A, 415, 391
                      & None detected    & 1, 11 \\
        47 UMa        & (2.1AU, $840M_\oplus$), (3.6AU, $180M_\oplus$), (11.8AU, $550M_\oplus$)
                      & None detected    & 2, 11 \\
% 14.0pc, G0V, 7.4Gyr, Gregory \& Fischer (2010)
        $\mu$ Ara     & (0.09AU, $11M_\oplus$), (0.92AU, $170M_\oplus$), (1.5AU, $560M_\oplus$),
                        (5.2AU, $600M_\oplus$)
                      & None detected    & 3, 11 \\
% 15.3pc, G3IV/V, 6.4Gyr, Santos et al. 2004, A\&A, 426, L19
        51 Peg        & (0.05AU, $160M_\oplus$)
                      & None detected    & 4, 11 \\
% 14.7pc, G2IV, 4Gyr, Mayor \& Queloz (1995), Nature, 378, 355
        HD190360      & (0.13AU, $18M_\oplus$), (3.9AU, $500M_\oplus$)
                      & None detected   & 5, 11 \\
% 15.9pc, G6IV, 12.1Gyr, Vogt et al. (2005), ApJ, 632, 638?
        \hline
        \multicolumn{2}{l}{\textbf{Low-mass planet systems}} & &  \\
%        \textbf{Low-mass planets} & & & \\
        HD20794       & (0.12AU, $2.8M_\oplus$), (0.20AU, $2.5M_\oplus$), (0.35AU, $5.0M_\oplus$)
                      & (19AU, $4 \times 10^{-6}$)    & 6, 12, 13 \\
% 6.1pc, G8V, 5.8Gyr, Pepe et al. 2011, A\&A, 534, A58
        61 Vir        & (0.05AU, $5.3M_\oplus$), (0.22AU, $19M_\oplus$), (0.48AU, $24M_\oplus$)
                      & (29-350AU, $3 \times 10^{-5}$)    & 7, 13 \\
% 8.5pc, G5V, 9.0Gyr, Vogt et al. (2010)
        HD102365      & (0.46AU, $17M_\oplus$)
                      & None detected$^\star$    & 8, 12, 13 \\
% 9.2pc, G2V, 9Gyr, Tinney et al. 2011, ApJ, 727, 103
        HD69830       & (0.08AU, $11M_\oplus$), (0.19AU, $13M_\oplus$), (0.63AU, $19M_\oplus$)
                      & (1AU, $2 \times 10^{-4}$)    & 9, 14 \\
% 12.6pc, K0V, 7Gyr, Lovis et al. (2006)
        HD136352      & (0.09AU, $5.5M_\oplus$), (0.17AU, $12M_\oplus$), (0.41AU, $10M_\oplus$)
                      & None detected$^\star$    & 10, 12 \\
% 14.8pc, G4V, Mayor et al. 2011, A\&A, submitted (astro-ph/1109.2497)
        HD38858       & (1.0AU, $32M_\oplus$)
                      & (102AU, $8 \times 10^{-5}$)    & 10, 12, 15 \\
% 15.2pc, G4V, Mayor et al. 2011
        \hline
    \end{tabular}
    \label{tab:g60}
  \end{center}
$^\star$=See Fig.~\ref{fig:detlims} for constraints on disk luminosity,
1=Mayor et al. (2004), 
2=Gregory \& Fischer (2010),
3=Santos et al. (2004),
4=Mayor \& Queloz (1995),
5=Vogt et al. (2005),
6=Pepe et al. (2011),
7=Vogt et al. (2010),
8=Tinney et al. (2011),
9=Lovis et al. (2006),
10=Mayor et al. (2011),
11=Bryden et al. (2009),
12=Beichman et al. (2006),
13=This work,
14=Beichman et al. (2011),
15=Lawler et al. (2009).
%\end{minipage}
\end{table*}

Of the high-mass planet sample, all have been observed with Spitzer at
70~$\mu$m, and all but HD190360 were also observed with Spitzer at 24~$\mu$m
(although fairly stringent limits do exist for this source from IRAS 25~$\mu$m observations).
None were found to have excess emission from circumstellar debris (Bryden et al. 2009).
A detection rate of 0/5 is consistent with the number of detections that would
be expected if the presence of debris is uncorrelated with the presence of
planets, since nearby stars have a $14-16$\% detection rate with Spitzer
(Trilling et al. 2008; Bryden et al. 2009).

Note that if the Sun had been included in this sample Jupiter would fall above
the detection threshold of radial velocity surveys, and so the Sun would have
been classified as a high-mass planet system.
Although we know the Sun has a debris disk, this is much lower in brightness
compared with the disks known around nearby G stars (e.g., Greaves \& Wyatt 2010),
and its thermal emission would not have been detectable in any current debris disk
survey.
This illustrates that the lack of debris disk detections around
high-mass planet systems does not necessarily mean that these stars are
completely clear of debris, since they could have disks below the detection
threshold.
Just as it was important to quantify what we mean by a system with high- and 
low-mass planets (see above, and Fig.~\ref{fig:harps}), so it is important
to also quantify what we mean by a system without debris;
this is discussed further in \S \ref{sss:summ}.

%%%%%%%%%%%%%%%%%%%%%%%%%%%%%%%%%%%%%%%%%%%%%%
%%%%%%%%%%%%%%%%%%%%%%%%%%%%%%%%%%%%%%%%%%%%%%
\subsection{Debris around the low-mass planet sample}
\label{ss:g60low}
In contrast to the high-mass planet sample, 4/6 of the low-mass planet sample are found
to have debris.
The debris toward three of these (61 Vir, HD69830, HD38858) was already known
about from Spitzer (Bryden et al. 2006; Beichman et al. 2006).
This is already a high detection fraction, because Poisson statistics show
that the detection of $\geq 3$ in a sample of 6 when the expected detection
rate is 15\% is a 6.5\% probability event. 
This was not noted previously, because the planetary systems were only discovered
since the Spitzer debris detections were reported.
Excess emission toward HD20794 was not reported from early analysis of the
Spitzer observations (Beichman et al. 2006), but was detected by Herschel as part
of the DEBRIS program, and this discovery is reported here for the first time.
Below we provide more details on the debris toward members of the low-mass planet
sample.

%%%%%%%%%%%%%%%%%%%%%%%%%%%%%%%%%%%%%%%%%%%%%%
\begin{figure*}
  \begin{center}
    \begin{tabular}{ccc}
      \hspace{-0.25in} \psfig{figure=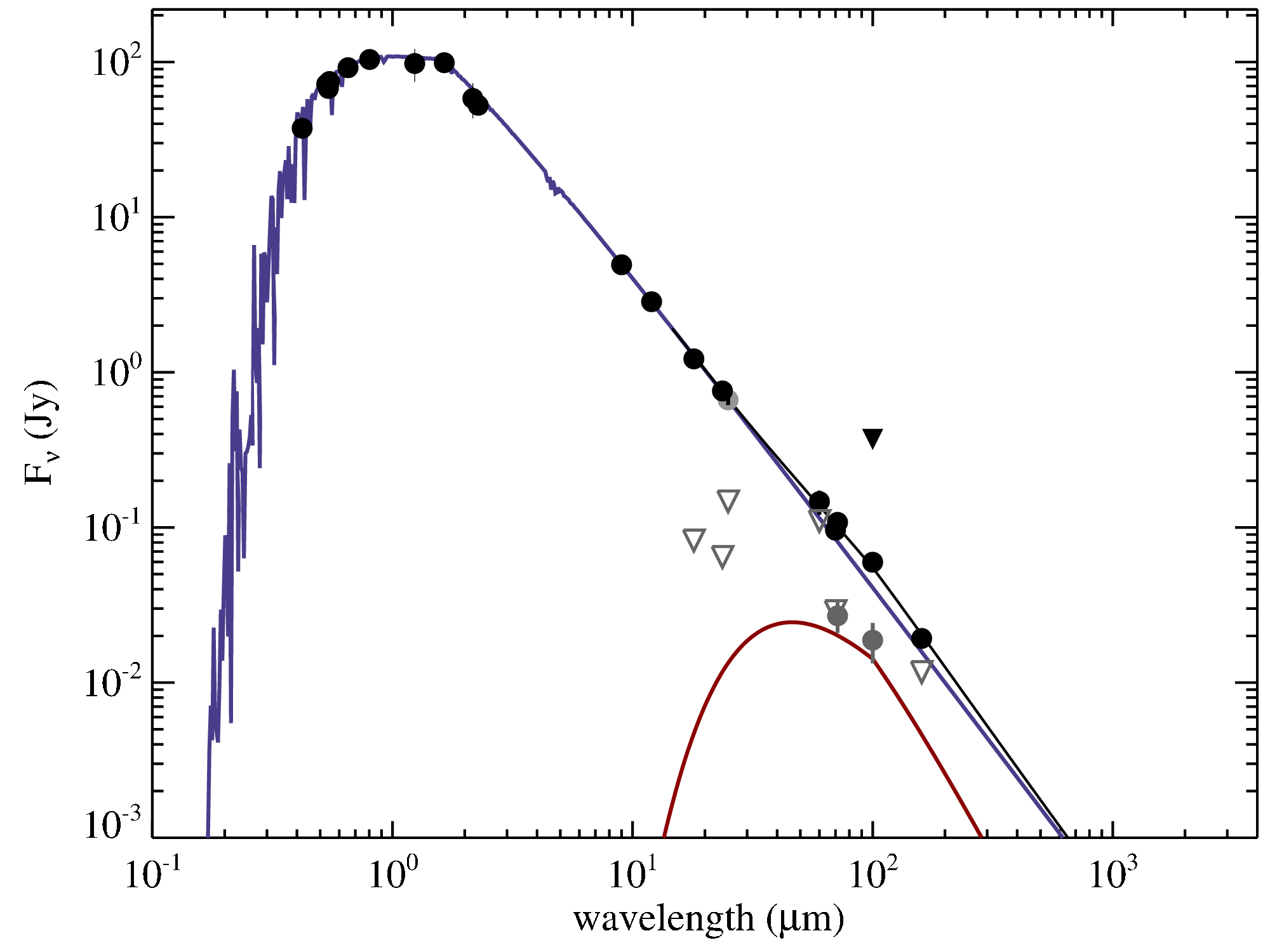,height=1.7in} &
      \hspace{-0.1in} \psfig{figure=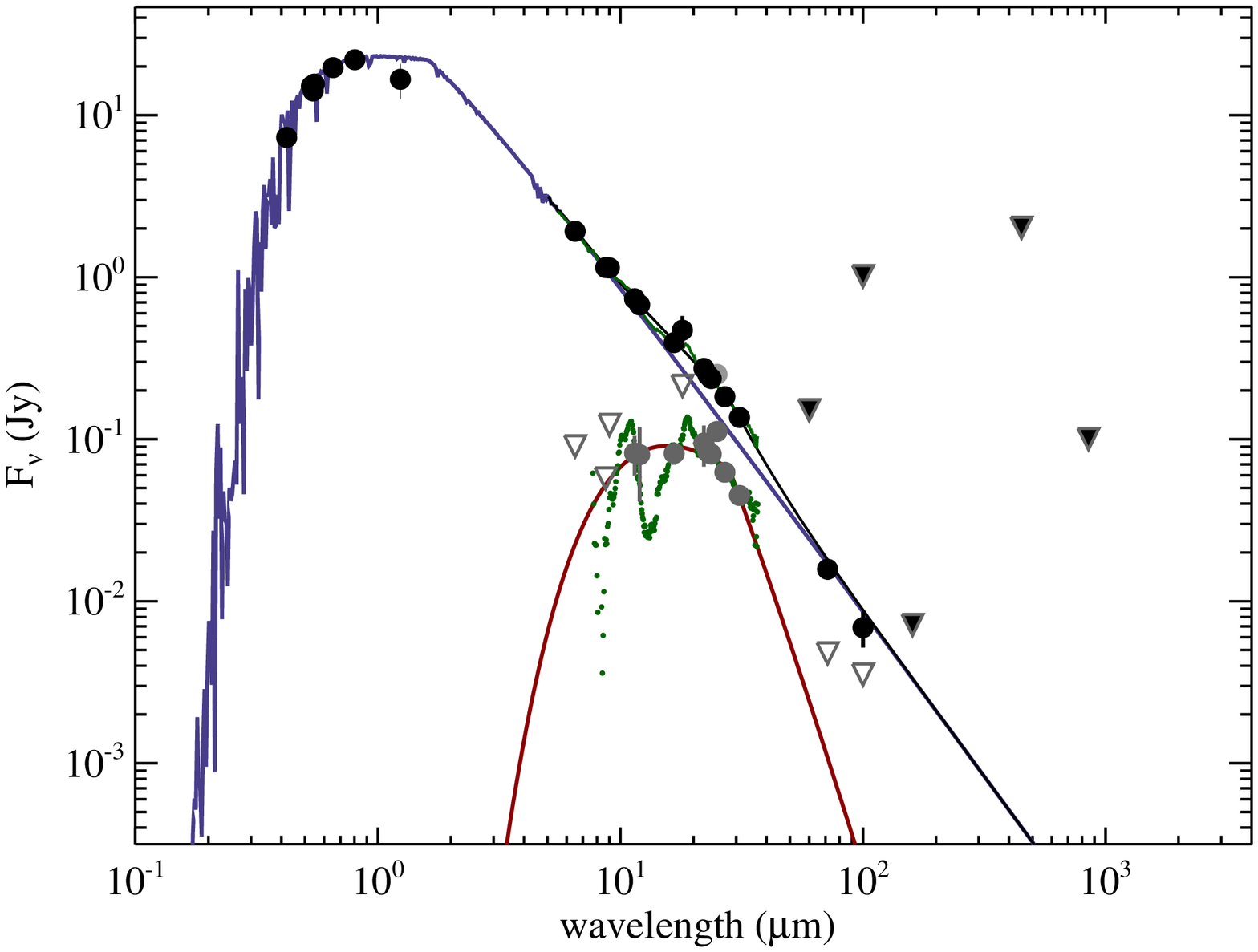,height=1.7in} &
      \hspace{-0.1in} \psfig{figure=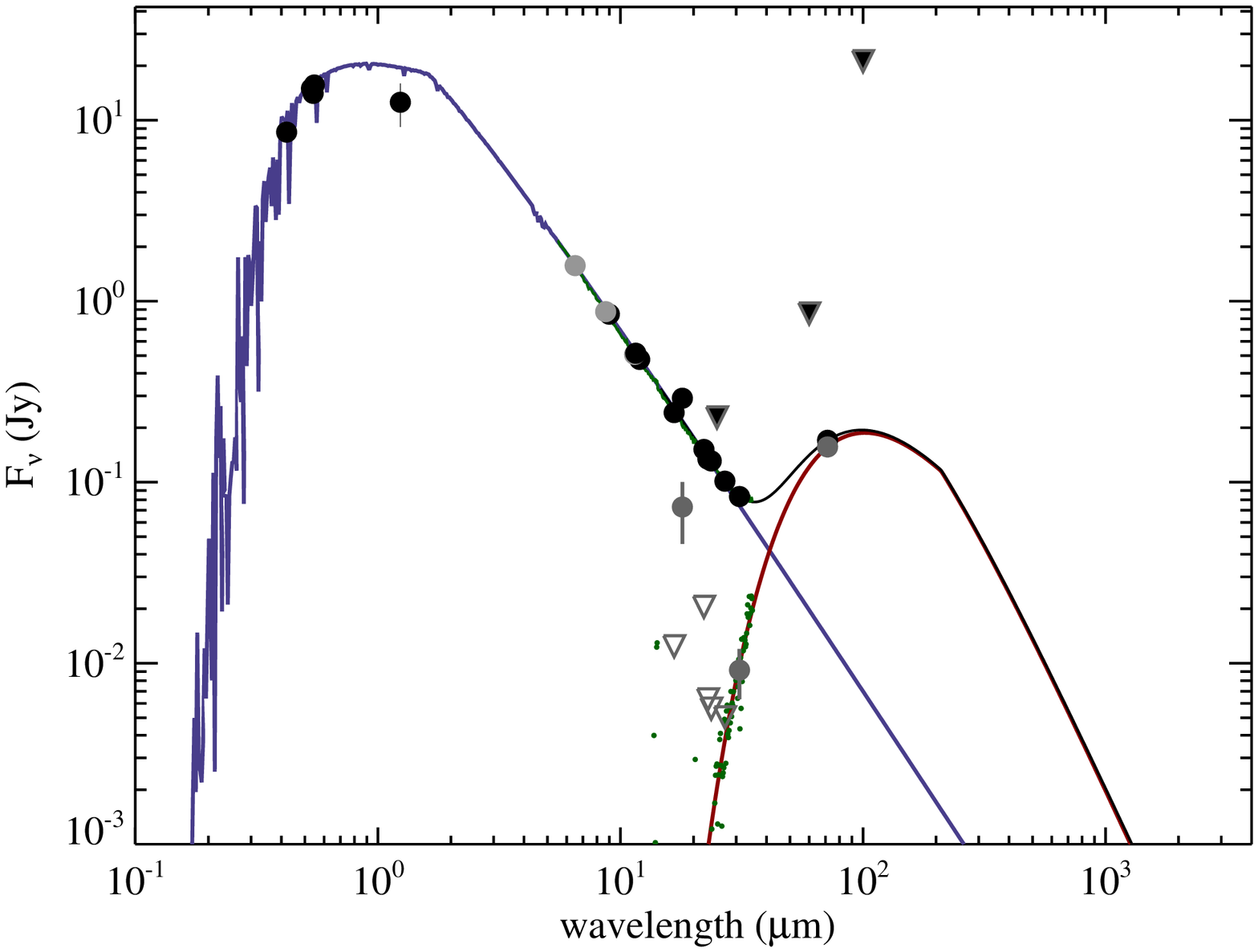,height=1.7in} 
    \end{tabular}
    \caption{
    Spectral energy distributions for low-mass planet systems with debris:
    (left) HD20794, (middle) HD69830, (right) HD38858.
    Total fluxes are black points, open circles are predicted photospheric fluxes,
    grey points are photosphere subtracted fluxes.
    The photospheric model is shown with a blue line, and is a fit to data at wavelengths
    short enough that no excess is present.
    Longer wavelength fluxes are fitted as a modified black body at the temperatures
    given in the text which is shown with a red line.
    The total flux from the models (photosphere + disk) is shown with a black solid
    line.
    }
   \label{fig:moreseds}
  \end{center}
\end{figure*}

%%%%%%%%%%%%%%%%%%%%%%%%%%%%%%%%%%%%%%%%%%%%%%
\subsubsection{HD20794}
\label{sss:20794}
At 6.0pc the G8V star HD20794 is the 5th nearest G star to the Sun;
its age deduced from chromospheric Ca II H\&K is 6.2Gyr with 50\% uncertainty (Vican 2012).
A non-detection with Spitzer was reported in Beichman et al. (2006), but a reanalysis
of Spitzer data, and a remodelling of the stellar photosphere, shows that at
70~$\mu$m the observed flux of $106.7 \pm 5.8$~mJy is in fact significantly higher than
the photospheric level of 80.9~mJy.
The PACS observations also show a marginal excess at 70~$\mu$m of $97 \pm 5.5$mJy
(the photosphere is 84.2mJy in this bandpass), and
a significant excess at 100~$\mu$m of $59.0 \pm 4.8$~mJy (compared with just
$41.0$~mJy of photospheric flux), both measured with a 15 arcsec aperture.
Although only marginal levels of excess were measured with MIPS at 24~$\mu$m ($758.8 \pm 7.6$~mJy 
compared with a 741.5~mJy photosphere) and with PACS at 160~$\mu$m
($20.6 \pm 2.8$~mJy from a PSF fit to the image, compared with a 15.8~mJy photosphere),
these observations provide additional constraints that help pin down the temperature of
the excess to $110 \pm 35$~K with a fractional luminosity of $4.3 \times 10^{-6}$
(see Fig.~\ref{fig:moreseds})\footnote{Note that the marginal excesses are plotted as upper
limits on Fig.~\ref{fig:moreseds}, but explain why the model fit for the excess flux on that
figure goes slightly under the two significant detections.}.
The stellar luminosity of $0.66L_\odot$ puts the dust at 5.2AU if it emits like a black body,
but we estimate its location using the same $f_{\rm{T}}=1.9$ factor derived for 61 Vir to predict
its location to be around 19AU, noting that the uncertainty in the temperature, as well as that
inherent in converting this to a radial location, means this location
cannot be considered well constrained until the disk has been
resolved.\footnote{The 70 and 100~$\mu$m PACS observations show tentative evidence for
extension at around the predicted scale, which is why fluxes were derived using an aperture
at these wavelengths.}

%%%%%%%%%%%%%%%%%%%%%%%%%%%%%%%%%%%%%%%%%%%%%%
\subsubsection{HD102365}
\label{sss:102365}
At 9.2pc the G2V star HD102365 is the 12th nearest G star to the Sun;
its age deduced from chromospheric Ca II H\&K is 5.7Gyr (Vican 2012).
The previous non-detection with Spitzer (Beichman et al. 2006) is confirmed in our
reanalysis of the data at 70~$\mu$m ($47.5 \pm 4.5$~mJy observed for 40.1~mJy
photosphere) and 24~$\mu$m ($363.1 \pm 3.6$~mJy observed for 366.9 photosphere).
This is further strengthened by a non-detection of excess in PACS images
at both 100~$\mu$m ($15.9 \pm 2.0$~mJy observed for $20.3$~mJy photosphere) and
160~$\mu$m ($3.0 \pm 4.4$~mJy observed for 7.8~mJy photosphere), where
both measurements were made by fitting template PSFs to the images.
This star has a binary companion at 23" projected separation, corresponding to an
expected semimajor axis of 297AU (Allen, Poveda \& Herrera 2000).
We fit the photometry available for the companion to estimate an effective temperature
of 2699K and note that its 0.7~mJy photosphere is not detected in the 100~$\mu$m
maps ($-1.7 \pm 2.0$~mJy was measured at the binary location).

%%%%%%%%%%%%%%%%%%%%%%%%%%%%%%%%%%%%%%%%%%%%%%
\subsubsection{HD69830}
\label{sss:69830}
At 12.6pc the G8V star HD69830 is the 22nd nearest G star to the Sun;
its age deduced from chromospheric Ca II H\&K is 5.7Gyr (Vican 2012).
The excess toward this star has been analysed in great detail, since its infrared
spectrum contains numerous spectral features that provide valuable information on the
composition of the dust.
The reader is referred to Beichman et al. (2011) for more details of the observations,
and here we summarise that the dust temperature puts it at $\sim 1$AU, a location
consistent with high resolution mid-IR imaging and interferometry (Smith, Wyatt \& Haniff
2009), and that its composition is similar to that of main-belt C-type asteroids in the Solar
System.
This location is at odds with the age of the star, since steady state collisional erosion
of a planetesimal belt at 1AU should have reduced the mass of such a belt to levels far
below that observed (Wyatt et al. 2007a).
This leads to the conclusion that the dust must be either transient (e.g.,
the product of a recent collision), or that it is replenished from a more
distant planetesimal belt where longer collisional timescales mean that significant mass can
survive (Wyatt et al. 2007a; Heng 2011).
Far-IR observations have so far failed to find evidence for any outer belt
(Beichman et al. 2011).

%%%%%%%%%%%%%%%%%%%%%%%%%%%%%%%%%%%%%%%%%%%%%%
\subsubsection{HD136352}
\label{sss:136352}
The G2V star HD136352 is 14.8pc from the Sun;
its age deduced from chromospheric Ca II H\&K is 6.4Gyr (L. Vican, personal communication).
No excess was reported toward this star with Spitzer at 70~$\mu$m ($17.5 \pm 5.1$~mJy
observed for 19.9~mJy photosphere; Beichman et al. 2006).
This star was not observed by Herschel as part of the DEBRIS survey because of the high background
level (Phillips et al. 2010).

%%%%%%%%%%%%%%%%%%%%%%%%%%%%%%%%%%%%%%%%%%%%%%
\subsubsection{HD38858}
\label{sss:38858}
The G2V star HD38858 is 15.2pc from the Sun;
its age deduced from chromospheric Ca II H\&K is 6.2Gyr (L. Vican, personal communication).
Excess at 70~$\mu$m was detected with Spitzer ($153.7 \pm 9.8$~mJy observed for
15.0~mJy photosphere; Beichman et al. 2006), and later excess in the mid-IR spectrum
was measured allowing the temperature to be determined and a dust location estimated
at 12-29AU (Lawler et al. 2009).
Like HD136352, this star was not observed by Herschel as part of the DEBRIS survey because
of the high background level (Phillips et al. 2010). 
For consistency with our interpretation of other sources, we repeat the analysis of its excess
spectrum (see Fig.~\ref{fig:moreseds}), and use the inferred dust temperature of 50K to
estimate a location from this $0.83L_\odot$ star as 102AU (again using the $f_{\rm{T}}=1.9$
factor inferred for 61 Vir).
This size is consistent with the marginal extension of the MIPS 70~$\mu$m emission that can
be fitted by a ring of size around 100~AU size (section 5.2.3 of Lawler et al. 2009).

%%%%%%%%%%%%%%%%%%%%%%%%%%%%%%%%%%%%%%%%%%%%%%
\subsubsection{Summary}
\label{sss:summ}
The detection rate of 4/6 is much higher than that seen in the background population.
Although the analysis included observations with Herschel, in fact all of the
detections were present in the Spitzer data, and so it may be reasonable to compare
this detection rate with the 15\% of Sun-like stars found to have excess by Spitzer.
Poisson statistics shows that detecting $\geq 4$ from a sample of 6 for an expected
detection rate of 15\% is a 1.4\% probability event.
If instead the 4/6 rate was compared with the detection rate of disks found by Herschel
in an unbiased population of Sun-like stars\footnote{For example, 38 of the nearest
60 G star sample have been observed by Herschel as part of either the DEBRIS or DUNES key
programmes (Matthews et al. 2010; Eiroa et al. 2010), of which 8 have disk detections.},
which is around 20-25\%, a rate this high would have been up to a 6\% probability event
(for a 25\% detection rate).
Thus, although this analysis is necessarily subject to small number statistics, 
it appears to be very unlikely that the low-mass planet sample has as low a detection
fraction as other Sun-like stars.
As one of the two low-mass planet systems without detected debris is a wide binary system,
the high disk fraction would be further enhanced if the sample was restricted to single
low-mass planet systems, though having a wide binary companion does not necessarily
preclude the presence of debris (Trilling et al. 2007; Rodriguez \& Zuckerman 2012).

Note that we are only claiming that the debris incidence rate is higher around stars with
low-mass planets (that are detectable in current radial velocity surveys), not that all
low-mass planet systems have detectable debris.
Indeed this appears unlikely given that low-mass planetary systems may be present around
30-50\% of stars (Mayor et al. 2011), and that only 15\% of stars have debris that was detectable
with Spitzer.
Also we cannot comment on any correlation with planets closer to Earth-mass
that would be below the detection threshold of radial velocity surveys (e.g.,
Fig.~\ref{fig:harps}), though such planets may be common
(e.g., Melis et al. 2010; Borucki et al. 2011).

%%%%%%%%%%%%%%%%%%%%%%%%%%%%%%%%%%%%%%%%%%%%%%
\begin{figure}
  \begin{center}
    \begin{tabular}{c}
      \psfig{figure=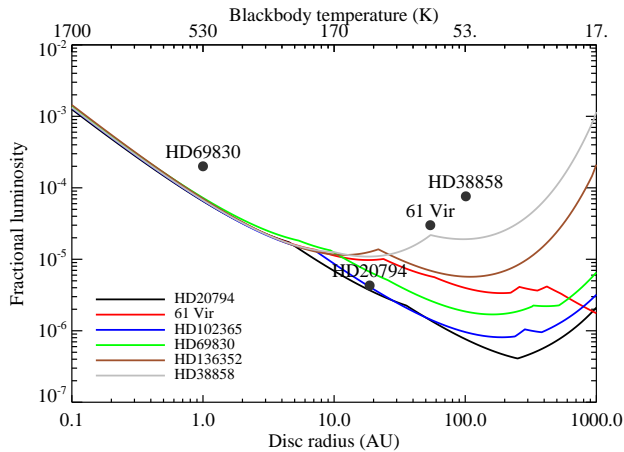,height=2.4in}
    \end{tabular}
    \caption{Parameter space for which excess emission from a debris disk
    would have been discovered in existing observations of the stars in the
    low-mass planet sample.
    The limits are derived by assuming that the star has a disk with a
    given fractional luminosity ($L_{\rm{ir}}/L_\star$) and temperature (which is
    translated into disk radius using $f_{\rm{T}}=1.9$), and
    determining whether that emission would have been detectable in the observations
    at any wavelength at which the star was observed using eq. 8 and 11 of Wyatt (2008)
    with the modification of Lestrade et al. (2009) for disks extended beyond the
    beam.
    The limits at different disk temperatures are dominated by different wavelengths
    (e.g., temperatures above 170K are limited by MIPS 24~$\mu$m observations, with
    longer wavelengths constraining lower temperatures).
    The limits for different stars in the sample are plotted with different colours,
    and the black dots denote the parameters derived for the disks detected around
    four of the sample.}
   \label{fig:detlims}
  \end{center}
\end{figure}

Likewise it is important to clarify that a detectable debris disk in this context
implies that the fractional luminosity and disk temperature lie above the appropriate
lines on Fig.~\ref{fig:detlims}.
Although the detection thresholds are similar for the different stars in the low-mass
planet sample, there is a dispersion in sensitivity to large cold disks ($<100$K), and it should be
acknowledged that a disk with the same fractional luminosity and temperature as that of
HD20794 would not have been detectable around all of the sample.
Nevertheless, the same would be true of disks that could have been detected around the rest
of the nearest 60 G star sample.

We defer a detailed consideration of the planet correlation to a later paper
(Moro-Mart\'{i}n et al., in prep.), but note that the discovery of
a debris disk around an M star with low-mass planets could suggest that this trend also
continues to lower mass stars (Lestrade et al., in prep.).

%%%%%%%%%%%%%%%%%%%%%%%%%%%%%%%%%%%%%%%%%%%%%%
%%%%%%%%%%%%%%%%%%%%%%%%%%%%%%%%%%%%%%%%%%%%%%
\subsection{Implications for planet formation}
\label{ss:plform}
The correlation summarised in \S \ref{sss:summ} suggests that, even if not
directly connected, the formation of a system that was conducive to the formation of
low-mass planets was also conducive to large quantities of planetesimals remaining
in the system after several Gyr of evolution.
This may be connected to a different correlation, in which stars with high-mass
planets tend to be metal-rich whereas those with low-mass planets have no bias in
metallicity (Sousa et al. 2011), and neither do stars that host debris disks
(Greaves et al. 2006; Maldonado et al. 2012).

Ideally the debris disks in low-mass planetary systems would share some characteristics
that could both point to a reason for the correlation between these phenomena,
and help identify systems as having low-mass planets through study of their debris disks.
However, Table~\ref{tab:g60} shows that debris disks in low-mass planetary systems have
a range of radii (1-350AU) that spans the range of known debris disks.
Thus, planetesimals in low-mass planetary systems
could either be at large distances from the planets (like 61 Vir, HD20794 and HD38858),
or some of those planetesimals could be in close proximity to the planets, at least at
some point in their evolution (like HD69830).

Why might such a correlation exist?
Here we consider the two scenarios that have been proposed for the
formation of low-mass planet systems:
(i) that the planets formed further out but then migrated
in, perhaps gaining mass as they migrate (Alibert et al. 2006, 
Terquem \& Papaloizou 2007, Ida \& Lin 2010), or
(ii) that they formed in situ in which case significant solid mass ($50-100M_\oplus$)
would have to be concentrated within 1AU (Hansen \& Murray 2011).

Although not without issues in comparison with exoplanet statistics
(McNeil \& Nelson 2010), formation scenario (i) seems to be perfectly suited to
leaving large quantities of outer debris.
Payne et al. (2009) showed that the migration of the planets through a
planetesimal disk is an inefficient process, in the sense that not all of the
planetesimals are accreted or ejected by the migrating planets.
Rather, many planetesimals are left stranded in the region through which the planets
recently migrated.
As long as there were no more distant planets that subsequently migrated in thus
depleting this region, these planetesimals could be dynamically stable over long
timescales, but would predominantly deplete through mutual collisions.

Formation scenario (i) could also be applied to the formation of close-in massive
planets.
However, in this case the stronger gravitational perturbations of the migrating planets
would result in the ejection or accretion of a larger fraction of planetesimals
(though not necessarily all of them, e.g., Raymond et al. 2006).
Moreover the migration of multiple high-mass planets may result in a dynamically
unstable configuration, unless this can be prevented by resonance locking, in
which case the ensuing instability could deplete any remaining planetesimals
(e.g., Gomes et al. 2005; Raymond et al. 2011).

Formation scenario (ii) may also be conducive to forming an outer planetesimal
belt, depending on how efficiently the solid mass is accreted onto planets.
If even a small fraction of the planetesimal mass is scattered onto highly
eccentric orbits, it would form a mini-scattered disk of planetesimals with
pericentres in the vicinity of the planets but apocentres extending out
to large distances.
If this happened during the protoplanetary disk phase, interaction with solid
or gaseous material in the outer planetary system could then pull the
pericentres of scattered planetesimals away from the planets (i.e., through gas
drag or dynamical friction).
This could prevent the planetesimals from being ejected and leave them in
a long-lived extended disk configuration.

Again, application of scenario (ii) to the formation of massive planets would
be less favourable to the formation of massive outer debris disks.
This is because planetesimals are more readily ejected by more massive planets
(e.g., Tremaine 1993), on timescales that would be shorter compared with the
timescale for the outer disk to decouple the planetesimals from the planets'
gravitational influence.

It thus seems plausible that high levels of outer debris could be a natural
outcome of systems that form only low-mass planets.
%However, the inclusion in this sample of a system with
%dust at 1AU (HD69830) merits further discussion,
%since such hot dust is an extremely rare (2-4\%), and potentially
%transient, occurrence amongst field stars (e.g., Trilling et al. 2008).
%It could be that the debris seen toward HD69830 is unrelated to its planets,
%since if transient HD69830-like events occurred independently of planetary system
%architecture, we would have a 10-20\% chance of finding one in our sample.
%However, the close proximity of the dust to the planets suggests otherwise.
%As mentioned in \S \ref{sss:69830}, various explanations have been proposed for
%the origin of this hot dust, and here we simply note that the correlation discussed
%in \S \ref{sss:summ} emphasises that the known planets are likely to play a key role
%in a successful interpretation of this system.
%If this is interpreted as a transient event, a detection rate of 1/6
%suggests that these rare events would have to be observable for a relatively
%large fraction of the star's life for us to be likely to witness one (small
%number statistics notwithstanding).
%An alternative interpretation is that the dust is fed from an outer disk,
%though the lack of cold dust is not promising (Booth et al. 2009), suggesting that
%any outer planetary system would have to be particularly efficient at
%scattering planetesimals inward. 

%%%%%%%%%%%%%%%%%%%%%%%%%%%%%%%%%%%%%%%%%%%%%%
%%%%%%%%%%%%%%%%%%%%%%%%%%%%%%%%%%%%%%%%%%%%%%
%%%%%%%%%%%%%%%%%%%%%%%%%%%%%%%%%%%%%%%%%%%%%%
\section{Conclusions}
\label{s:conc}
Observations from the DEBRIS Key Programme are presented that resolve the
structure of the debris disk around the exoplanet host star 61 Vir
(\S \ref{s:obs}).
Modelling shows that the dust extends from 30AU out to at least 100AU in a nearly edge-on
configuration (\S \ref{s:mod}).
There is likely little interaction between the disk and the known planets,
which are at $<0.5$AU.
The lack of planetesimals in the $<30$AU region could be explained by the existence
of planets in this region, but the depletion can also be explained
by collisional erosion (\S \ref{s:pl}).
Considering a sample of the nearest 60 G stars there is an emerging trend
that stars which, like 61 Vir, only have low-mass planets,
are more likely to have detectable debris (\S \ref{s:stat}).
We attribute this trend to the fact that the formation processes that make low-mass
planets are likely to also result in large quantities of distant debris.

%%%%%%%%%%%%%%%%%%%%%%%%%%%%%%%%%%%%%%%%%%%%%%%%%%%%%%
%%%%%%%%%%%%%%%%%%%%%%%%%%%%%%%%%%%%%%%%%%%%%%%%%%%%%%
\section*{Acknowledgments}
The authors are grateful to Ben Zuckerman for helpful comments on the paper.
This work was supported by the European Union through ERC grant number 279973.

%%%%%%%%%%%%%%%%%%%%%%%%%%%%%%%%%%%%%%%%%%%%%%
%%%%%%%%%%%%%%%%%%%%%%%%%%%%%%%%%%%%%%%%%%%%%%
%%%%%%%%%%%%%%%%%%%%%%%%%%%%%%%%%%%%%%%%%%%%%%%%%%%%%%

\end{document}